\documentclass[aps,floats,prd,twocolumn,preprintnumbers,nofootinbib,superscriptaddress,longbibliography]{revtex4-1}

\usepackage{amssymb}
\usepackage{amsmath}

\usepackage{graphicx}
\usepackage{graphics}
\usepackage{dcolumn}
\usepackage[dvipsnames]{xcolor}
\usepackage{fancyhdr} 
\usepackage{graphicx}
\usepackage{float}
\usepackage{multirow}

\usepackage{subfig}
\usepackage[colorlinks=true,linktocpage=true,urlcolor=blue,citecolor=blue,linkcolor=blue]{hyperref}

\usepackage[utf8]{inputenc}

\usepackage{dsfont}

\usepackage[justification=centerlast, format=plain, labelfont=bf]{caption}

\makeatletter
\renewcommand\tableofcontents{%
    \@starttoc{toc}%
}
\makeatother

\DeclareCaptionJustification{justified}{\leftskip=0pt \rightskip=0pt \parfillskip=0pt plus 1fil}

    \newcommand{\be}{\begin{equation}}
  \newcommand{\ee}{\end{equation}}
    \newcommand{\ba}{\begin{align}}
  \newcommand{\ea}{\end{align}}

\newcommand{\Msun}{M_{\odot}}
\newcommand{\Mpcinv}{ {\rm Mpc}^{-1} }
\newcommand{\fNL}{ f_{\rm NL} }

\makeatletter
\def\doauthor#1#2#3{%
  \ignorespaces#1\unskip
  \begingroup
   #3%
  \@if@empty{#2}{\@listcomma\endgroup{}{}}{\endgroup{\comma@space}{}\frontmatter@footnote{#2}}%
  \space \@listand
}%
\makeatother

\makeatletter
\def\@ssect@ltx#1#2#3#4#5#6[#7]#8{%
  \def\H@svsec{\phantomsection}%
  \@tempskipa #5\relax
  \@ifdim{\@tempskipa>\z@}{%
    \begingroup
      \interlinepenalty \@M
      #6{%
       \@ifundefined{@hangfroms@#1}{\@hang@froms}{\csname @hangfroms@#1\endcsname}%
       {\hskip#3\relax\H@svsec}{#8}%
      }%
      \@@par
    \endgroup
    \@ifundefined{#1smark}{\@gobble}{\csname #1smark\endcsname}{#7}%
  }{%
    \def\@svsechd{%
      #6{%
       \@ifundefined{@runin@tos@#1}{\@runin@tos}{\csname @runin@tos@#1\endcsname}%
       {\hskip#3\relax\H@svsec}{#8}%
      }%
      \@ifundefined{#1smark}{\@gobble}{\csname #1smark\endcsname}{#7}%
      \addcontentsline{toc}{#1}{\protect\numberline{}#8}%
    }%
  }%
  \@xsect{#5}%
}%
\makeatother

\begin{document}

\preprint{KCL-2020-41}

\title{First Constraints on Small-Scale Non-Gaussianity from\\ UV Galaxy Luminosity Functions}

\author{Nashwan Sabti$^{\mathds{S},}$}
\affiliation{Department of Physics, King's College London, Strand, London WC2R 2LS, UK}
\author{Julian B. Mu\~{n}oz$^{\mathds{M},}$}
\affiliation{Harvard-Smithsonian Center for Astrophysics, 60 Garden St., Cambridge, MA 02138, USA}
\affiliation{Department of Physics, Harvard University, 17 Oxford St., Cambridge, MA 02138, USA}
\author{Diego Blas$^{\mathds{B},}$}
\affiliation{Department of Physics, King's College London, Strand, London WC2R 2LS, UK}

\def\thefootnote{$\mathds{S}$\hspace{0.7pt}}\footnotetext{\href{mailto:nashwan.sabti@kcl.ac.uk}{nashwan.sabti@kcl.ac.uk}}
\def\thefootnote{$\mathds{M}$\hspace{-0.9pt}}\footnotetext{\href{mailto:julianmunoz@fas.harvard.edu}{julianmunoz@fas.harvard.edu}}
\def\thefootnote{$\mathds{B}$}\footnotetext{\href{mailto:diego.blas@kcl.ac.uk}{diego.blas@kcl.ac.uk}}
\setcounter{footnote}{0}
\def\thefootnote{\arabic{footnote}}

\begin{abstract}
    \noindent UV luminosity functions provide a wealth of information on the physics of galaxy formation in the early Universe. Given that this probe indirectly tracks the evolution of the mass function of dark matter halos, it has the potential to constrain alternative theories of structure formation. One of such scenarios is the existence of primordial non-Gaussianity at scales beyond those probed by observations of the Cosmic Microwave Background. Through its impact on the halo mass function, such small-scale non-Gaussianity would alter the abundance of galaxies at high redshifts. 
    In this work we present an application of UV luminosity functions as measured by the Hubble Space Telescope to constrain the non-Gaussianity parameter $f_\mathrm{NL}$ for wavenumbers above a cut-off scale $k_{\rm cut}$.
    After marginalizing over the unknown astrophysical parameters and accounting for potential systematic errors, we arrive at a $2\sigma$ bound of $f_{\rm NL}=71^{+426}_{-237}$ for a cut-off scale  $k_{\rm cut}=0.1\,\mathrm{Mpc}^{-1}$ in the bispectrum of the primordial gravitational potential. 
    Moreover, we perform forecasts for the James Webb Space Telescope and the Nancy Grace Roman Space Telescope, finding an expected improvement of a factor $3-4$ upon the current bound.
\end{abstract}

\maketitle

\tableofcontents

\section{Introduction}
\vspace{-0.05cm}

Cosmological surveys over the last few decades have provided us with an unprecedented understanding of the Universe.
These include measurements of the Cosmic Microwave Background (CMB)~\cite{Akrami:2018vks}, as well as of the large-scale structure (LSS) of the Universe~\cite{Abazajian:2008wr, Abbott:2005bi}.
Nevertheless, a large swath of our cosmos, corresponding to the cosmic dawn and reionization eras, remains largely unprobed.
These two eras are the next frontier of precision cosmology.
\vspace{6pt}

Progress has been made by obtaining indirect information on the epoch of reionization (EoR) through its effect on the CMB anisotropies~\cite{Hu:1999vq, Adam:2016hgk}, the spectra of distant quasars~\cite{Barkana:2000fd, Becker:2001ee}, as well as the redshifted 21-cm line~\cite{Morales:2009gs}.
These observables track the transition from a mostly neutral intergalactic medium to an ionized one.
A more direct approach, however, involves observing the redshifted emission of the galaxies at that time.
For this, our main handle is the (rest-frame) UV luminosity function (LF) observed by the {\it Hubble Space Telescope} (HST)~\cite{Bouwens:2014fua,Finkelstein_2015,Atek:2015axa,Livermore:2016mbs,Bouwens_2017asdasd,Mehta_2017,Ishigaki_2018,Oesch_2018,Atek:2018nsc}.
Data collected by the HST over the last decades have provided us with an increasingly detailed galactic census at high redshifts, which has dramatically enhanced our understanding of early stellar formation~\cite{Tacchella:2012ih}. 
Besides providing key insights on the astrophysics of reionization, these LFs open a window towards probing different aspects of our cosmological models.
In particular, the UV LFs probe cosmological {\it small} scales, which are otherwise difficult to access by current data sets.
New features of the fundamental model of our Universe may lie at these scales, e.g.~\cite{Chevallard:2014sxa, Dayal:2014nva, Corasaniti:2016epp, Menci:2017nsr, Yue:2017hbz, Lovell:2017eec, Unal:2018yaa, Irsic:2019iff, Yoshiura:1809192}. The main purpose of this work is to illustrate these exciting possibilities. We do this by exploiting LF observations to learn about the physics of inflation in the form of primordial non-Gaussianity~\cite{Maldacena:2002vr, Celoria:2018euj}.
\newpage

The most accepted paradigm to explain the currently observed features of the Universe is that it went through an inflationary period at early times~\cite{Guth:1980zm, Linde:1981mu, Baumann:2009ds}. This framework is, however, quite broad in terms of determining which fundamental mechanism was actually operating.
A promising strategy to unearth the physics of the inflationary era consists of exploring observables that can differentiate between families of inflationary models, grouped for instance according to effective-field-theory criteria~\cite{Cheung:2007st, Arkani-Hamed:2015bza}.  
A key feature of many non-minimal models is a deviation in the primordial fluctuations from the simplest Gaussian prediction, 
a feature known as
primordial non-Gaussianity (PNG, see e.g.~\cite{Biagetti:2019bnp} for a recent review). 
This PNG can be scale dependent, for instance in models in which there is a relevant scale during the inflationary period.  
{\it Scale-dependent non-Gaussianity} is thus a powerful probe into the physics of the primordial Universe~\cite{Verde:2000vr, Komatsu:2009kd, Byrnes:2010ft}.
\vspace{6pt}

A departure from Gaussianity in the primordial fluctuations alters the abundance of halos, and thus the UV LF measured by the HST. In particular, local-type PNG has been shown to affect the rarest objects (such as the heaviest halos), as they lie the furthest from the peak of the distribution of overdensities (see e.g.~\cite{Pillepich:2008ka} and references therein). It is in this region where deviations from Gaussianity would be more apparent. 
This makes galaxy clusters a good probe of local-type PNG in the local ($z\sim 0$) Universe~\cite{Mana:2013qba, LoVerde:2007ri,Jimenez:2009us,LoVerde:2011iz,Shandera:2012ke}.
Interestingly, however, the galaxies that the HST observes are hosted in halos which were very rare at their own redshift. This is because, despite their lower overall mass (thousands of times smaller than those of clusters today), they corresponded to large overdensities due to the smaller size of matter fluctuations at that time. 
Here we show that this makes the HST UV LFs a powerful probe of local-type PNG, enabling us to search for it at scales corresponding to wavenumbers $k\gtrsim 0.1\, \Mpcinv$, which are difficult to access by CMB~\cite{Akrami:2019izv} and LSS observations~\cite{Shirasaki:2012sx,Leistedt:2014zqa}.
PNG at even smaller scales can be accessed through other probes, for instance, through spectral distortions of the CMB anisotropies~\cite{Naruko:2015pva,Emami:2015xqa,Khatri:2015tla,Cabass:2018jgj} (although current bounds are at the level of $f_{\rm NL}\lesssim 10^5$ for scale-independent PNG).
\vspace{6pt}

\begin{figure}[h!]
    \centering
    \includegraphics[width=\linewidth]{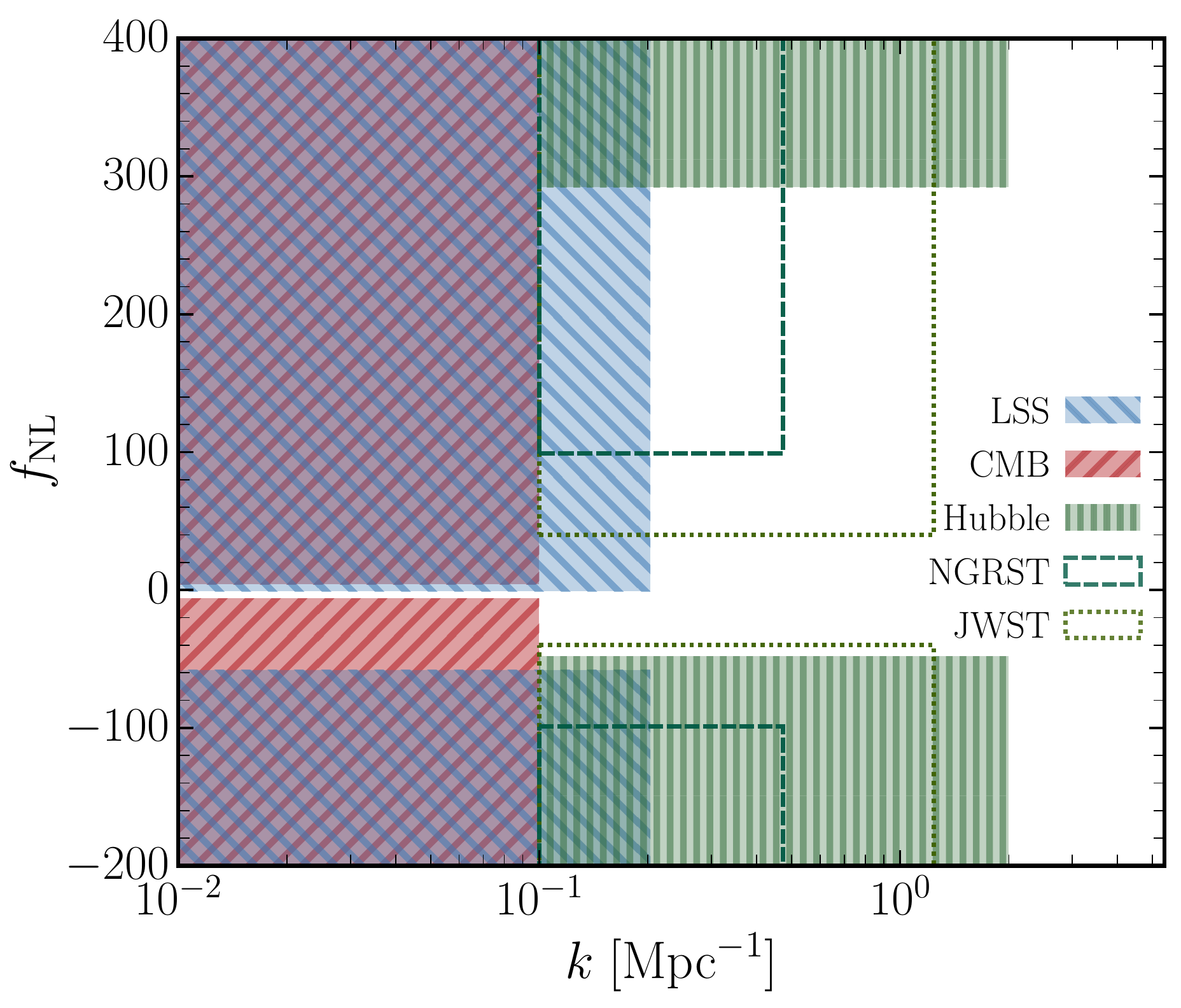}
    \caption{Illustration of the current $1\sigma$ constraints on $f_\mathrm{NL}$ as a function of comoving wavenumber $k$ from LSS~\cite{Castorina:2019wmr}, CMB~\cite{Akrami:2019izv} and LF observations, together with forecasts for JWST and NGRST. The smallest scale in our bounds, $k_\mathrm{max}\sim 2\,\mathrm{Mpc}^{-1}$, corresponds to the smallest halo mass probed by the Hubble fields.
    We set the cut-offs for the LSS and CMB observables following~\cite{LoVerde:2007ri,Castorina:2019wmr}, although we note that these are approximate and for illustration purposes only. The forecasts for JWST and NGRST are based on a wide-field and high-latitude survey mode respectively (see Section~\ref{sec:forecasts}). While forecasts for 21-cm experiments are not included, they possibly have the ability to  reach scales $k\sim 50\,\mathrm{Mpc}^{-1}$ (see Section~\ref{subsec:21cm_forecast}).}
    \label{fig:current_status}
\end{figure}

In our main analysis we use the LFs from the Hubble Legacy Fields (HLF) catalog~\cite{Bouwens:2014fua}. In particular, we cover the redshift range $z=4-8$ and rest-frame UV magnitudes $M_{\rm UV}$ between $-22.7$ and $-16.4$ to find constraints on the amplitude $f_{\rm NL}$ of primordial non-Gaussianities at small scales $k > k_\mathrm{cut} = 0.1\,\mathrm{Mpc}^{-1}$.  
We fit a semi-analytical model to the shape of the UV LFs based on that of~\cite{Gillet:2019fjd} and use corrections to the halo mass function induced by primordial non-Gaussianity. By accounting for possible systematic errors in the UV LF data and marginalizing over the astrophysical parameters in our model, we find a bound of $f_{\rm NL}=71^{+426}_{-237}$ at $2\sigma$ for $k_{\rm cut}=0.1\,\Mpcinv$. This is the first constraint on primordial non-Gaussianities from LF data and covers smaller scales than currently probed, as illustrated in Figure~\ref{fig:current_status}.
Our approach is complementary to forecasts using future CMB spectral distortion data~\cite{Emami:2015xqa, Dimastrogiovanni:2016aul}, as well as those proposed for observations of fast radio bursts~\cite{Reischke:2020cgd}.
As a cross-check, we have derived constraints using different UV LFs, including those from the lensing-based Hubble Frontier Fields (HFF)~\cite{Atek:2015axa,Livermore:2016mbs,Bouwens_2017asdasd,Ishigaki_2018,Oesch_2018,Atek:2018nsc}, where we find comparable results. 
Moreover, we perform forecasts for the upcoming {\it James Webb Space Telescope} (JWST) and {\it Nancy Grace Roman Space Telescope} (NGRST), showing that they will improve upon our HST constraints by a factor of $3-4$.

\vspace{6pt}

In what follows, we will assume a cosmological model with base parameters as measured by Planck~\cite{Aghanim:2018eyx}: $h = 0.6727,\, \Omega_\mathrm{b}h^2 = 0.02236,\, \Omega_\mathrm{c}h^2 = 0.1202,\, n_\mathrm{s} = 0.9649,\, \tau = 0.0544,\, A_\mathrm{s} = 2.101\times 10^{-9}\ \mathrm{and}\ k_\mathrm{pivot}=0.05\,\mathrm{Mpc}^{-1}$.
This paper is structured as follows: Section~\ref{sec:UVLF} lays out our semi-analytical model for the UV LF and the HST data used in the analysis. In Section~\ref{sec:primordial_nonGauss} we summarize the formalism of small-scale non-Gaussianity and its impact on the UV LF. In Section~\ref{sec:results} the results of this work are presented. In Section~\ref{sec:forecasts} we make forecasts for JWST and NGRST. Finally, we present our conclusions in Section~\ref{sec:conclusions}. Complementary details are included in the Appendices~\ref{app:comparison_previous_literature}
$-$\ref{subsec:21cm_forecast}.

\section{The UV Luminosity Function}
\label{sec:UVLF}

\subsection{UV LF Model}
\label{subsec:LF_definitions}

The abundance of galaxies in the early Universe can be tracked through their luminosity function, which describes the relation between the observed number density of galaxies and their flux (or magnitude) in a particular band.
In the early Universe, galaxies contain young stars that emit in the UV part of the spectrum. 
This radiation gets redshifted due to the expansion of the Universe and can be observed today with optical or IR-band telescopes, such as the HST. 
An interesting application of the UV emission is to track the star-formation rate (SFR) across the cosmos~\cite{Kennicutt:1998zb, Robertson:2010an}.

We are interested in the UV LF of galaxies around the epoch of reionization.
In order to employ HST data, we ought to model the abundance of such galaxies, as well as their properties.
This process can be separated into two parts.
The first is the halo mass function, which describes how many halos of each mass there are, and is chiefly influenced by cosmology.
The second is the halo-galaxy connection, driven by astrophysical processes, which allows us to relate the halo mass to the observed emission.
While different parts of this calculation can be directly simulated (see e.g.~\cite{2016MNRAS.462..235L,Tacchella:2018qny,Yung_2018}), throughout this work we will use a semi-analytical numerical method based on simulation results.

\vspace{-0.1cm}
\subsubsection{Halo Mass Function}

Massive, high-luminous galaxies tend to be hosted by heavy halos. While massive halos are more likely able to form such galaxies, they are more rarely found than lower mass halos. The abundance of halos has been extensively studied in the literature, e.g.~\cite{Jenkins:2000bv, Reed:2006rw}.
Here we follow the excursion-set approach based on ellipsoidal gravitational collapse of dark matter halos (which results in a better agreement with numerical simulations than spherical-collapse models). In such models the barrier height, i.e., the threshold above which a density perturbation will collapse, depends on the mass of the object. We adapt the collapse formalism developed by Sheth\,\&\,Tormen~\cite{Sheth:2001dp}, where the halo mass function is of the following form:
\begin{align}
    \label{eq:ST_HMF}
    \frac{\mathrm{d}n}{\mathrm{d}M_\mathrm{h}} = \frac{\rho_\mathrm{m}}{M_\mathrm{h}}\frac{\mathrm{d}\ln\sigma^{-1}}{\mathrm{d}M_\mathrm{h}}f_\mathrm{ST}\ ,
\end{align}
with 
\begin{align}
    f_\mathrm{ST} = & A_\mathrm{ST}\sqrt{\frac{2a_\mathrm{ST}}{\pi}}\left[1+\left(\frac{\sigma_M^2}{a_\mathrm{ST}\delta_\mathrm{ST}^2}\right)^{p_\mathrm{ST}}\right]\frac{\delta_\mathrm{ST}}{\sigma_M}\times\nonumber\\
    &\times\exp\left(-\frac{a_\mathrm{ST}\delta_\mathrm{ST}^2}{2\sigma_M^2}\right)\ ,
\end{align}
and where $A_\mathrm{ST} = 0.3222$, $a_\mathrm{ST} = 0.707$, $p_\mathrm{ST} = 0.3$, $\delta_\mathrm{ST} = 1.686$ and $\sigma_M$ is the root mean square of the density field smoothed over a mass scale $M$ (see Eq.~\eqref{eq:sigmasq_M}).

\subsubsection{Halo-galaxy Connection}

We follow a simple phenomenological approach to link host halos to the properties of galaxies that reside in them.
We assume that each dark-matter halo hosts one galaxy on average (see, e.g.,~\cite{Wechsler:2018pic} for a detailed review on the halo occupation distribution).
The efficiency at which this galaxy will form stars depends on the mass of the host halo and is expected to exhibit a peak at halo masses $10^{11}-10^{12}\, M_\odot$ (at $z = 4$)~\cite{Tacchella:2018qny}, similar to that of our own Milky Way. A simple analytic model that captures this behaviour relates the mass of the host halo $M_\mathrm{h}$ to the typical stellar mass $M_*$ inside the halo via a double power-law\footnote{Note that the usual power-law is expressed in terms of $M_\mathrm{h}$, and our expression can be seen as an approximation to the inverse of that function.}:
\begin{align}
    \label{eq:Mh_Mstar_doublepower_approx}
    M_\mathrm{h} = \left(\frac{\epsilon_*M_\mathrm{c}^{\alpha_*}}{M_*}\right)^{\frac{1}{\alpha_*-1}} + \left(\frac{\epsilon_*M_\mathrm{c}^{\beta_*}}{M_*}\right)^{\frac{1}{\beta_*-1}}\ ,
\end{align}
where $\epsilon_*$, $\alpha_*$ and $\beta_*$ are free parameters that we will fit for with data and $M_\mathrm{c} = 1.6\times10^{11}\, M_\odot$.
We take the fitting parameters to be redshift-independent, as suggested by the results of~\cite{Tacchella:2018qny} (see also~\cite{Trenti:2010sz, Sitwell:2013fpa, Mason:2015cna, Yung_2018}). 
We have explicitly tested that varying these parameters independently at each redshift does not change our constraints significantly.
The UV emission is dominated by massive, young stars and thus tracks the SFR ($\dot M_*$), rather than $M_*$.
These two quantities can be related via~\cite{Gillet:2019fjd}:
\begin{align}
    \label{eq:Mstardot_Mstar_relation}
    \dot{M}_* = \frac{M_*}{t_*H^{-1}(z)}\ ,
\end{align}
where $t_*$ is a (dimensionless) parameter that corrects the stellar-formation time-scale with respect to the cosmic Hubble rate $H(z)$. 
While this parameter ought to be fit from data, in practice $t_*$ and $\epsilon_*$ have identical effects on the UV LF, and thus we will fix $t_*$ to unity hereafter without any loss of generality. 
The star formation rate in the rest-frame can be expressed in terms of the UV luminosity $L_\mathrm{UV}$ as~\cite{Sun_2016}:
\begin{align}
    \dot{M}_* = \kappa_\mathrm{UV}L_\mathrm{UV}\ ,
\end{align}
where $\kappa_\mathrm{UV} = 1.15\times 10^{-28}\ M_\odot \, \mathrm{s}\,\mathrm{erg^{-1} yr^{-1}}$ is a conversion factor, and 
\begin{align}
\label{eq:LUV}
\log_{10}\left(\frac{L_\mathrm{UV}}{\mathrm{erg \, s^{-1}}}\right) = 0.4(51.63 + \langle A_\mathrm{UV}\rangle-M_\mathrm{UV})\ ,
\end{align}
with $M_\mathrm{UV}$ the absolute UV magnitude and $\langle A_\mathrm{UV}\rangle$ a dust correction term. The observed UV luminosity can experience significant attenuation by dust extinction, especially at high luminosities and low redshifts~\cite{Yung_2018}. 
We model this extinction following~\cite{Tacchella:2018qny} (similar to the case of Lyman-break galaxies).
For galaxies with a spectrum given by $f\sim\lambda^\beta$, the attenuation is assumed to follow $A_\mathrm{UV} = 4.43 + 1.99\beta$ \cite{Meurer:1999jj,Smit:2012nf}. We use the observations of the $\beta$ parameter at $z \leq 8$ reported in~\cite{Bouwens:2013hxa} and fit it following the prescription in~\cite{Trenti:2014hka}:
\begin{align}
    \langle\beta(z,M_\mathrm{UV})\rangle =
    \begin{cases}
    a(z)e^{-\frac{b(z)}{a(z)}}+c & M_\mathrm{UV}\geq M_0\\
    a(z) + b(z) + c & M_\mathrm{UV} < M_0
    \end{cases}\ ,
\end{align}
where $a(z) = \beta_{M_0}(z)-c$, $b(z) = \frac{\mathrm{d}\beta}{\mathrm{d}M_0}(z)(M_\mathrm{UV}-M_0)$, $c = -2.33$, $M_0 = -19.5$ and the values for $\beta_{M_0}$ and $\mathrm{d}\beta/\mathrm{d}M_0$ are taken from~\cite{Bouwens:2013hxa}. The exponential fit at $M_\mathrm{UV}\geq M_0$ prevents the dust extinction from becoming negative. At any given $M_\mathrm{UV}$ a Gaussian distribution with standard deviation $\sigma_\beta = 0.34$~\cite{Bouwens:2011yy} is assigned to $\beta$, which then gives the desired average extinction~\cite{Smit:2012nf}:
\begin{align}
    \label{eq:dust}
    \langle A_\mathrm{UV}\rangle = 4.43 + 0.79\ln(10)\sigma_\beta^2+1.99\langle\beta\rangle\ .
\end{align}
At $z > 8$ the dust extinction quickly vanishes~\cite{Yung_2018} and thus we neglect it. In Appendix~\ref{app:dust}, we explore the impact of alternative fitting parameters for the dust extinction on our results. 

\vspace{6pt}

Finally, with all the ingredients combined, the luminosity function can be computed as:
\begin{align}
    \label{eq:UVLF_definition}
    \phi_\mathrm{UV} = \frac{\mathrm{d}n}{\mathrm{d}M_\mathrm{UV}} = \frac{\mathrm{d}n}{\mathrm{d}M_\mathrm{h}}\frac{\mathrm{d}M_\mathrm{h}}{\mathrm{d}M_\mathrm{UV}}\ .
\end{align}

\begin{figure}
    \centering
    \includegraphics[width=\linewidth]{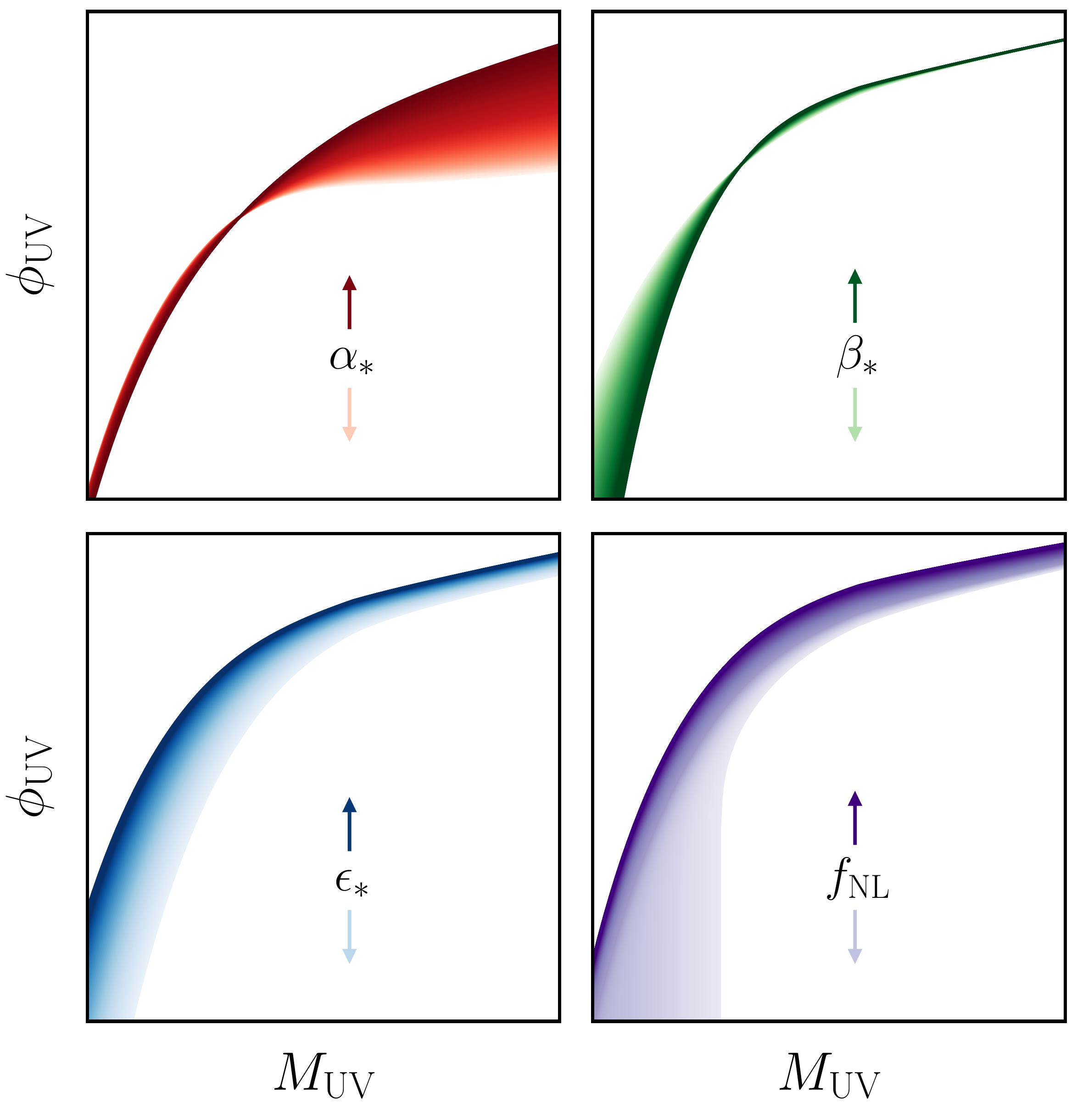}
    \caption{An illustration of the dependence of the UV luminosity function on the fitting parameters in Eq.~\eqref{eq:Mh_Mstar_doublepower_approx} and the amplitude $f_\mathrm{NL}$ of the small-scale PNG (for $k_{\rm cut}=0.1\,\Mpcinv$). The changes in the UV LF are exaggerated for descriptive purposes only. The dark (light) shades indicate an increase (decrease) of the corresponding parameter. The plots cover the magnitude range $-24\leq M_\mathrm{UV}\leq-16$ and parameter ranges $-5\leq \alpha_*\leq 0$, $0\leq\beta_*\leq 0.35$, $0.08\leq\epsilon_*\leq 0.31$ and $-3000\leq f_\mathrm{NL}\leq 2000$.}
    \label{fig:UVLF_params_dependence}
\end{figure}

Note that in this approach the stellar properties of galaxies only depend on the halo mass, rather than the unique formation history of the host halo. As such, this model is not applicable at the level of each individual galaxy, but should be thought of as describing the {\it average} evolution of stellar properties in galaxies.
We illustrate the dependence of the LF on the different parameters in Figure~\ref{fig:UVLF_params_dependence}. It is clear that the effect of $\epsilon_*$ and $f_\mathrm{NL}$ (Section~\ref{subsec:HMF_fNL_corrections}) on the UV LF are strongly degenerate. However, as we will show later on, using a combination of UV LF data at different redshifts will break this degeneracy to a reasonable degree, allowing for the UV LF to be a strong probe of primordial non-Gaussianity.

\subsection{UV LF Data}
\label{subsec:UVLF_data}

The high-redshift UV LF has been observed by the Hubble Space Telescope over a decades-long endeavour. 
This has resulted in two main data catalogs dubbed the Hubble Legacy Fields (HLF) and the Hubble Frontier Fields (HFF). The first consists of several deep-field surveys and has robustly probed the UV LF at the bright end, while the latter consists of observations of six cluster lenses, where faint background galaxies are magnified enough to become observable. Both methods have their own advantages and systematics~\cite{Maizy:2009df}.
For instance, the HFF can reach fainter objects, as those are strongly magnified by the cluster lenses, whereas lensing can introduce important uncertainties~\cite{Bouwens_2017asdasd}. 
On the other hand, the deep blank fields from the HLF catalog have the advantage of being easier to model and, in addition, can better probe the bright end of the LF given the relatively large observed areas~\cite{Maizy:2009df}. 
As will be discussed in Section~\ref{sec:primordial_nonGauss}, and can be readily seen in the lower right panel of Figure~\ref{fig:UVLF_params_dependence}, the impact of primordial non-Gaussianities will be mainly visible at the bright end of the LF. 
Therefore, we perform our main analysis with the data obtained from the HLF (data set 1 below) and summarize the results obtained from other data sets in Appendix~\ref{app:alternative_dataset_results}. 
In particular, we make use of the measured LFs reported by the following references:

\begin{itemize}
    \item \textbf{Data set 1}: Bouwens et al. 2015 ($z = 4+5+6+7+8$)~\cite{Bouwens:2014fua}.
    \item \textbf{Data set 2}: Atek et al. 2018 ($z = 6$)~\cite{Atek:2018nsc}, Atek et al. 2015 ($z = 7$)~\cite{Atek:2015axa}, Ishigaki et al. 2018 ($z = 8$)~\cite{Ishigaki_2018} and Oesch et al. 2018 ($z = 10$)~\cite{Oesch_2018}.
    \item \textbf{Data set 3}: Livermore et al. 2017 ($z = 6+7+8$)~\cite{Livermore:2016mbs} and Oesch et al. 2018 ($z = 10$)~\cite{Oesch_2018}.
\end{itemize}

\begin{figure}[t!]
    \centering
    \includegraphics[width=\linewidth]{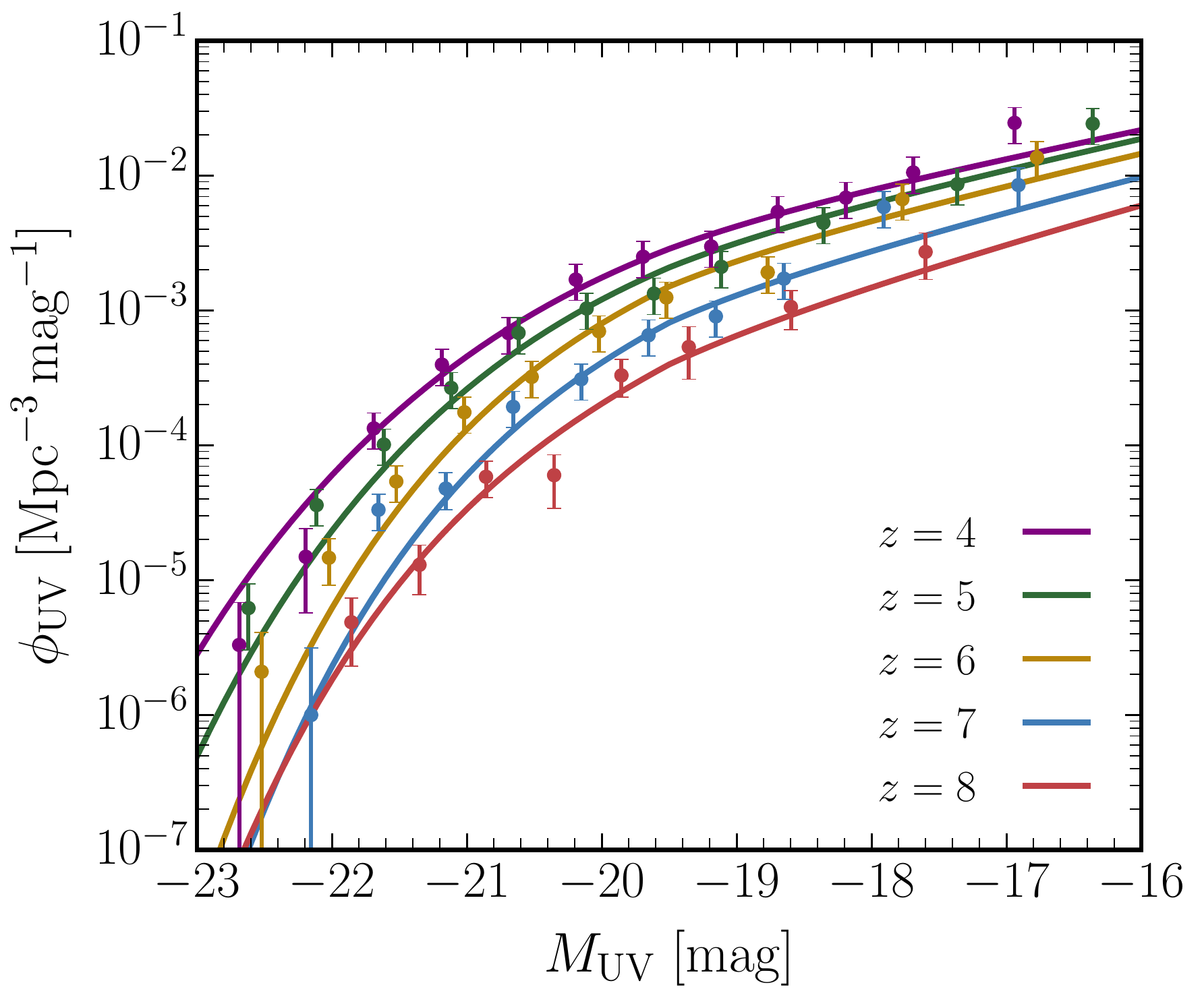}
    \caption{Global fits of our UV luminosity function model to the data from~\cite{Bouwens:2014fua} in the absence of non-Gaussianity (i.e., fixing $f_\mathrm{NL}=0$). A minimum relative error of 30\% is imposed on the data (see Section~\ref{sec:results} for details). The best-fit parameters are $\{\alpha_*, \beta_*, \epsilon_*\} = \{-1.14, 0.20, 0.23\}$.}
    \label{fig:UVLF_Bouwens2015_fit}
\end{figure}

The first data set derives the UV LF from the HLF catalog, while the latter two use HFF data. Note that in some references the UV LF is reported using either a 1500 or {1600\,\AA} UV band filter. This induces a shift of $|M_{1500} - M_{1600}| \lesssim 0.05$~\cite{Williams_2018}, which we have explicitly checked to not change our results. Hence, from this point onward, we will simply use $M_\mathrm{UV}$ to denote the UV magnitude. 
Next, we note that while the UV LF at $z = 10$ is also reported in~\cite{Bouwens:2014fua}, 
we do not include it in our analyses, as nearly all search fields contain zero galaxy candidates at that redshift. A final important point to bear in mind is that the faint end of the quasar LF and the bright end of the galaxy LF overlap, see e.g.~\cite{Matsuoka:2017frx}. The subtracted result, i.e. the galaxy LF, may then present a power-law feature at the bright end~\cite{Ono:2017wjz}. This remains an experimental challenge, as it is difficult to cleanly separate the two contributions.
We show data set 1 in Figure~\ref{fig:UVLF_Bouwens2015_fit}, along with our best-fit model (in the absence of primordial non-Gaussianity).

\section{Primordial Non-Gaussianity}
\label{sec:primordial_nonGauss}

Here we explore the phenomenology of the primordial non-Gaussianity models considered in this work and how they affect the distribution of matter and thus the halo mass function.

\subsection{Formalism and Models}

We begin by laying down our formalism.
Assuming statistical homogeneity and isotropy, we can generically write the 2- and 3-point correlation functions of the gravitational potential $\Phi$ as:
\begin{align}
    \label{eq:correlation_function}
    \langle\Phi(k_1)\Phi(k_2)\rangle =&\ P_\Phi(k_1)(2\pi)^3\delta_D^3(\mathbf{k_1}+\mathbf{k_2})\\
    \label{eq:bispectrum}
    \langle\Phi(k_1)\Phi(k_2)\Phi(k_3)\rangle =&\ B_\Phi (k_1,k_2,k_3) (2\pi)^3 \nonumber\\
    &\ \times \delta_D^3(\mathbf{k_1}+\mathbf{k_2}+\mathbf{k_3})\ , 
\end{align}
where
\begin{align}
 P_\Phi(k) = \frac{2\pi^2\Delta_\Phi^2(k)}{k^3} = \frac{2\pi^2}{k^3}\frac{9}{25}A_\mathrm{s}\left(\frac{k}{k_\mathrm{pivot}}\right)^{n_\mathrm{s}-1}
\end{align}
is the power spectrum of $\Phi$ and $B_\Phi$ its bispectrum. The factor ${9}/{25}$ comes from the relation $\Phi = {3}\zeta/{5}$ between the gravitational potential and the comoving curvature perturbation $\zeta$. Both the power spectrum and bispectrum have been measured to great precision by CMB observations, as well as galaxy surveys, confirming that the large-scale fluctuations of the Universe are Gaussian (and $B_\Phi$ is consistent with zero within current errors), e.g.  \cite{Akrami:2019izv,Castorina:2019wmr}.
The situation is different at smaller scales, however, where much less is known.
We will, therefore, consider deviations from Gaussianity at small scales only, which affect the formation of the galaxies that we will probe, as they have relatively low masses ($M_\mathrm{h}\lesssim 10^{11} M_\odot$).
In order to probe non-Gaussianity at those smaller scales, while not altering the CMB predictions, we introduce a cut-off scale $k_\mathrm{cut} = 0.1\, \mathrm{Mpc^{-1}}$ in the bispectrum, below which it vanishes. 

More concretely, throughout this work we will focus on local-type primordial non-Gaussianity for simplicity, although our analysis could be extended to other models.
The most minimal model of this family simply alters the initial gravitational perturbation $\Phi$ by a series expansion around a Gaussian field $\Phi_\mathrm{G}$, which to linear order reads:
\begin{align}
    \label{eq:phi_nonGaussian}
    \Phi(x) = \Phi_\mathrm{G}(x) + f_\mathrm{NL}\left(\Phi_\mathrm{G}^2(x)-\langle\Phi_\mathrm{G}^2\rangle\right).
\end{align}
In this case, the bispectrum is given by:
\begin{align}
\label{eq:phi_nonGaussian_bispectrum}
B_\Phi = f_\mathrm{NL}P_\Phi(k_1)P_\Phi(k_2) \prod_{i=1}^3 K(k_i) + \left(5\ \mathrm{perms.} \right)\ ,
\end{align}
where $K(k_i) = \Theta(k_i-k_{\rm cut})$ ensures that only modes above the cut-off contribute to the bispectrum. This is the main shape we will use throughout this work. Examples of theoretical models that may generate small-scale non-Gaussianity\footnote{These models may not generate a sharp cut-off at large scales. As such, our approach should be considered as a phenomenological first step to approximate small-scale PNG.} 
could involve inflationary scenarios with a changing speed of sound (see e.g.~\cite{LoVerde:2007ri} and references therein). Moreover, it has been shown that small-scale PNG may impact the formation and abundance of primordial black holes~\cite{Young:2013oia, Atal:2018neu, Atal:2019erb, DeLuca:2019qsy}, opening up a door to exciting new possibilities.
We note that we are not considering changes to the power spectrum $P_\Phi$ due to primordial non-Gaussianity, as those vanish to first order in $\fNL$, although some inflationary models might directly affect $P_\Phi$ and produce a richer phenomenology. 

\pagebreak


\subsection{Cumulants}

We are mostly interested in quantities that are coarse-grained over a region which will collapse into a halo of mass $M$.
We define the density perturbation $\delta_M$ smoothed over mass scale $M\equiv M_\mathrm{h}$ as:
\begin{align}
\label{eq:density_perturbation}
    \delta_M(z) = \int\frac{d^3k}{(2\pi)^3}W_M(k)T_\Phi(k,z)\Phi(k)\ ,
\end{align}
where $T_\Phi(k,z)$ is the linear transfer function and
\begin{align}
    W_M(k) =& \frac{3\sin\left(kR\right)}{\left(kR\right)^3} - \frac{3\cos\left(kR\right)}{\left(kR\right)^2}\ 
\end{align}
is a top-hat window function with comoving radius $R(M) = (3M/(4\pi\rho_\mathrm{m}))^{1/3}$.
Note that $k$ and $\rho_\mathrm{m}$ are also comoving quantities. 
The transfer function $T_\Phi(k,z)$ is computed as $T_\Phi(k,z) = 5D(z)T_\zeta(k,0)/3$, where $D(z)$ is the linear growth factor and $T_\zeta(k,0)$ is the transfer function for the curvature perturbation at redshift $z = 0$, which we obtain from the \texttt{CLASS} code~\cite{Blas:2011rf}.

The deviation from Gaussianity is usually parametrized in terms of higher-order cumulants of the field $\Phi$. We will work to first order in $f_\mathrm{NL}$, where only the skewness is relevant, and we define:
\begin{align}
    \label{eq:kappathree}
    \kappa_3(M) &= \frac{\langle\delta_M^3\rangle}{\sigma_M^3}\ ,
\end{align}
with $\sigma_M^2$ the mass variance smoothed over mass scale $M$:
\begin{align}
    \label{eq:sigmasq_M}
    \sigma^2_M &= \int\frac{d^3k}{(2\pi)^3}W_M^2(k)T^2_\Phi(k,z)P_\Phi(k)\ .
\end{align}
It is obvious from Eq.~\eqref{eq:bispectrum} that $\kappa_3$ itself is proportional to $f_\mathrm{NL}$.
In practice, we make use of a fitting function for $\kappa_3$ to ease the computational load, which we calibrate explicitly for $k_\mathrm{cut}=0.1\,\mathrm{Mpc}^{-1}$ in Appendix~\ref{app:kappa3_fit}.

\begin{figure}[h]
    \centering
    \includegraphics[width=\linewidth]{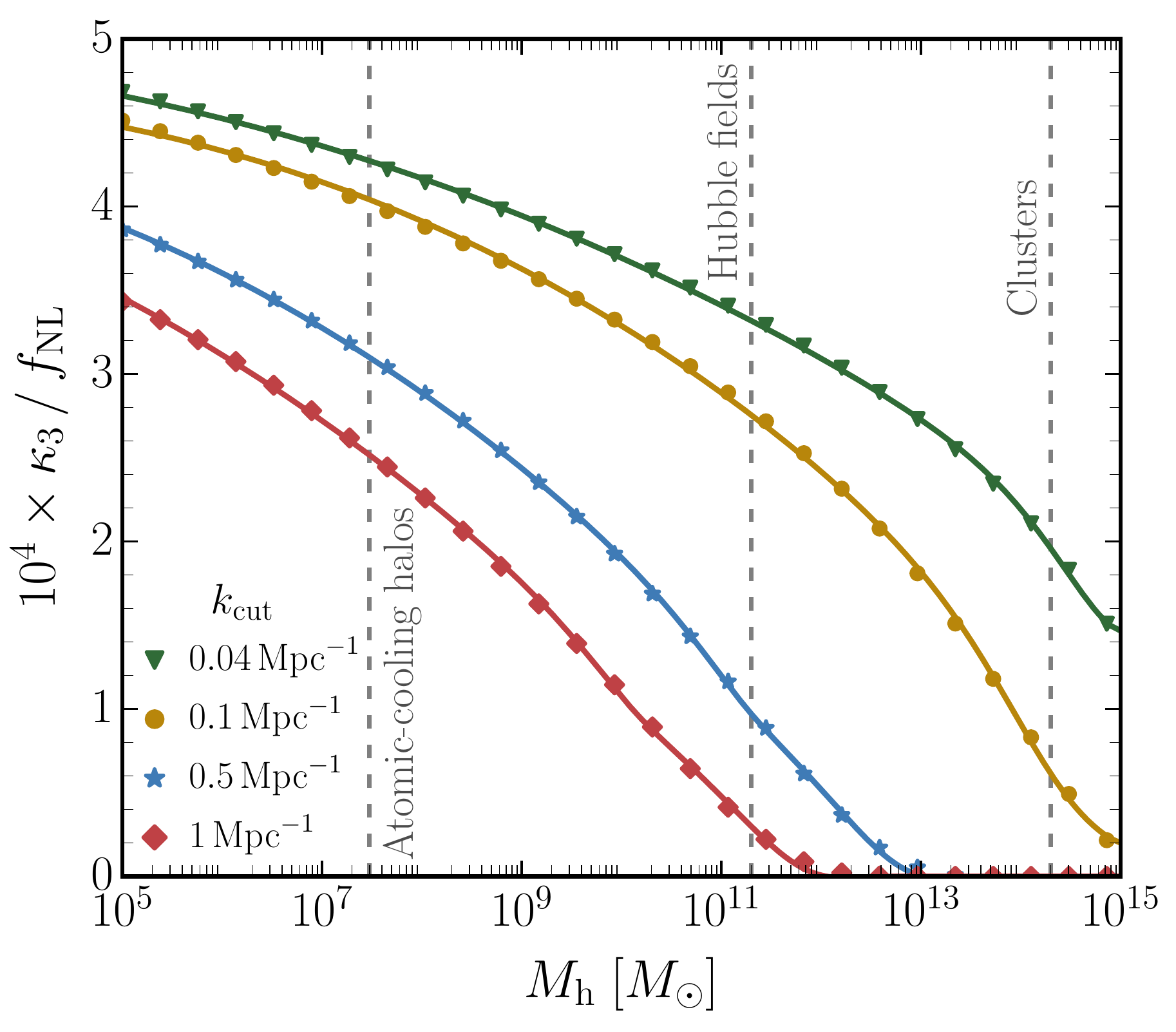}
    \caption{The normalized skewness as a function of halo mass in the presence of local-type non-Gaussianity for different cut-off scales $k_\mathrm{cut}$. The scatter points are obtained directly from Eq.~\eqref{eq:kappathree}, whereas the solid lines are based on the fit in Eq.~\eqref{eq:kappa3_fit}. The dashed vertical lines roughly represent -- from left to right --  the mass of atomic cooling halos (relevant to the 21-cm signal from cosmic dawn), the heaviest halos probed in the Hubble Legacy/Frontier Fields and halos in which clusters reside.}
    \label{fig:kappa3_fit}
\end{figure}

We show $\kappa_3$ as a function of halo mass in Figure~\ref{fig:kappa3_fit} for different choices of the  cut-off scale  $k_{\rm cut}$.
Increasing $k_{\rm cut}$ produces an overall suppression of $\kappa_3$.
The most striking effect is, however, the vanishing of $\kappa_3$ for halos much heavier than $M_{\rm cut} = 4\pi \rho_\mathrm{m} k_{\rm cut}^{-3}/3$.
For $k_{\rm cut}=0.1\,\Mpcinv$ this corresponds to $M_{\rm cut}\approx 2 \times 10^{14}\, \Msun$, roughly the mass of galaxy clusters.
Furthermore, a more stringent cut of $k_{\rm cut}=1\,\Mpcinv$ does not alter the abundance of halos above ${\sim}2 \times 10^{11}\, \Msun$, which encompasses halos smaller than those hosting our own Milky Way.
This shows that the PNG models that we consider would leave no signature in usual searches, such as cluster abundance or CMB analyses, whereas they will affect the UV LF, as well as the 21-cm signal during cosmic dawn.

\subsection{Effect on the HMF}
\label{subsec:HMF_fNL_corrections}

The deviation from Gaussianity in the PDF for $\delta_M$ will become imprinted onto the abundance and distribution of galaxies~\cite{LoVerde:2007ri, LoVerde:2011iz}.
For the derivation of the correction to the HMF induced by non-Gaussianities, we will use the Press-Schechter (PS) formalism\footnote{While the HMF we use is not the PS HMF, the moving barrier in the ST formalism adds an {\it extra} term to the constant barrier in the PS formalism~\cite{Sheth:2001dp}, which in turn would make our bounds stronger. Since we do not include this extra term in the computation of the HMF correction, this will make our constraints conservative.}~\cite{LoVerde:2007ri}. In this framework, the volume fraction that has collapsed into halos of mass $M$ is given by the integral of the 1-point PDF of the scaled density perturbation $\nu \equiv \delta_M/\sigma_M$:
\begin{align}
    \label{eq:collapse_fraction}
    F(M) &= \int_{\nu_\mathrm{c}(M)}^{\infty}\mathrm{d}\nu\rho(\nu,M)\ ,
\end{align}
where $\nu_\mathrm{c}(M) = \sqrt{a_{\rm ST}} \, \delta_\mathrm{crit}/\sigma_M = 1.42 / \sigma_M$.  
The (differential) halo mass function can then be directly obtained from:
\begin{align}
    \label{eq:HMF_definition}
    \frac{\mathrm{d}n(M)}{\mathrm{d}M} &= -2\frac{\rho_\mathrm{m}}{M}\frac{\mathrm{d}F(M)}{\mathrm{d}M} = -2\frac{\rho_\mathrm{m}}{M}F'(M)\ ,
\end{align}
where the prime denotes differentiation with respect to halo mass $M$. In a Gaussian cosmology, the PDF is given by $\rho(\nu,M) = \left(2\pi\right)^{-1/2}\exp(-\nu^2/2)$. Now, in a non-Gaussian cosmology any deviations from this distribution can be described by making use of the Edgeworth series expansion~\cite{Juszkiewicz:1993hm, Bernardeau:1994aq}. In this case, the PDF is written as a series in the higher-order cumulants of the distribution. Given that non-Gaussianities manifest themselves in deviations of these cumulants, this makes the Edgeworth expansion a useful tool. Note that even though this is an asymptotic series and its convergence is not guaranteed, we have checked the required truncation order for parameter ranges relevant to our purpose: modelling UV luminosity functions. As we will show in Appendix~\ref{app:higher_order_terms}, we find that a reasonable accuracy (up to $|f_\mathrm{NL}|\sim\mathcal{O}(10^3)$) can be obtained by cutting off the series already at first order, and our constraints do not change when including higher-order corrections. 
In explicit form, the 1-point PDF then reads:
\begin{align}
    \label{eq:PDF}
    \rho(\nu, M) =& \frac{\exp(-\nu^2/2)}{(2\pi)^{1/2}}\left(1 + \frac{\kappa_3(M)H_3(\nu)}{6}\right)\ ,
\end{align}
with 
\be
H_n(x) = (-1)^n\exp(\nu^2/2)\frac{\mathrm{d}^n}{\mathrm{d}\nu^n}\exp(-\nu^2/2)
\ee
the (probabilists') Hermite polynomials. This expression can then be inserted into Eq.~\eqref{eq:collapse_fraction} to obtain the collapse fraction to first order in $f_\mathrm{NL}$, which can be written as $F_\mathrm{NG}(M) = F_0(M) + F_1(M)$ and with:
\begin{align}
    F_0 &= \frac{1}{2}\mathrm{erfc}\left(\frac{\nu_\mathrm{c}}{\sqrt{2}}\right)\\
    F_1 &= \frac{\exp(-\nu_\mathrm{c}^2/2)}{(2\pi)^{1/2}}\frac{\kappa_3H_2(\nu_\mathrm{c})}{6}\ .
\end{align}
The derivatives of these quantities with respect to halo mass $M_\mathrm{h}$ read:
\begin{align}
    F_0' &= -\frac{\exp(-\nu_\mathrm{c}^2/2)}{(2\pi)^{1/2}}\nu_\mathrm{c}'\\
    \frac{F_1'}{F_0'} &= \frac{\kappa_3H_3(\nu_\mathrm{c})}{6} - \frac{H_2(\nu_\mathrm{c})}{6}\frac{\kappa_3'}{\nu_\mathrm{c}'}\ .
\end{align}

The non-Gaussian mass function up to first order in $f_\mathrm{NL}$ is then:
\begin{align}
    \label{eq:Edgeworth_correction}
    \frac{n_\mathrm{NG}'}{n_\mathrm{G}'} = \frac{F'_\mathrm{NG}}{F'_\mathrm{G}} \approx 1 + \frac{F_1'}{F_0'}\ ,
\end{align}
where $n_\mathrm{G}$ indicates the Gaussian HMF. The Sheth-Tormen HMF in Eq.~\eqref{eq:ST_HMF} is then multiplied by this correction to obtain the luminosity function dependence on $f_\mathrm{NL}$. For negative values of $f_\mathrm{NL}$ one must proceed with caution, as the correction can lead to an unphysical (negative) value of the HMF.
Instead, we set the correction equal to 0 for all masses where it is negative. 
As discussed in~\cite{LoVerde:2011iz}, this issue can be circumvented by using the log-Edgeworth expansion. 
However, while for negative $f_\mathrm{NL}$ this can prove to be a useful trick, we find that its convergence for positive $f_\mathrm{NL}$ is far worse. 
We also compared with the Edgeworth mass function in this same reference at low redshifts and found good agreement with both the semi-analytical results and the results from N-body simulations. 
We are not aware of any simulations of the HMF at high redshifts including PNG. Nevertheless, as the halos we probe at high redshift are comparable in rarity to the clusters studied in~\cite{LoVerde:2011iz}, and our HMF has been tested (assuming Gaussianity) against simulations in~\cite{Tacchella:2018qny}, we expect our approximations to be valid in the entire redshift range we consider.

\section{Results}
\label{sec:results}

With the UV LF and non-Gaussianity formalisms established in the previous sections, we present here constraints on the amplitude $f_\mathrm{NL}$ of non-Gaussianity at small scales. We focus mainly on the results obtained by using data set 1 and include results obtained from using the two other data sets with additional remarks in Appendix~\ref{app:alternative_dataset_results}.

We start by constructing a $\chi^2$ to assess deviations from the data due to the presence of small-scale non-Gaussianity:
\begin{align}
    \label{eq:chisq}
    \chi^2(z,\boldsymbol{\theta}) = \underset{M_\mathrm{UV}}{\sum}\left(\frac{\phi_\mathrm{model}(z,M_\mathrm{UV};\boldsymbol{\theta})-\phi_\mathrm{data}(z,M_\mathrm{UV})}{\sigma_\phi^\mathrm{data}(z,M_\mathrm{UV})}\right)^2\ ,
\end{align}
where $\boldsymbol{\theta} = \{\alpha_*,\beta_*,\epsilon_*,f_\mathrm{NL}\}$ represents a vector of the free parameters in our model and the sum goes over all data points. 
In order to account for cosmic variance, as well as any potential systematic errors in estimations of the UV LF, we impose a minimum relative error of 30\% in the data for all data sets (i.e., $\sigma_\phi^\mathrm{data}$ is at least $0.3\times \phi_\mathrm{data}$ at each $z$ and $M_{\rm UV}$).
This is a more conservative approach than used in~\cite{Gillet:2019fjd, Bouwens_2017asdasd},
where the minimum error was set at 20\%.
We determined this noise floor by solving for $\chi^2_\text{best-fit} / g_\mathrm{dof}\approx 1$, where $\chi^2_\text{best-fit}$ is the best-fit value of the $\chi^2$ distribution and $g_\mathrm{dof}$ is its number  of degrees of freedom.

We begin by reproducing the analysis of~\cite{Gillet:2019fjd} in the absence of primordial non-Gaussianity.
That work only used a subset of our data, and under the same assumptions we find good agreement.
Details of the comparison can be found 
in Appendix~\ref{app:comparison_previous_literature}, which acts as a consistency check of our model assumptions.
\vspace{6pt}

Next, we obtain constraints on $\fNL$ in two different ways.
The first is by directly marginalizing Eq.~\eqref{eq:chisq} over the parameters $\{\alpha_*,\beta_*,\epsilon_*\}$ for each $f_\mathrm{NL}$, and thus finding the marginalized $\chi^2_{\rm marg}(\fNL)$.
The second is by performing a joint MCMC analysis of all parameters.
While both methods result in similar $\fNL$ constraints, they have different benefits.
The first method will allow us to quickly find constraints on $\fNL$ under different assumptions.
The advantage of the second method is that any correlations between the different parameters will be clear. 

We impose broad flat priors of $\alpha_*= [-2,0]$, $\beta_*= [0, 0.9]$, and $\fNL= [-2000,1000]$.
The negative prior on $\alpha_*$ (and positive one on $\beta_*$) reflects our understanding of feedback, where both the low- and high-mass limits of galaxies are less efficient at forming stars than in between~\cite{Tacchella:2018qny,Yung_2018}.
Moreover, since the parameter $\epsilon_*$ determines the fraction of baryons in a dark matter halo, we include an upper limit on its prior of $\epsilon_*^\mathrm{max} = 2{M_*}/{M_\mathrm{h}} \sim 2{\Omega_\mathrm{b}}/{\Omega_\mathrm{m}}\approx 0.31$~\cite{Tacchella:2018qny, Aghanim:2018eyx}, whereas its lower bound we set equal to 0.001 for convenience. 
Additionally we fix our cosmological parameters to the Planck 2018 best fits.

\vspace{6pt}

Our constraints, obtained through the marginalized-$\chi^2$ method, are summarized in Table~\ref{tab:HLF_bounds}. The $1\sigma$ and $2\sigma$ limits are obtained by determining for which $f_\mathrm{NL}$ the quantity $\Delta \chi^2(\fNL) \equiv \chi_{\rm marg}^2 - \chi_\text{best-fit}^2$ is equal to 1 and 4 respectively. The full $\Delta\chi^2(\fNL)$ curves are included in Appendix~\ref{app:alternative_dataset_results}. Using data set 1 we find that $\fNL$ is consistent with zero, and has a one-sided error of $\sigma(\fNL)=235$ at 1$\sigma$ (and 343 at 2$\sigma$). This error is not symmetric around the mean, as negative values of $\fNL$ have a more marked effect on the HMF (Section~\ref{sec:UVLF}).
While this error is significantly larger than the one obtained with Planck data (where $\sigma(\fNL)=5.1$ for local-type PNG~\cite{Akrami:2019izv}), it places a constraint on smaller scales, beyond where CMB data can naturally access, and is thus complementary to such bounds.

If instead of a 30\% minimum relative error in the data we set this error equal to 20\% (as in~\cite{Gillet:2019fjd, Bouwens_2017asdasd}), the bounds would become stronger by approximately 25\%. In the case such error is not included at all, the improvement would be roughly 50\%.

\begin{figure*}[hbtp!]
    \centering
    \includegraphics[width=0.7\textwidth]{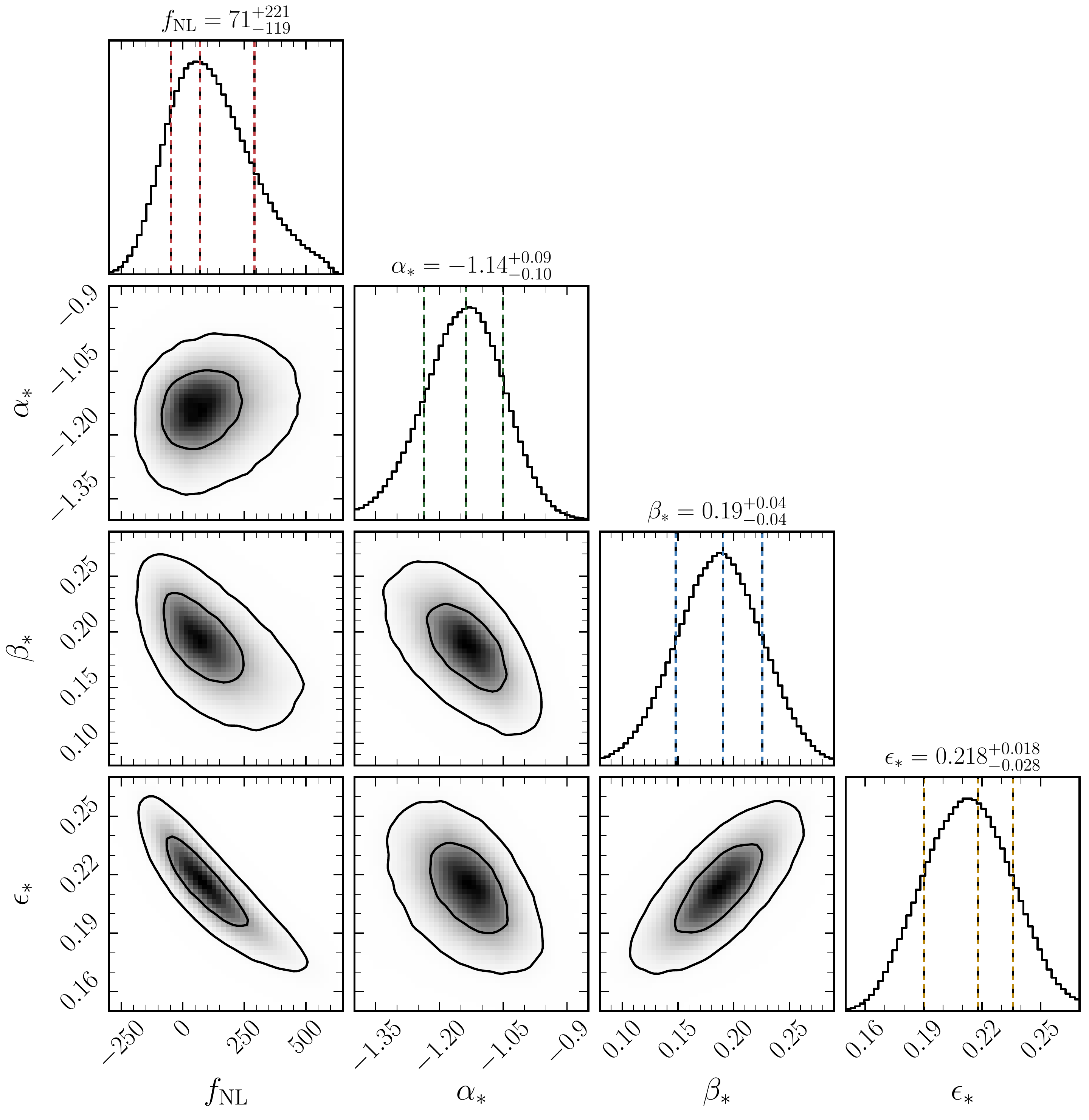}
    \caption{Posteriors for $\alpha_*,\, \beta_*,\,  \epsilon_*$ and $f_\mathrm{NL}$ using data set 1 (Section~\ref{subsec:UVLF_data}) and a cut-off scale in the bispectrum of $k_\mathrm{cut} = 0.1\,\mathrm{Mpc}^{-1}$. A minimal error of 30\% is imposed in the UV LF data points to account for cosmic variance and other systematic errors.
    The 2D contours depict the $1\sigma$ and $2\sigma$ confidence levels. The titles and vertical lines in the 1D posteriors represent the maximum-likelihood best fit (central line) and the $\pm1\sigma$ quantiles (outer lines).}
    \label{fig:MCMC_Bouwens2015zAll}
\end{figure*}

Table~\ref{tab:HLF_bounds} also shows the results for the other two data sets we consider, which use the HFF instead of the HLF.
These are also consistent with no small-scale PNG within 2$\sigma$ and have roughly comparable error-bars, showing that the specific data used, including the range of redshifts and magnitudes accessible, does not dramatically alter our conclusions.

\begin{table}[h!]
    \centering
{\def\arraystretch{1.35}
    \begin{tabular}{c|c|c}
        \hline\hline
         \textbf{Data set} & $\boldsymbol{1\sigma}$& $\boldsymbol{2\sigma}$ \\
        \hline\hline
        \textbf{1}  (HLF)  & $73^{+277}_{-192}$ & $73^{+430}_{-256}$\\
        \hline
        \textbf{2} (HFF) & $-155^{+185}_{-126}$ & $-155^{+297}_{-169}$ \\
        \hline
        \textbf{3} (HFF)& $-302^{+262}_{-351}$ & $-302^{+412}_{-662}$ \\
        \hline\hline
    \end{tabular}
}
    \caption{Constraints on $f_\mathrm{NL}$ at 68\% and 95\% C.L.~using the HLF and HFF data sets described in Section~\ref{subsec:UVLF_data}. A minimum relative error of 30\% is used in the data and the cut-off scale in the bispectrum is set equal to $k_\mathrm{cut} = 0.1\,\mathrm{Mpc}^{-1}$. These bounds are obtained by directly marginalizing the $\chi^2$ in Eq.~\eqref{eq:chisq} over the parameters $\alpha_*,\,\beta_*$ and $\epsilon_*$, although a very similar result is obtained with a direct MCMC search.
    }
    \label{tab:HLF_bounds}
\end{table}

In order to study degeneracies between the parameters, we now perform an MCMC analysis using data set 1 and show the posteriors in Figure~\ref{fig:MCMC_Bouwens2015zAll}. Note that while at a single redshift the impact of $f_\mathrm{NL}$ and $\epsilon_*$ on the UV LF is highly degenerate (see Figure~\ref{fig:UVLF_params_dependence}), this degeneracy is lifted when combining data at different redshifts, as is clear 
in Figure~\ref{fig:MCMC_Bouwens2015zAll}. This is because different redshift slices have slightly different $f_\mathrm{NL}-\epsilon_*$ degeneracy directions, making their combination break the degeneracy and yielding a nearly Gaussian posterior.
The MCMC best fit at $2\sigma$ reads:
\begin{align}
    \label{eq:main_bound}
    f_\mathrm{NL} = 71^{+426}_{-237}\ ,
\end{align}
in excellent agreement with our result reported in Table~\ref{tab:HLF_bounds}. At $1\sigma$ (see top panels in Figure~\ref{fig:MCMC_Bouwens2015zAll}), the agreement is{\parfillskip=0pt\par}

reasonable and the deviations could be due to the implicit assumption in the marginalized-$\chi^2$ method that the data is Gaussian distributed. Since the MCMC method is free of any such assumptions, we consider Eq.~\eqref{eq:main_bound} our main result. 

\subsection*{Results for other cut-off scales}
\label{subsec:other_kcuts}

The main analysis in this work uses a cut-off scale of $k_\mathrm{cut} = 0.1\,\mathrm{Mpc}^{-1}$, which roughly denotes the smallest scale that can be probed by the CMB and below which we set the bispectrum equal to zero. While in principle $k_\mathrm{cut}$ ought to be included as a free parameter in the analysis, it is computationally expensive to do so. Therefore, we devote this section to illustrate the sensitivity of our bounds to the cut-off scale $k_\mathrm{cut}$ for a few cases.

\begin{figure}[h]
    \centering
    \includegraphics[width=\linewidth]{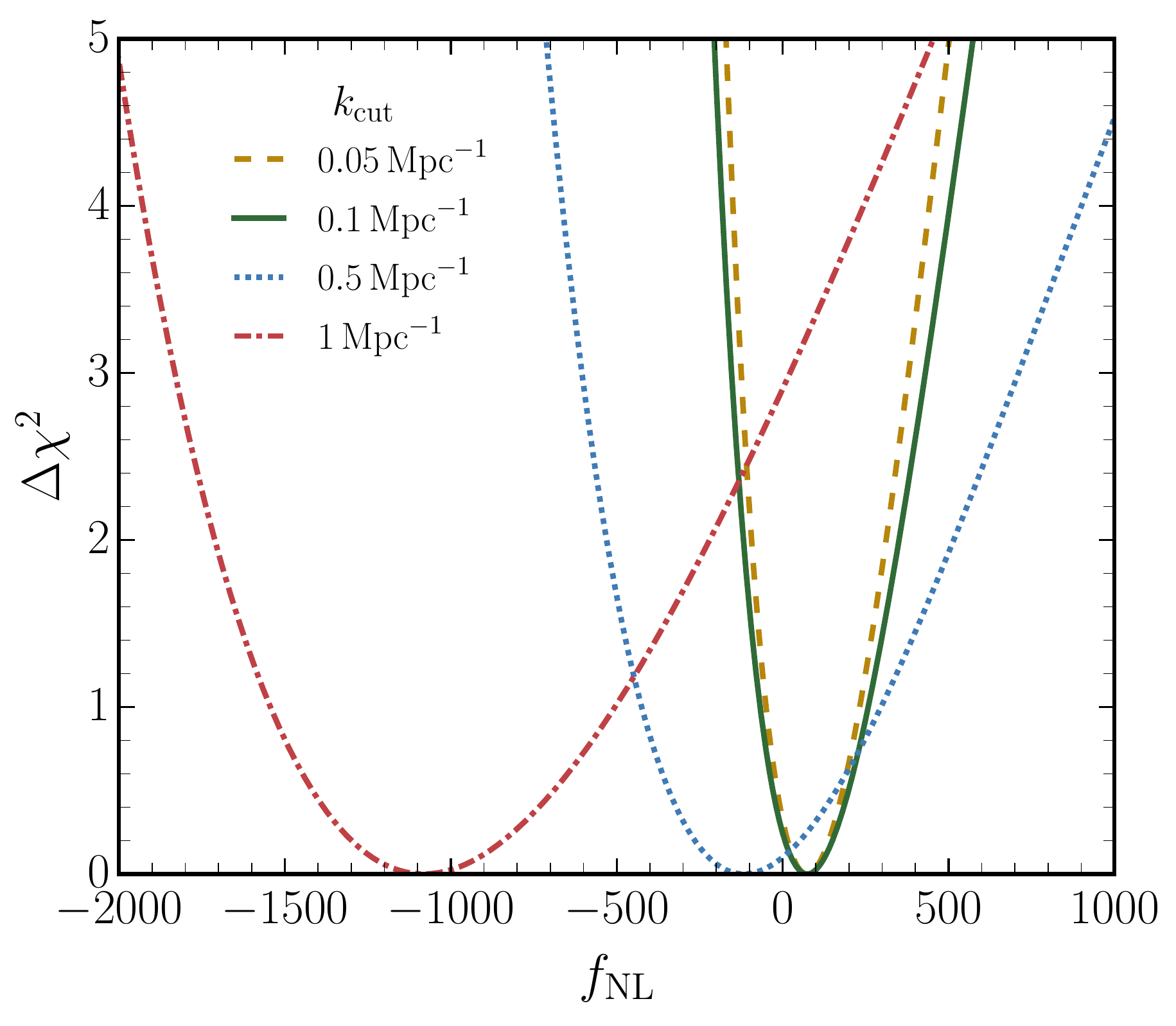}
    \caption{Marginalized $\Delta\chi^2$ as a function of $f_\mathrm{NL}$ and its dependence on the cut-off scale $k_\mathrm{cut}$ of the PNG. These constraints are obtained using data set 1 in Section~\ref{subsec:UVLF_data} at redshifts $z=4+5+6+7+8$ with a minimum relative error of 30\% in the data.}
    \label{fig:bounds_kcut}
\end{figure}

In a similar fashion as before, we calculate the marginalized $\chi^2$ using different values for $k_\mathrm{cut}$ and display these in Figure~\ref{fig:bounds_kcut}. This figure shows that the bounds on small-scale non-Gaussianity mainly come from scales between $0.1 - 1\,\mathrm{Mpc}^{-1}$. The choice of $k_\mathrm{cut} = 0.1\,\mathrm{Mpc}^{-1}$ exemplifies a scenario in which small-scale non-Gaussianity nearly exploits the UV LF to the fullest extent. This is because such scales correspond to masses around $10^{11}-10^{13}\,M_\odot$, which coincide with mass scales at the bright end of the UV LF as probed by the HST. Therefore, when increasing the cut-off to smaller scales, the UV LF quickly loses its constraining power. 
\vspace{6pt}

Interestingly, however, for $k_\mathrm{cut} = 1\,\mathrm{Mpc}^{-1}$ the best-fit value of $f_\mathrm{NL}$ moves away from 0 (and we note that for larger $k_{\rm cut}$ the constraints widen significantly).
This hints to the existence of a bump-like feature in the data (possibly due to non-Gaussian HMF corrections only present at $M_\mathrm{h}\lesssim 10^{11}\,M_\odot$), which would favour a negative $f_\mathrm{NL}$ over $f_\mathrm{NL} = 0$ by ${\sim}1.7\,\sigma$. This behaviour is also present when performing an MCMC analysis, even with redshift-dependent astrophysical parameters. Moreover, we find this deviation in HFF data as well, at the level of ${\sim}1.9\sigma$ for data set 2 and ${\sim}2.5\sigma$ for data set 3. It should be noted, however, that this deviation from Gaussianity can disappear if a different dust correction is employed (see Appendix~\ref{app:dust}).
In the next section we will study the potential of the upcoming JWST and NGRST in resolving whether this excess has a physical origin.

\section{Future Data}
\label{sec:forecasts}
Here we study how well future data from the epochs of cosmic dawn and reionization will be able to constrain small-scale PNG. We will focus on two experiments: the James Webb Space Telescope and the Nancy Grace Roman Space Telescope, which both will significantly improve upon the UV LFs of HST.
We also briefly explore how small-scale non-Gaussianity would affect complementary 21-cm experiments in Appendix~\ref{subsec:21cm_forecast}.


The JWST is expected to improve upon Hubble mainly at the faint end~\cite{Yung_2018, Behroozi:2020jhj}, leaving the reach at the bright end (at fixed coverage) mostly unchanged.
We will consider the Wide Field survey mode from~\cite{Mason:2015cna} (see also~\cite{Williams_2018} for a discussion), which covers an area of ${\sim}4000\,\mathrm{arcmin}^2$ and has a total exposure time of 800 hours. This is a much smaller timescale than those of the surveys available in the HLF catalog~\cite{Beckwith:2006qi}. As such, with this configuration less faint galaxies can be observed, although galaxies at higher redshifts will be found more efficiently due to a different wavelength coverage~\cite{Gardner:2006ky}. With a higher exposure time, we expect JWST to probe fainter galaxies and thus smaller scales. Moreover, it will allow to refine our phenomenological model, since it will help to accurately determine the star formation efficiency of high redshift galaxies~\cite{Tacchella_2020}. Here we assume that JWST observes galaxies with magnitudes above $M_\mathrm{UV}^\mathrm{min} = -22.75$, which roughly corresponds to the brightest galaxies in the HLF catalog~\cite{Bouwens:2014fua}. The lowest-brightness galaxies the JWST can detect correspond to an apparent magnitude of approximately $m_\mathrm{lim} = 29.3$~\cite{Mason:2015cna}. This quantity is translated into an absolute magnitude $M_\mathrm{UV}^\mathrm{max}$ via $M_\mathrm{UV}^\mathrm{max} = m_\mathrm{lim}+ 5-5\log_{10}\left({D_\mathrm{L}}/{\mathrm{pc}}\right)$, where $D_\mathrm{L}$ is the luminosity distance.

\vspace{6pt}

The NGRST, on the other hand, is expected to cover a significantly larger area than both the HST and JWST. In particular, the High-Latitude Survey~\cite{Spergel:2015sza} that we consider here will image an area of ${\sim}2000\,\mathrm{deg}^2$ over a 2-year period, which will allow us to further probe the bright end of the UV LF. We determine $M_\mathrm{UV}^\mathrm{min}$ for NGRST by requiring that at least one galaxy is contained in the brightest magnitude bin. This gives a value of $-25 \lesssim M_\mathrm{UV}^\mathrm{min} \lesssim -24$\footnote{The dust attenuation at such bright magnitudes can suppress the UV LF significantly. Given that the dust extinction parameters have not been measured at $z > 8$ yet, the absence of this effect at these redshifts and magnitudes is an important caveat in our method of generating mock data}. The galaxies of lowest brightness observable correspond to an apparent magnitude of $m_\mathrm{lim} = 26.5$~\cite{Mason:2015cna} and is translated to an absolute magnitude using the formula above.

We make simple forecasts for both JWST and NGRST by generating a set of mock data at redshifts $z=\{4,5,6,7,8,9,10\}$ through the following procedure:
\begin{enumerate}
\item We define a fiducial luminosity function based on the global fit to data set 1 at $z=4+5+6+7+8$ (fixing $f_\mathrm{NL} = 0$ here, see also Figure~\ref{fig:UVLF_Bouwens2015_fit}). The fiducial parameters of this model are $\{\alpha_*,\beta_*,\epsilon_* ,f_\mathrm{NL}\} = \{-1.14,0.20,0.23,0\}$.

\item Next, we bin the luminosity function from $M_\mathrm{UV}^\mathrm{min}$ to $M_\mathrm{UV}^\mathrm{max}$. 
We use a bin size of $\Delta M_\mathrm{UV}\approx 0.5$ and calculate the average (comoving) number of galaxies inside each bin at each redshift. For the NGRST mock data, the value of $M_\mathrm{UV}^\mathrm{min}$ is determined by requiring that at least one galaxy is contained in the brightest magnitude bin at each redshift.

\item Then we draw from a Poisson distribution with average the number of galaxies inside each bin. The central values of the mock data are then these random sampled numbers, placed at the center of each bin. If the Poisson sampling gives 0, we set as average value 0.5 galaxies (so as to obtain a rough upper bound of 1 galaxy).

\item The errors are obtained in two steps: Firstly, we draw from a Poisson distribution with mean the average number of galaxies inside each bin. If in the previous step 0 was obtained, then the error is also set to 0.5 galaxies. Secondly, like in our main analysis, we impose a minimal relative error of 10\%, to account for cosmic variance~\cite{Trenti:2007dh}. Since 10\% could be regarded as a conservative value, depending on the sky coverage, we also report forecasts using 0\% and 5\% for comparison. The errors are placed symmetrically around each mock data point.

\item Lastly, we translate back to the luminosity function by dividing by volume and the bin size $\Delta M_\mathrm{UV}$.
\end{enumerate}

\begin{table}[b!]
    \centering
{\def\arraystretch{1.35}
    \begin{tabular}{c|c|c|c|c|c|c}
        \hline\hline
         \multirow{2}{*}{\textbf{Min. error}} & \multicolumn{2}{c|}{\textbf{JWST}} & \multicolumn{2}{c|}{\textbf{NGRST}} & \multicolumn{2}{c}{\textbf{Combined}}\\ 
         & \multicolumn{1}{c}{$1\sigma$} & $2\sigma$ & \multicolumn{1}{c}{$1\sigma$} & $2\sigma$ & \multicolumn{1}{c}{$1\sigma$} & \multicolumn{1}{c}{$2\sigma$} \\
        \hline\hline
        \textbf{None}$^\dagger$& $\pm 20$ & $\pm 28$ & $\pm 9$ & $\pm 13$ & $\pm 14$ & $\pm 20$ \\
        \hline
        \textbf{5\%}& $\pm 37$ & $\pm 53$ & $\pm 49$  & $\pm 70$ & $\pm 24$ & $\pm 34$ \\
        \hline
        \textbf{10\%}& $\pm 56$ & $\pm 79$ & $\pm 99$ & $\pm 156$ & $\hspace{0.79mm} \pm 39$\hspace{0.79mm} & $\pm 56$\\
        \hline\hline
    \end{tabular}
}
    \caption{Forecasted sensitivities to $f_\mathrm{NL}$ by JWST and NGRST, using $k_\mathrm{cut}=0.1\,\mathrm{Mpc}^{-1}$ and different minimal errors in the mock data. These bounds are obtained by marginalizing the $\chi^2$ in Eq.~\eqref{eq:chisq} over the parameters $\alpha_*,\,\beta_*$ and $\epsilon_*$. The $\dagger$ indicates that only Poisson error is included (although unrealistic due to cosmic variance).
    }
    \label{tab:JWST}
\end{table}

Given the JWST and NGRST mock data, we construct a $\chi^2$ in a similar fashion as for the HST data and follow the same two approaches as in Section~\ref{sec:results} to forecast constraints on $\fNL$. Here, we again use a cut-off scale in the bispectrum of $k_\mathrm{cut} = 0.1\,\mathrm{Mpc}^{-1}$. We summarise our forecasts in Table~\ref{tab:JWST}, where we directly marginalized the $\chi^2$ to obtain the bounds on $\fNL$. The values in this table are calculated with respect to the median marginalized $\Delta\chi^2$.
NGRST is found to give slightly weaker constraints than JWST, which is mainly due to the lower value of $m_\mathrm{lim}$, i.e., only the brightest galaxies can be observed at high redshifts. When combining the two mock data sets, without overlapping the covered magnitude ranges, the bounds can be further improved. We also ran an MCMC simulation with the JWST mock data (since it is expected to give the strongest constraints) and show the posteriors in Appendix~\ref{app:JWST_posteriors}. All in all, JWST and NGRST would be able to improve upon our current bounds based on HST observations roughly by a factor $3-4$ under conservative assumptions (10\% minimum error) and up to an order of magnitude for more optimistic assumptions.
Moreover, given this factor of $3-4$ improvement, JWST and NGRST will be able to either alleviate or further strengthen (to ${\sim}3\sigma$) the deviation from zero in our bounds on $f_\mathrm{NL}$ for $k_\mathrm{cut} \sim 1\,\mathrm{Mpc}^{-1}$ (see Figure~\ref{fig:bounds_kcut} and Section~\ref{subsec:other_kcuts} for details).

\section{Conclusions}
\label{sec:conclusions}
In this work we have demonstrated the ability of UV luminosity functions to probe small-scale non-Gaussianity. This opens a window into the physics of the highest energies known, cosmic inflation,
as well as other primordial phenomena happening at {\it small scales}, such as the production of PBHs. 
We focused on non-Gaussianity manifested at scales smaller than those probed by the CMB and LSS, for which there are no other current bounds, cf. Figure~\ref{fig:current_status}.
We have shown that constraints can already be obtained from HST observations. By using UV LF data from the Hubble Legacy Fields and Hubble Frontier Fields catalogs, we have put bounds on the non-Gaussianity parameter $f_\mathrm{NL}$ and examined its robustness with regards to several assumptions in our analysis. The approach of this work is described in Sections~\ref{sec:UVLF} and \ref{sec:primordial_nonGauss}, the results are presented in Section~\ref{sec:results} and forecasts are made in Section~\ref{sec:forecasts}. We conclude that:

\begin{itemize}
    \item Small-scale non-Gaussianity affects the UV luminosity function mostly at the bright end. While there are degeneracies between $f_\mathrm{NL}$ and some astrophysical parameters, these can be broken by combining data at different redshifts.
    \item Current observations of the UV luminosity function can provide robust bounds on small-scale non-Gaussianity. Our main analysis is performed by using UV LF data from the HLF catalog and assuming a cut-off scale in the bispectrum of $0.1\,\mathrm{Mpc}^{-1}$. We obtain constraints on $f_\mathrm{NL}$ of $71^{+221}_{-119}$ at $1\sigma$ and $71^{+426}_{-237}$ at $2\sigma$.
    These are comparable to the results obtained with HFF data or under different assumptions regarding the astrophysical parameters.
    \item JWST and NGRST can further improve upon these bounds by a factor $3-4$. A set of forecasts shows that such experiments would be able to reduce the error on $f_\mathrm{NL}$ down to $\Delta f_\mathrm{NL}\sim100$ at $2\sigma$.
\end{itemize}

Having established the formalism of the UV luminosity function as a probe of small-scale non-Gaussianity, it is important to consider the origin of the non-zero best-fit amplitude that we find for a cut-off scale of $k_\mathrm{cut} = 1\,\mathrm{Mpc}^{-1}$ in the bispectrum (Section~\ref{subsec:other_kcuts}).
This anomaly persists for all UV LF data sets considered.
We have shown that JWST and NGRST are able to address this issue. Another promising way forward would be to study the impact of such small-scale PNG on the 21-cm cosmic-dawn signal measured by upcoming global-signal and interferometric 21-cm experiments (see Appendix~\ref{subsec:21cm_forecast}).
In addition, a more sophisticated forecast for JWST would give a better picture of the smallest scales that can be affected by PNG. This is particularly the case when a combination of different observational configurations and longer exposure times is considered. 

\vspace{6pt}

In conclusion, our work establishes the UV LF as a powerful probe of the fundamental processes that were at play in the early Universe. Upcoming surveys will offer an exciting possibility to unveil the origin of structures in our cosmos and in which process the UV LF will play a prominent role.

\section*{acknowledgements}
We thank Sandro Tacchella and Andrei Mesinger for insightful comments on the draft of this work. We are also grateful to Gonzalo Palma for discussions on PNG, and the anonymous referee for providing useful feedback on this paper. We acknowledge the use of the packages \texttt{emcee}~\cite{ForemanMackey:2012ig} and \texttt{corner}~\cite{corner}. NS is a recipient of a King's College London NMS Faculty Studentship.
JBM is supported by NSF grant AST-1813694
at Harvard and the Clay Fellowship at the Smithsonian
Astrophysical Observatory.

\bibliographystyle{apsrev4-1}
\bibliography{biblio}

\begin{thebibliography}{124}%
\makeatletter
\providecommand \@ifxundefined [1]{%
 \@ifx{#1\undefined}
}%
\providecommand \@ifnum [1]{%
 \ifnum #1\expandafter \@firstoftwo
 \else \expandafter \@secondoftwo
 \fi
}%
\providecommand \@ifx [1]{%
 \ifx #1\expandafter \@firstoftwo
 \else \expandafter \@secondoftwo
 \fi
}%
\providecommand \natexlab [1]{#1}%
\providecommand \enquote  [1]{``#1''}%
\providecommand \bibnamefont  [1]{#1}%
\providecommand \bibfnamefont [1]{#1}%
\providecommand \citenamefont [1]{#1}%
\providecommand \href@noop [0]{\@secondoftwo}%
\providecommand \href [0]{\begingroup \@sanitize@url \@href}%
\providecommand \@href[1]{\@@startlink{#1}\@@href}%
\providecommand \@@href[1]{\endgroup#1\@@endlink}%
\providecommand \@sanitize@url [0]{\catcode `\\12\catcode `\$12\catcode
  `\&12\catcode `\#12\catcode `\^12\catcode `\_12\catcode `\%12\relax}%
\providecommand \@@startlink[1]{}%
\providecommand \@@endlink[0]{}%
\providecommand \url  [0]{\begingroup\@sanitize@url \@url }%
\providecommand \@url [1]{\endgroup\@href {#1}{\urlprefix }}%
\providecommand \urlprefix  [0]{URL }%
\providecommand \Eprint [0]{\href }%
\providecommand \doibase [0]{http://dx.doi.org/}%
\providecommand \selectlanguage [0]{\@gobble}%
\providecommand \bibinfo  [0]{\@secondoftwo}%
\providecommand \bibfield  [0]{\@secondoftwo}%
\providecommand \translation [1]{[#1]}%
\providecommand \BibitemOpen [0]{}%
\providecommand \bibitemStop [0]{}%
\providecommand \bibitemNoStop [0]{.\EOS\space}%
\providecommand \EOS [0]{\spacefactor3000\relax}%
\providecommand \BibitemShut  [1]{\csname bibitem#1\endcsname}%
\let\auto@bib@innerbib\@empty
\bibitem [{\citenamefont {Akrami}\ \emph {et~al.}(2018)\citenamefont {Akrami}
  \emph {et~al.}}]{Akrami:2018vks}%
  \BibitemOpen
  \bibfield  {author} {\bibinfo {author} {\bibfnamefont {Y.}~\bibnamefont
  {Akrami}} \emph {et~al.} (\bibinfo {collaboration} {Planck}),\ }\href@noop {}
  {\  (\bibinfo {year} {2018})},\ \Eprint {http://arxiv.org/abs/1807.06205}
  {arXiv:1807.06205 [astro-ph.CO]} \BibitemShut {NoStop}%
\bibitem [{\citenamefont {Abazajian}\ \emph {et~al.}(2009)\citenamefont
  {Abazajian} \emph {et~al.}}]{Abazajian:2008wr}%
  \BibitemOpen
  \bibfield  {author} {\bibinfo {author} {\bibfnamefont {K.~N.}\ \bibnamefont
  {Abazajian}} \emph {et~al.} (\bibinfo {collaboration} {SDSS}),\ }\href
  {\doibase 10.1088/0067-0049/182/2/543} {\bibfield  {journal} {\bibinfo
  {journal} {Astrophys. J. Suppl.}\ }\textbf {\bibinfo {volume} {182}},\
  \bibinfo {pages} {543} (\bibinfo {year} {2009})},\ \Eprint
  {http://arxiv.org/abs/0812.0649} {arXiv:0812.0649 [astro-ph]} \BibitemShut
  {NoStop}%
\bibitem [{\citenamefont {Abbott}\ \emph {et~al.}(2005)\citenamefont {Abbott}
  \emph {et~al.}}]{Abbott:2005bi}%
  \BibitemOpen
  \bibfield  {author} {\bibinfo {author} {\bibfnamefont {T.}~\bibnamefont
  {Abbott}} \emph {et~al.} (\bibinfo {collaboration} {DES}),\ }\href@noop {} {\
   (\bibinfo {year} {2005})},\ \Eprint {http://arxiv.org/abs/astro-ph/0510346}
  {arXiv:astro-ph/0510346} \BibitemShut {NoStop}%
\bibitem [{\citenamefont {Hu}(2000)}]{Hu:1999vq}%
  \BibitemOpen
  \bibfield  {author} {\bibinfo {author} {\bibfnamefont {W.}~\bibnamefont
  {Hu}},\ }\href {\doibase 10.1086/308279} {\bibfield  {journal} {\bibinfo
  {journal} {Astrophys. J.}\ }\textbf {\bibinfo {volume} {529}},\ \bibinfo
  {pages} {12} (\bibinfo {year} {2000})},\ \Eprint
  {http://arxiv.org/abs/astro-ph/9907103} {arXiv:astro-ph/9907103} \BibitemShut
  {NoStop}%
\bibitem [{\citenamefont {Adam}\ \emph {et~al.}(2016)\citenamefont {Adam} \emph
  {et~al.}}]{Adam:2016hgk}%
  \BibitemOpen
  \bibfield  {author} {\bibinfo {author} {\bibfnamefont {R.}~\bibnamefont
  {Adam}} \emph {et~al.} (\bibinfo {collaboration} {Planck}),\ }\href {\doibase
  10.1051/0004-6361/201628897} {\bibfield  {journal} {\bibinfo  {journal}
  {Astron. Astrophys.}\ }\textbf {\bibinfo {volume} {596}},\ \bibinfo {pages}
  {A108} (\bibinfo {year} {2016})},\ \Eprint {http://arxiv.org/abs/1605.03507}
  {arXiv:1605.03507 [astro-ph.CO]} \BibitemShut {NoStop}%
\bibitem [{\citenamefont {Barkana}\ and\ \citenamefont
  {Loeb}(2001)}]{Barkana:2000fd}%
  \BibitemOpen
  \bibfield  {author} {\bibinfo {author} {\bibfnamefont {R.}~\bibnamefont
  {Barkana}}\ and\ \bibinfo {author} {\bibfnamefont {A.}~\bibnamefont {Loeb}},\
  }\href {\doibase 10.1016/S0370-1573(01)00019-9} {\bibfield  {journal}
  {\bibinfo  {journal} {Phys. Rept.}\ }\textbf {\bibinfo {volume} {349}},\
  \bibinfo {pages} {125} (\bibinfo {year} {2001})},\ \Eprint
  {http://arxiv.org/abs/astro-ph/0010468} {arXiv:astro-ph/0010468} \BibitemShut
  {NoStop}%
\bibitem [{\citenamefont {Becker}\ \emph {et~al.}(2001)\citenamefont {Becker}
  \emph {et~al.}}]{Becker:2001ee}%
  \BibitemOpen
  \bibfield  {author} {\bibinfo {author} {\bibfnamefont {R.~H.}\ \bibnamefont
  {Becker}} \emph {et~al.} (\bibinfo {collaboration} {SDSS}),\ }\href {\doibase
  10.1086/324231} {\bibfield  {journal} {\bibinfo  {journal} {Astron. J.}\
  }\textbf {\bibinfo {volume} {122}},\ \bibinfo {pages} {2850} (\bibinfo {year}
  {2001})},\ \Eprint {http://arxiv.org/abs/astro-ph/0108097}
  {arXiv:astro-ph/0108097} \BibitemShut {NoStop}%
\bibitem [{\citenamefont {Morales}\ and\ \citenamefont
  {Wyithe}(2010)}]{Morales:2009gs}%
  \BibitemOpen
  \bibfield  {author} {\bibinfo {author} {\bibfnamefont {M.~F.}\ \bibnamefont
  {Morales}}\ and\ \bibinfo {author} {\bibfnamefont {J.~B.}\ \bibnamefont
  {Wyithe}},\ }\href {\doibase 10.1146/annurev-astro-081309-130936} {\bibfield
  {journal} {\bibinfo  {journal} {Ann. Rev. Astron. Astrophys.}\ }\textbf
  {\bibinfo {volume} {48}},\ \bibinfo {pages} {127} (\bibinfo {year} {2010})},\
  \Eprint {http://arxiv.org/abs/0910.3010} {arXiv:0910.3010 [astro-ph.CO]}
  \BibitemShut {NoStop}%
\bibitem [{\citenamefont {Bouwens}\ \emph {et~al.}(2015)\citenamefont {Bouwens}
  \emph {et~al.}}]{Bouwens:2014fua}%
  \BibitemOpen
  \bibfield  {author} {\bibinfo {author} {\bibfnamefont {R.}~\bibnamefont
  {Bouwens}} \emph {et~al.},\ }\href {\doibase 10.1088/0004-637X/803/1/34}
  {\bibfield  {journal} {\bibinfo  {journal} {Astrophys. J.}\ }\textbf
  {\bibinfo {volume} {803}},\ \bibinfo {pages} {34} (\bibinfo {year} {2015})},\
  \Eprint {http://arxiv.org/abs/1403.4295} {arXiv:1403.4295 [astro-ph.CO]}
  \BibitemShut {NoStop}%
\bibitem [{\citenamefont {Finkelstein}\ \emph {et~al.}(2015)\citenamefont
  {Finkelstein} \emph {et~al.}}]{Finkelstein_2015}%
  \BibitemOpen
  \bibfield  {author} {\bibinfo {author} {\bibfnamefont {S.~L.}\ \bibnamefont
  {Finkelstein}} \emph {et~al.},\ }\href {\doibase 10.1088/0004-637x/810/1/71}
  {\bibfield  {journal} {\bibinfo  {journal} {The Astrophysical Journal}\
  }\textbf {\bibinfo {volume} {810}},\ \bibinfo {pages} {71} (\bibinfo {year}
  {2015})}\BibitemShut {NoStop}%
\bibitem [{\citenamefont {Atek}\ \emph {et~al.}(2015)\citenamefont {Atek} \emph
  {et~al.}}]{Atek:2015axa}%
  \BibitemOpen
  \bibfield  {author} {\bibinfo {author} {\bibfnamefont {H.}~\bibnamefont
  {Atek}} \emph {et~al.},\ }\href {\doibase 10.1088/0004-637X/814/1/69}
  {\bibfield  {journal} {\bibinfo  {journal} {Astrophys. J.}\ }\textbf
  {\bibinfo {volume} {814}},\ \bibinfo {pages} {69} (\bibinfo {year} {2015})},\
  \Eprint {http://arxiv.org/abs/1509.06764} {arXiv:1509.06764 [astro-ph.GA]}
  \BibitemShut {NoStop}%
\bibitem [{\citenamefont {Livermore}\ \emph {et~al.}(2017)\citenamefont
  {Livermore}, \citenamefont {Finkelstein},\ and\ \citenamefont
  {Lotz}}]{Livermore:2016mbs}%
  \BibitemOpen
  \bibfield  {author} {\bibinfo {author} {\bibfnamefont {R.}~\bibnamefont
  {Livermore}}, \bibinfo {author} {\bibfnamefont {S.}~\bibnamefont
  {Finkelstein}}, \ and\ \bibinfo {author} {\bibfnamefont {J.}~\bibnamefont
  {Lotz}},\ }\href {\doibase 10.3847/1538-4357/835/2/113} {\bibfield  {journal}
  {\bibinfo  {journal} {Astrophys. J.}\ }\textbf {\bibinfo {volume} {835}},\
  \bibinfo {pages} {113} (\bibinfo {year} {2017})},\ \Eprint
  {http://arxiv.org/abs/1604.06799} {arXiv:1604.06799 [astro-ph.GA]}
  \BibitemShut {NoStop}%
\bibitem [{\citenamefont {Bouwens}\ \emph {et~al.}(2017)\citenamefont
  {Bouwens}, \citenamefont {Oesch}, \citenamefont {Illingworth}, \citenamefont
  {Ellis},\ and\ \citenamefont {Stefanon}}]{Bouwens_2017asdasd}%
  \BibitemOpen
  \bibfield  {author} {\bibinfo {author} {\bibfnamefont {R.~J.}\ \bibnamefont
  {Bouwens}}, \bibinfo {author} {\bibfnamefont {P.~A.}\ \bibnamefont {Oesch}},
  \bibinfo {author} {\bibfnamefont {G.~D.}\ \bibnamefont {Illingworth}},
  \bibinfo {author} {\bibfnamefont {R.~S.}\ \bibnamefont {Ellis}}, \ and\
  \bibinfo {author} {\bibfnamefont {M.}~\bibnamefont {Stefanon}},\ }\href
  {\doibase 10.3847/1538-4357/aa70a4} {\bibfield  {journal} {\bibinfo
  {journal} {The Astrophysical Journal}\ }\textbf {\bibinfo {volume} {843}},\
  \bibinfo {pages} {129} (\bibinfo {year} {2017})}\BibitemShut {NoStop}%
\bibitem [{\citenamefont {Mehta}\ \emph {et~al.}(2017)\citenamefont {Mehta}
  \emph {et~al.}}]{Mehta_2017}%
  \BibitemOpen
  \bibfield  {author} {\bibinfo {author} {\bibfnamefont {V.}~\bibnamefont
  {Mehta}} \emph {et~al.},\ }\href {\doibase 10.3847/1538-4357/aa6259}
  {\bibfield  {journal} {\bibinfo  {journal} {The Astrophysical Journal}\
  }\textbf {\bibinfo {volume} {838}},\ \bibinfo {pages} {29} (\bibinfo {year}
  {2017})},\ \Eprint {http://arxiv.org/abs/1702.06953} {arXiv:1702.06953
  [astro-ph.GA]} \BibitemShut {NoStop}%
\bibitem [{\citenamefont {Ishigaki}\ \emph {et~al.}(2018)\citenamefont
  {Ishigaki}, \citenamefont {Kawamata}, \citenamefont {Ouchi}, \citenamefont
  {Oguri}, \citenamefont {Shimasaku},\ and\ \citenamefont
  {Ono}}]{Ishigaki_2018}%
  \BibitemOpen
  \bibfield  {author} {\bibinfo {author} {\bibfnamefont {M.}~\bibnamefont
  {Ishigaki}}, \bibinfo {author} {\bibfnamefont {R.}~\bibnamefont {Kawamata}},
  \bibinfo {author} {\bibfnamefont {M.}~\bibnamefont {Ouchi}}, \bibinfo
  {author} {\bibfnamefont {M.}~\bibnamefont {Oguri}}, \bibinfo {author}
  {\bibfnamefont {K.}~\bibnamefont {Shimasaku}}, \ and\ \bibinfo {author}
  {\bibfnamefont {Y.}~\bibnamefont {Ono}},\ }\href {\doibase
  10.3847/1538-4357/aaa544} {\bibfield  {journal} {\bibinfo  {journal} {The
  Astrophysical Journal}\ }\textbf {\bibinfo {volume} {854}},\ \bibinfo {pages}
  {73} (\bibinfo {year} {2018})}\BibitemShut {NoStop}%
\bibitem [{\citenamefont {Oesch}\ \emph {et~al.}(2018)\citenamefont {Oesch},
  \citenamefont {Bouwens}, \citenamefont {Illingworth}, \citenamefont
  {Labb\'{e}},\ and\ \citenamefont {Stefanon}}]{Oesch_2018}%
  \BibitemOpen
  \bibfield  {author} {\bibinfo {author} {\bibfnamefont {P.~A.}\ \bibnamefont
  {Oesch}}, \bibinfo {author} {\bibfnamefont {R.~J.}\ \bibnamefont {Bouwens}},
  \bibinfo {author} {\bibfnamefont {G.~D.}\ \bibnamefont {Illingworth}},
  \bibinfo {author} {\bibfnamefont {I.}~\bibnamefont {Labb\'{e}}}, \ and\
  \bibinfo {author} {\bibfnamefont {M.}~\bibnamefont {Stefanon}},\ }\href
  {\doibase 10.3847/1538-4357/aab03f} {\bibfield  {journal} {\bibinfo
  {journal} {The Astrophysical Journal}\ }\textbf {\bibinfo {volume} {855}},\
  \bibinfo {pages} {105} (\bibinfo {year} {2018})}\BibitemShut {NoStop}%
\bibitem [{\citenamefont {Atek}\ \emph {et~al.}(2018)\citenamefont {Atek},
  \citenamefont {Richard}, \citenamefont {Kneib},\ and\ \citenamefont
  {Schaerer}}]{Atek:2018nsc}%
  \BibitemOpen
  \bibfield  {author} {\bibinfo {author} {\bibfnamefont {H.}~\bibnamefont
  {Atek}}, \bibinfo {author} {\bibfnamefont {J.}~\bibnamefont {Richard}},
  \bibinfo {author} {\bibfnamefont {J.-P.}\ \bibnamefont {Kneib}}, \ and\
  \bibinfo {author} {\bibfnamefont {D.}~\bibnamefont {Schaerer}},\ }\href
  {\doibase 10.1093/mnras/sty1820} {\bibfield  {journal} {\bibinfo  {journal}
  {Mon. Not. Roy. Astron. Soc.}\ }\textbf {\bibinfo {volume} {479}},\ \bibinfo
  {pages} {5184} (\bibinfo {year} {2018})},\ \Eprint
  {http://arxiv.org/abs/1803.09747} {arXiv:1803.09747 [astro-ph.GA]}
  \BibitemShut {NoStop}%
\bibitem [{\citenamefont {Tacchella}\ \emph {et~al.}(2013)\citenamefont
  {Tacchella}, \citenamefont {Trenti},\ and\ \citenamefont
  {Carollo}}]{Tacchella:2012ih}%
  \BibitemOpen
  \bibfield  {author} {\bibinfo {author} {\bibfnamefont {S.}~\bibnamefont
  {Tacchella}}, \bibinfo {author} {\bibfnamefont {M.}~\bibnamefont {Trenti}}, \
  and\ \bibinfo {author} {\bibfnamefont {C.}~\bibnamefont {Carollo}},\ }\href
  {\doibase 10.1088/2041-8205/768/2/L37} {\bibfield  {journal} {\bibinfo
  {journal} {Astrophys. J. Lett.}\ }\textbf {\bibinfo {volume} {768}},\
  \bibinfo {pages} {L37} (\bibinfo {year} {2013})},\ \Eprint
  {http://arxiv.org/abs/1211.2825} {arXiv:1211.2825 [astro-ph.CO]} \BibitemShut
  {NoStop}%
\bibitem [{\citenamefont {Chevallard}\ \emph {et~al.}(2015)\citenamefont
  {Chevallard}, \citenamefont {Silk}, \citenamefont {Nishimichi}, \citenamefont
  {Habouzit}, \citenamefont {Mamon},\ and\ \citenamefont
  {Peirani}}]{Chevallard:2014sxa}%
  \BibitemOpen
  \bibfield  {author} {\bibinfo {author} {\bibfnamefont {J.}~\bibnamefont
  {Chevallard}}, \bibinfo {author} {\bibfnamefont {J.}~\bibnamefont {Silk}},
  \bibinfo {author} {\bibfnamefont {T.}~\bibnamefont {Nishimichi}}, \bibinfo
  {author} {\bibfnamefont {M.}~\bibnamefont {Habouzit}}, \bibinfo {author}
  {\bibfnamefont {G.~A.}\ \bibnamefont {Mamon}}, \ and\ \bibinfo {author}
  {\bibfnamefont {S.}~\bibnamefont {Peirani}},\ }\href {\doibase
  10.1093/mnras/stu2280} {\bibfield  {journal} {\bibinfo  {journal} {Mon. Not.
  Roy. Astron. Soc.}\ }\textbf {\bibinfo {volume} {446}},\ \bibinfo {pages}
  {3235} (\bibinfo {year} {2015})},\ \Eprint {http://arxiv.org/abs/1410.7768}
  {arXiv:1410.7768 [astro-ph.CO]} \BibitemShut {NoStop}%
\bibitem [{\citenamefont {Dayal}\ \emph {et~al.}(2015)\citenamefont {Dayal},
  \citenamefont {Mesinger},\ and\ \citenamefont {Pacucci}}]{Dayal:2014nva}%
  \BibitemOpen
  \bibfield  {author} {\bibinfo {author} {\bibfnamefont {P.}~\bibnamefont
  {Dayal}}, \bibinfo {author} {\bibfnamefont {A.}~\bibnamefont {Mesinger}}, \
  and\ \bibinfo {author} {\bibfnamefont {F.}~\bibnamefont {Pacucci}},\ }\href
  {\doibase 10.1088/0004-637X/806/1/67} {\bibfield  {journal} {\bibinfo
  {journal} {Astrophys. J.}\ }\textbf {\bibinfo {volume} {806}},\ \bibinfo
  {pages} {67} (\bibinfo {year} {2015})},\ \Eprint
  {http://arxiv.org/abs/1408.1102} {arXiv:1408.1102 [astro-ph.GA]} \BibitemShut
  {NoStop}%
\bibitem [{\citenamefont {Corasaniti}\ \emph {et~al.}(2017)\citenamefont
  {Corasaniti}, \citenamefont {Agarwal}, \citenamefont {Marsh},\ and\
  \citenamefont {Das}}]{Corasaniti:2016epp}%
  \BibitemOpen
  \bibfield  {author} {\bibinfo {author} {\bibfnamefont {P.}~\bibnamefont
  {Corasaniti}}, \bibinfo {author} {\bibfnamefont {S.}~\bibnamefont {Agarwal}},
  \bibinfo {author} {\bibfnamefont {D.}~\bibnamefont {Marsh}}, \ and\ \bibinfo
  {author} {\bibfnamefont {S.}~\bibnamefont {Das}},\ }\href {\doibase
  10.1103/PhysRevD.95.083512} {\bibfield  {journal} {\bibinfo  {journal} {Phys.
  Rev. D}\ }\textbf {\bibinfo {volume} {95}},\ \bibinfo {pages} {083512}
  (\bibinfo {year} {2017})},\ \Eprint {http://arxiv.org/abs/1611.05892}
  {arXiv:1611.05892 [astro-ph.CO]} \BibitemShut {NoStop}%
\bibitem [{\citenamefont {Menci}\ \emph {et~al.}(2017)\citenamefont {Menci},
  \citenamefont {Merle}, \citenamefont {Totzauer}, \citenamefont {Schneider},
  \citenamefont {Grazian}, \citenamefont {Castellano},\ and\ \citenamefont
  {Sanchez}}]{Menci:2017nsr}%
  \BibitemOpen
  \bibfield  {author} {\bibinfo {author} {\bibfnamefont {N.}~\bibnamefont
  {Menci}}, \bibinfo {author} {\bibfnamefont {A.}~\bibnamefont {Merle}},
  \bibinfo {author} {\bibfnamefont {M.}~\bibnamefont {Totzauer}}, \bibinfo
  {author} {\bibfnamefont {A.}~\bibnamefont {Schneider}}, \bibinfo {author}
  {\bibfnamefont {A.}~\bibnamefont {Grazian}}, \bibinfo {author} {\bibfnamefont
  {M.}~\bibnamefont {Castellano}}, \ and\ \bibinfo {author} {\bibfnamefont
  {N.~G.}\ \bibnamefont {Sanchez}},\ }\href {\doibase
  10.3847/1538-4357/836/1/61} {\bibfield  {journal} {\bibinfo  {journal}
  {Astrophys. J.}\ }\textbf {\bibinfo {volume} {836}},\ \bibinfo {pages} {61}
  (\bibinfo {year} {2017})},\ \Eprint {http://arxiv.org/abs/1701.01339}
  {arXiv:1701.01339 [astro-ph.CO]} \BibitemShut {NoStop}%
\bibitem [{\citenamefont {Yue}\ \emph {et~al.}(2018)\citenamefont {Yue} \emph
  {et~al.}}]{Yue:2017hbz}%
  \BibitemOpen
  \bibfield  {author} {\bibinfo {author} {\bibfnamefont {B.}~\bibnamefont
  {Yue}} \emph {et~al.},\ }\href {\doibase 10.3847/1538-4357/aae77f} {\bibfield
   {journal} {\bibinfo  {journal} {Astrophys. J.}\ }\textbf {\bibinfo {volume}
  {868}},\ \bibinfo {pages} {115} (\bibinfo {year} {2018})},\ \Eprint
  {http://arxiv.org/abs/1711.05130} {arXiv:1711.05130 [astro-ph.GA]}
  \BibitemShut {NoStop}%
\bibitem [{\citenamefont {Lovell}\ \emph {et~al.}(2018)\citenamefont {Lovell}
  \emph {et~al.}}]{Lovell:2017eec}%
  \BibitemOpen
  \bibfield  {author} {\bibinfo {author} {\bibfnamefont {M.~R.}\ \bibnamefont
  {Lovell}} \emph {et~al.},\ }\href {\doibase 10.1093/mnras/sty818} {\bibfield
  {journal} {\bibinfo  {journal} {Mon. Not. Roy. Astron. Soc.}\ }\textbf
  {\bibinfo {volume} {477}},\ \bibinfo {pages} {2886} (\bibinfo {year}
  {2018})},\ \Eprint {http://arxiv.org/abs/1711.10497} {arXiv:1711.10497
  [astro-ph.CO]} \BibitemShut {NoStop}%
\bibitem [{\citenamefont {Unal}(2019)}]{Unal:2018yaa}%
  \BibitemOpen
  \bibfield  {author} {\bibinfo {author} {\bibfnamefont {C.}~\bibnamefont
  {Unal}},\ }\href {\doibase 10.1103/PhysRevD.99.041301} {\bibfield  {journal}
  {\bibinfo  {journal} {Phys. Rev. D}\ }\textbf {\bibinfo {volume} {99}},\
  \bibinfo {pages} {041301} (\bibinfo {year} {2019})},\ \Eprint
  {http://arxiv.org/abs/1811.09151} {arXiv:1811.09151 [astro-ph.CO]}
  \BibitemShut {NoStop}%
\bibitem [{\citenamefont {Ir\v{si}\v{c}}\ \emph {et~al.}(2020)\citenamefont
  {Ir\v{si}\v{c}}, \citenamefont {Xiao},\ and\ \citenamefont
  {McQuinn}}]{Irsic:2019iff}%
  \BibitemOpen
  \bibfield  {author} {\bibinfo {author} {\bibfnamefont {V.}~\bibnamefont
  {Ir\v{si}\v{c}}}, \bibinfo {author} {\bibfnamefont {H.}~\bibnamefont {Xiao}},
  \ and\ \bibinfo {author} {\bibfnamefont {M.}~\bibnamefont {McQuinn}},\ }\href
  {\doibase 10.1103/PhysRevD.101.123518} {\bibfield  {journal} {\bibinfo
  {journal} {Phys. Rev. D}\ }\textbf {\bibinfo {volume} {101}},\ \bibinfo
  {pages} {123518} (\bibinfo {year} {2020})},\ \Eprint
  {http://arxiv.org/abs/1911.11150} {arXiv:1911.11150 [astro-ph.CO]}
  \BibitemShut {NoStop}%
\bibitem [{\citenamefont {Yoshiura}\ \emph {et~al.}(2020)\citenamefont
  {Yoshiura}, \citenamefont {Oguri}, \citenamefont {Takahashi},\ and\
  \citenamefont {Takahashi}}]{Yoshiura:1809192}%
  \BibitemOpen
  \bibfield  {author} {\bibinfo {author} {\bibfnamefont {S.}~\bibnamefont
  {Yoshiura}}, \bibinfo {author} {\bibfnamefont {M.}~\bibnamefont {Oguri}},
  \bibinfo {author} {\bibfnamefont {K.}~\bibnamefont {Takahashi}}, \ and\
  \bibinfo {author} {\bibfnamefont {T.}~\bibnamefont {Takahashi}},\ }\href@noop
  {} {\  (\bibinfo {year} {2020})},\ \Eprint {http://arxiv.org/abs/2007.14695}
  {arXiv:2007.14695 [astro-ph.CO]} \BibitemShut {NoStop}%
\bibitem [{\citenamefont {Maldacena}(2003)}]{Maldacena:2002vr}%
  \BibitemOpen
  \bibfield  {author} {\bibinfo {author} {\bibfnamefont {J.~M.}\ \bibnamefont
  {Maldacena}},\ }\href {\doibase 10.1088/1126-6708/2003/05/013} {\bibfield
  {journal} {\bibinfo  {journal} {JHEP}\ }\textbf {\bibinfo {volume} {05}},\
  \bibinfo {pages} {013} (\bibinfo {year} {2003})},\ \Eprint
  {http://arxiv.org/abs/astro-ph/0210603} {arXiv:astro-ph/0210603} \BibitemShut
  {NoStop}%
\bibitem [{\citenamefont {Celoria}\ and\ \citenamefont
  {Matarrese}(2018)}]{Celoria:2018euj}%
  \BibitemOpen
  \bibfield  {author} {\bibinfo {author} {\bibfnamefont {M.}~\bibnamefont
  {Celoria}}\ and\ \bibinfo {author} {\bibfnamefont {S.}~\bibnamefont
  {Matarrese}}\ }(\bibinfo {year} {2018})\ \Eprint
  {http://arxiv.org/abs/1812.08197} {arXiv:1812.08197 [astro-ph.CO]}
  \BibitemShut {NoStop}%
\bibitem [{\citenamefont {Guth}(1987)}]{Guth:1980zm}%
  \BibitemOpen
  \bibfield  {author} {\bibinfo {author} {\bibfnamefont {A.~H.}\ \bibnamefont
  {Guth}},\ }\href {\doibase 10.1103/PhysRevD.23.347} {\bibfield  {journal}
  {\bibinfo  {journal} {Adv. Ser. Astrophys. Cosmol.}\ }\textbf {\bibinfo
  {volume} {3}},\ \bibinfo {pages} {139} (\bibinfo {year} {1987})}\BibitemShut
  {NoStop}%
\bibitem [{\citenamefont {Linde}(1987)}]{Linde:1981mu}%
  \BibitemOpen
  \bibfield  {author} {\bibinfo {author} {\bibfnamefont {A.~D.}\ \bibnamefont
  {Linde}},\ }\href {\doibase 10.1016/0370-2693(82)91219-9} {\bibfield
  {journal} {\bibinfo  {journal} {Adv. Ser. Astrophys. Cosmol.}\ }\textbf
  {\bibinfo {volume} {3}},\ \bibinfo {pages} {149} (\bibinfo {year}
  {1987})}\BibitemShut {NoStop}%
\bibitem [{\citenamefont {Baumann}(2011)}]{Baumann:2009ds}%
  \BibitemOpen
  \bibfield  {author} {\bibinfo {author} {\bibfnamefont {D.}~\bibnamefont
  {Baumann}}\ }(\bibinfo {year} {2011})\ pp.\ \bibinfo {pages} {523--686},\
  \Eprint {http://arxiv.org/abs/0907.5424} {arXiv:0907.5424 [hep-th]}
  \BibitemShut {NoStop}%
\bibitem [{\citenamefont {Cheung}\ \emph {et~al.}(2008)\citenamefont {Cheung},
  \citenamefont {Creminelli}, \citenamefont {Fitzpatrick}, \citenamefont
  {Kaplan},\ and\ \citenamefont {Senatore}}]{Cheung:2007st}%
  \BibitemOpen
  \bibfield  {author} {\bibinfo {author} {\bibfnamefont {C.}~\bibnamefont
  {Cheung}}, \bibinfo {author} {\bibfnamefont {P.}~\bibnamefont {Creminelli}},
  \bibinfo {author} {\bibfnamefont {A.}~\bibnamefont {Fitzpatrick}}, \bibinfo
  {author} {\bibfnamefont {J.}~\bibnamefont {Kaplan}}, \ and\ \bibinfo {author}
  {\bibfnamefont {L.}~\bibnamefont {Senatore}},\ }\href {\doibase
  10.1088/1126-6708/2008/03/014} {\bibfield  {journal} {\bibinfo  {journal}
  {JHEP}\ }\textbf {\bibinfo {volume} {03}},\ \bibinfo {pages} {014} (\bibinfo
  {year} {2008})},\ \Eprint {http://arxiv.org/abs/0709.0293} {arXiv:0709.0293
  [hep-th]} \BibitemShut {NoStop}%
\bibitem [{\citenamefont {Arkani-Hamed}\ and\ \citenamefont
  {Maldacena}(2015)}]{Arkani-Hamed:2015bza}%
  \BibitemOpen
  \bibfield  {author} {\bibinfo {author} {\bibfnamefont {N.}~\bibnamefont
  {Arkani-Hamed}}\ and\ \bibinfo {author} {\bibfnamefont {J.}~\bibnamefont
  {Maldacena}},\ }\href@noop {} {\  (\bibinfo {year} {2015})},\ \Eprint
  {http://arxiv.org/abs/1503.08043} {arXiv:1503.08043 [hep-th]} \BibitemShut
  {NoStop}%
\bibitem [{\citenamefont {Biagetti}(2019)}]{Biagetti:2019bnp}%
  \BibitemOpen
  \bibfield  {author} {\bibinfo {author} {\bibfnamefont {M.}~\bibnamefont
  {Biagetti}},\ }\href {\doibase 10.3390/galaxies7030071} {\bibfield  {journal}
  {\bibinfo  {journal} {Galaxies}\ }\textbf {\bibinfo {volume} {7}},\ \bibinfo
  {pages} {71} (\bibinfo {year} {2019})},\ \Eprint
  {http://arxiv.org/abs/1906.12244} {arXiv:1906.12244 [astro-ph.CO]}
  \BibitemShut {NoStop}%
\bibitem [{\citenamefont {Verde}\ \emph {et~al.}(2001)\citenamefont {Verde},
  \citenamefont {Jimenez}, \citenamefont {Kamionkowski},\ and\ \citenamefont
  {Matarrese}}]{Verde:2000vr}%
  \BibitemOpen
  \bibfield  {author} {\bibinfo {author} {\bibfnamefont {L.}~\bibnamefont
  {Verde}}, \bibinfo {author} {\bibfnamefont {R.}~\bibnamefont {Jimenez}},
  \bibinfo {author} {\bibfnamefont {M.}~\bibnamefont {Kamionkowski}}, \ and\
  \bibinfo {author} {\bibfnamefont {S.}~\bibnamefont {Matarrese}},\ }\href
  {\doibase 10.1046/j.1365-8711.2001.04459.x} {\bibfield  {journal} {\bibinfo
  {journal} {Mon. Not. Roy. Astron. Soc.}\ }\textbf {\bibinfo {volume} {325}},\
  \bibinfo {pages} {412} (\bibinfo {year} {2001})},\ \Eprint
  {http://arxiv.org/abs/astro-ph/0011180} {arXiv:astro-ph/0011180} \BibitemShut
  {NoStop}%
\bibitem [{\citenamefont {Komatsu}\ \emph {et~al.}(2009)\citenamefont {Komatsu}
  \emph {et~al.}}]{Komatsu:2009kd}%
  \BibitemOpen
  \bibfield  {author} {\bibinfo {author} {\bibfnamefont {E.}~\bibnamefont
  {Komatsu}} \emph {et~al.},\ }\href@noop {} {\  (\bibinfo {year} {2009})},\
  \Eprint {http://arxiv.org/abs/0902.4759} {arXiv:0902.4759 [astro-ph.CO]}
  \BibitemShut {NoStop}%
\bibitem [{\citenamefont {Byrnes}\ \emph {et~al.}(2010)\citenamefont {Byrnes},
  \citenamefont {Gerstenlauer}, \citenamefont {Nurmi}, \citenamefont
  {Tasinato},\ and\ \citenamefont {Wands}}]{Byrnes:2010ft}%
  \BibitemOpen
  \bibfield  {author} {\bibinfo {author} {\bibfnamefont {C.~T.}\ \bibnamefont
  {Byrnes}}, \bibinfo {author} {\bibfnamefont {M.}~\bibnamefont
  {Gerstenlauer}}, \bibinfo {author} {\bibfnamefont {S.}~\bibnamefont {Nurmi}},
  \bibinfo {author} {\bibfnamefont {G.}~\bibnamefont {Tasinato}}, \ and\
  \bibinfo {author} {\bibfnamefont {D.}~\bibnamefont {Wands}},\ }\href
  {\doibase 10.1088/1475-7516/2010/10/004} {\bibfield  {journal} {\bibinfo
  {journal} {JCAP}\ }\textbf {\bibinfo {volume} {10}},\ \bibinfo {pages} {004}
  (\bibinfo {year} {2010})},\ \Eprint {http://arxiv.org/abs/1007.4277}
  {arXiv:1007.4277 [astro-ph.CO]} \BibitemShut {NoStop}%
\bibitem [{\citenamefont {Pillepich}\ \emph {et~al.}(2010)\citenamefont
  {Pillepich}, \citenamefont {Porciani},\ and\ \citenamefont
  {Hahn}}]{Pillepich:2008ka}%
  \BibitemOpen
  \bibfield  {author} {\bibinfo {author} {\bibfnamefont {A.}~\bibnamefont
  {Pillepich}}, \bibinfo {author} {\bibfnamefont {C.}~\bibnamefont {Porciani}},
  \ and\ \bibinfo {author} {\bibfnamefont {O.}~\bibnamefont {Hahn}},\ }\href
  {\doibase 10.1111/j.1365-2966.2009.15914.x} {\bibfield  {journal} {\bibinfo
  {journal} {Mon. Not. Roy. Astron. Soc.}\ }\textbf {\bibinfo {volume} {402}},\
  \bibinfo {pages} {191} (\bibinfo {year} {2010})},\ \Eprint
  {http://arxiv.org/abs/0811.4176} {arXiv:0811.4176 [astro-ph]} \BibitemShut
  {NoStop}%
\bibitem [{\citenamefont {Mana}\ \emph {et~al.}(2013)\citenamefont {Mana},
  \citenamefont {Giannantonio}, \citenamefont {Weller}, \citenamefont {Hoyle},
  \citenamefont {Huetsi},\ and\ \citenamefont {Sartoris}}]{Mana:2013qba}%
  \BibitemOpen
  \bibfield  {author} {\bibinfo {author} {\bibfnamefont {A.}~\bibnamefont
  {Mana}}, \bibinfo {author} {\bibfnamefont {T.}~\bibnamefont {Giannantonio}},
  \bibinfo {author} {\bibfnamefont {J.}~\bibnamefont {Weller}}, \bibinfo
  {author} {\bibfnamefont {B.}~\bibnamefont {Hoyle}}, \bibinfo {author}
  {\bibfnamefont {G.}~\bibnamefont {Huetsi}}, \ and\ \bibinfo {author}
  {\bibfnamefont {B.}~\bibnamefont {Sartoris}},\ }\href {\doibase
  10.1093/mnras/stt1062} {\bibfield  {journal} {\bibinfo  {journal} {Mon. Not.
  Roy. Astron. Soc.}\ }\textbf {\bibinfo {volume} {434}},\ \bibinfo {pages}
  {684} (\bibinfo {year} {2013})},\ \Eprint {http://arxiv.org/abs/1303.0287}
  {arXiv:1303.0287 [astro-ph.CO]} \BibitemShut {NoStop}%
\bibitem [{\citenamefont {LoVerde}\ \emph {et~al.}(2008)\citenamefont
  {LoVerde}, \citenamefont {Miller}, \citenamefont {Shandera},\ and\
  \citenamefont {Verde}}]{LoVerde:2007ri}%
  \BibitemOpen
  \bibfield  {author} {\bibinfo {author} {\bibfnamefont {M.}~\bibnamefont
  {LoVerde}}, \bibinfo {author} {\bibfnamefont {A.}~\bibnamefont {Miller}},
  \bibinfo {author} {\bibfnamefont {S.}~\bibnamefont {Shandera}}, \ and\
  \bibinfo {author} {\bibfnamefont {L.}~\bibnamefont {Verde}},\ }\href
  {\doibase 10.1088/1475-7516/2008/04/014} {\bibfield  {journal} {\bibinfo
  {journal} {JCAP}\ }\textbf {\bibinfo {volume} {04}},\ \bibinfo {pages} {014}
  (\bibinfo {year} {2008})},\ \Eprint {http://arxiv.org/abs/0711.4126}
  {arXiv:0711.4126 [astro-ph]} \BibitemShut {NoStop}%
\bibitem [{\citenamefont {Jimenez}\ and\ \citenamefont
  {Verde}(2009)}]{Jimenez:2009us}%
  \BibitemOpen
  \bibfield  {author} {\bibinfo {author} {\bibfnamefont {R.}~\bibnamefont
  {Jimenez}}\ and\ \bibinfo {author} {\bibfnamefont {L.}~\bibnamefont
  {Verde}},\ }\href {\doibase 10.1103/PhysRevD.80.127302} {\bibfield  {journal}
  {\bibinfo  {journal} {Phys. Rev. D}\ }\textbf {\bibinfo {volume} {80}},\
  \bibinfo {pages} {127302} (\bibinfo {year} {2009})},\ \Eprint
  {http://arxiv.org/abs/0909.0403} {arXiv:0909.0403 [astro-ph.CO]} \BibitemShut
  {NoStop}%
\bibitem [{\citenamefont {LoVerde}\ and\ \citenamefont
  {Smith}(2011)}]{LoVerde:2011iz}%
  \BibitemOpen
  \bibfield  {author} {\bibinfo {author} {\bibfnamefont {M.}~\bibnamefont
  {LoVerde}}\ and\ \bibinfo {author} {\bibfnamefont {K.~M.}\ \bibnamefont
  {Smith}},\ }\href {\doibase 10.1088/1475-7516/2011/08/003} {\bibfield
  {journal} {\bibinfo  {journal} {JCAP}\ }\textbf {\bibinfo {volume} {1108}},\
  \bibinfo {pages} {003} (\bibinfo {year} {2011})},\ \Eprint
  {http://arxiv.org/abs/1102.1439} {arXiv:1102.1439 [astro-ph.CO]} \BibitemShut
  {NoStop}%
\bibitem [{\citenamefont {Shandera}\ \emph {et~al.}(2013)\citenamefont
  {Shandera}, \citenamefont {Erickcek}, \citenamefont {Scott},\ and\
  \citenamefont {Galarza}}]{Shandera:2012ke}%
  \BibitemOpen
  \bibfield  {author} {\bibinfo {author} {\bibfnamefont {S.}~\bibnamefont
  {Shandera}}, \bibinfo {author} {\bibfnamefont {A.~L.}\ \bibnamefont
  {Erickcek}}, \bibinfo {author} {\bibfnamefont {P.}~\bibnamefont {Scott}}, \
  and\ \bibinfo {author} {\bibfnamefont {J.~Y.}\ \bibnamefont {Galarza}},\
  }\href {\doibase 10.1103/PhysRevD.88.103506} {\bibfield  {journal} {\bibinfo
  {journal} {Phys. Rev. D}\ }\textbf {\bibinfo {volume} {88}},\ \bibinfo
  {pages} {103506} (\bibinfo {year} {2013})},\ \Eprint
  {http://arxiv.org/abs/1211.7361} {arXiv:1211.7361 [astro-ph.CO]} \BibitemShut
  {NoStop}%
\bibitem [{\citenamefont {Akrami}\ \emph {et~al.}(2019)\citenamefont {Akrami}
  \emph {et~al.}}]{Akrami:2019izv}%
  \BibitemOpen
  \bibfield  {author} {\bibinfo {author} {\bibfnamefont {Y.}~\bibnamefont
  {Akrami}} \emph {et~al.} (\bibinfo {collaboration} {Planck}),\ }\href@noop {}
  {\  (\bibinfo {year} {2019})},\ \Eprint {http://arxiv.org/abs/1905.05697}
  {arXiv:1905.05697 [astro-ph.CO]} \BibitemShut {NoStop}%
\bibitem [{\citenamefont {Shirasaki}\ \emph {et~al.}(2012)\citenamefont
  {Shirasaki}, \citenamefont {Yoshida}, \citenamefont {Hamana},\ and\
  \citenamefont {Nishimichi}}]{Shirasaki:2012sx}%
  \BibitemOpen
  \bibfield  {author} {\bibinfo {author} {\bibfnamefont {M.}~\bibnamefont
  {Shirasaki}}, \bibinfo {author} {\bibfnamefont {N.}~\bibnamefont {Yoshida}},
  \bibinfo {author} {\bibfnamefont {T.}~\bibnamefont {Hamana}}, \ and\ \bibinfo
  {author} {\bibfnamefont {T.}~\bibnamefont {Nishimichi}},\ }\href {\doibase
  10.1088/0004-637X/760/1/45} {\bibfield  {journal} {\bibinfo  {journal}
  {Astrophys. J.}\ }\textbf {\bibinfo {volume} {760}},\ \bibinfo {pages} {45}
  (\bibinfo {year} {2012})},\ \Eprint {http://arxiv.org/abs/1204.4981}
  {arXiv:1204.4981 [astro-ph.CO]} \BibitemShut {NoStop}%
\bibitem [{\citenamefont {Leistedt}\ \emph {et~al.}(2014)\citenamefont
  {Leistedt}, \citenamefont {Peiris},\ and\ \citenamefont
  {Roth}}]{Leistedt:2014zqa}%
  \BibitemOpen
  \bibfield  {author} {\bibinfo {author} {\bibfnamefont {B.}~\bibnamefont
  {Leistedt}}, \bibinfo {author} {\bibfnamefont {H.~V.}\ \bibnamefont
  {Peiris}}, \ and\ \bibinfo {author} {\bibfnamefont {N.}~\bibnamefont
  {Roth}},\ }\href {\doibase 10.1103/PhysRevLett.113.221301} {\bibfield
  {journal} {\bibinfo  {journal} {Phys. Rev. Lett.}\ }\textbf {\bibinfo
  {volume} {113}},\ \bibinfo {pages} {221301} (\bibinfo {year} {2014})},\
  \Eprint {http://arxiv.org/abs/1405.4315} {arXiv:1405.4315 [astro-ph.CO]}
  \BibitemShut {NoStop}%
\bibitem [{\citenamefont {Naruko}\ \emph {et~al.}(2015)\citenamefont {Naruko},
  \citenamefont {Ota},\ and\ \citenamefont {Yamaguchi}}]{Naruko:2015pva}%
  \BibitemOpen
  \bibfield  {author} {\bibinfo {author} {\bibfnamefont {A.}~\bibnamefont
  {Naruko}}, \bibinfo {author} {\bibfnamefont {A.}~\bibnamefont {Ota}}, \ and\
  \bibinfo {author} {\bibfnamefont {M.}~\bibnamefont {Yamaguchi}},\ }\href
  {\doibase 10.1088/1475-7516/2015/05/049} {\bibfield  {journal} {\bibinfo
  {journal} {JCAP}\ }\textbf {\bibinfo {volume} {05}},\ \bibinfo {pages} {049}
  (\bibinfo {year} {2015})},\ \Eprint {http://arxiv.org/abs/1503.03722}
  {arXiv:1503.03722 [astro-ph.CO]} \BibitemShut {NoStop}%
\bibitem [{\citenamefont {Emami}\ \emph {et~al.}(2015)\citenamefont {Emami},
  \citenamefont {Dimastrogiovanni}, \citenamefont {Chluba},\ and\ \citenamefont
  {Kamionkowski}}]{Emami:2015xqa}%
  \BibitemOpen
  \bibfield  {author} {\bibinfo {author} {\bibfnamefont {R.}~\bibnamefont
  {Emami}}, \bibinfo {author} {\bibfnamefont {E.}~\bibnamefont
  {Dimastrogiovanni}}, \bibinfo {author} {\bibfnamefont {J.}~\bibnamefont
  {Chluba}}, \ and\ \bibinfo {author} {\bibfnamefont {M.}~\bibnamefont
  {Kamionkowski}},\ }\href {\doibase 10.1103/PhysRevD.91.123531} {\bibfield
  {journal} {\bibinfo  {journal} {Phys. Rev. D}\ }\textbf {\bibinfo {volume}
  {91}},\ \bibinfo {pages} {123531} (\bibinfo {year} {2015})},\ \Eprint
  {http://arxiv.org/abs/1504.00675} {arXiv:1504.00675 [astro-ph.CO]}
  \BibitemShut {NoStop}%
\bibitem [{\citenamefont {Khatri}\ and\ \citenamefont
  {Sunyaev}(2015)}]{Khatri:2015tla}%
  \BibitemOpen
  \bibfield  {author} {\bibinfo {author} {\bibfnamefont {R.}~\bibnamefont
  {Khatri}}\ and\ \bibinfo {author} {\bibfnamefont {R.}~\bibnamefont
  {Sunyaev}},\ }\href {\doibase 10.1088/1475-7516/2015/9/026} {\bibfield
  {journal} {\bibinfo  {journal} {JCAP}\ }\textbf {\bibinfo {volume} {09}},\
  \bibinfo {pages} {026} (\bibinfo {year} {2015})},\ \Eprint
  {http://arxiv.org/abs/1507.05615} {arXiv:1507.05615 [astro-ph.CO]}
  \BibitemShut {NoStop}%
\bibitem [{\citenamefont {Cabass}\ \emph {et~al.}(2018)\citenamefont {Cabass},
  \citenamefont {Pajer},\ and\ \citenamefont {van~der Woude}}]{Cabass:2018jgj}%
  \BibitemOpen
  \bibfield  {author} {\bibinfo {author} {\bibfnamefont {G.}~\bibnamefont
  {Cabass}}, \bibinfo {author} {\bibfnamefont {E.}~\bibnamefont {Pajer}}, \
  and\ \bibinfo {author} {\bibfnamefont {D.}~\bibnamefont {van~der Woude}},\
  }\href {\doibase 10.1088/1475-7516/2018/08/050} {\bibfield  {journal}
  {\bibinfo  {journal} {JCAP}\ }\textbf {\bibinfo {volume} {08}},\ \bibinfo
  {pages} {050} (\bibinfo {year} {2018})},\ \Eprint
  {http://arxiv.org/abs/1805.08775} {arXiv:1805.08775 [astro-ph.CO]}
  \BibitemShut {NoStop}%
\bibitem [{\citenamefont {Gillet}\ \emph {et~al.}(2020)\citenamefont {Gillet},
  \citenamefont {Mesinger},\ and\ \citenamefont {Park}}]{Gillet:2019fjd}%
  \BibitemOpen
  \bibfield  {author} {\bibinfo {author} {\bibfnamefont {N.~J.~F.}\
  \bibnamefont {Gillet}}, \bibinfo {author} {\bibfnamefont {A.}~\bibnamefont
  {Mesinger}}, \ and\ \bibinfo {author} {\bibfnamefont {J.}~\bibnamefont
  {Park}},\ }\href {\doibase 10.1093/mnras/stz2988} {\bibfield  {journal}
  {\bibinfo  {journal} {Mon. Not. Roy. Astron. Soc.}\ }\textbf {\bibinfo
  {volume} {491}},\ \bibinfo {pages} {1980} (\bibinfo {year} {2020})},\ \Eprint
  {http://arxiv.org/abs/1906.06296} {arXiv:1906.06296 [astro-ph.GA]}
  \BibitemShut {NoStop}%
\bibitem [{\citenamefont {Dimastrogiovanni}\ and\ \citenamefont
  {Emami}(2016)}]{Dimastrogiovanni:2016aul}%
  \BibitemOpen
  \bibfield  {author} {\bibinfo {author} {\bibfnamefont {E.}~\bibnamefont
  {Dimastrogiovanni}}\ and\ \bibinfo {author} {\bibfnamefont {R.}~\bibnamefont
  {Emami}},\ }\href {\doibase 10.1088/1475-7516/2016/12/015} {\bibfield
  {journal} {\bibinfo  {journal} {JCAP}\ }\textbf {\bibinfo {volume} {12}},\
  \bibinfo {pages} {015} (\bibinfo {year} {2016})},\ \Eprint
  {http://arxiv.org/abs/1606.04286} {arXiv:1606.04286 [astro-ph.CO]}
  \BibitemShut {NoStop}%
\bibitem [{\citenamefont {Reischke}\ \emph {et~al.}(2020)\citenamefont
  {Reischke}, \citenamefont {Hagstotz},\ and\ \citenamefont
  {Lilow}}]{Reischke:2020cgd}%
  \BibitemOpen
  \bibfield  {author} {\bibinfo {author} {\bibfnamefont {R.}~\bibnamefont
  {Reischke}}, \bibinfo {author} {\bibfnamefont {S.}~\bibnamefont {Hagstotz}},
  \ and\ \bibinfo {author} {\bibfnamefont {R.}~\bibnamefont {Lilow}},\
  }\href@noop {} {\  (\bibinfo {year} {2020})},\ \Eprint
  {http://arxiv.org/abs/2007.04054} {arXiv:2007.04054 [astro-ph.CO]}
  \BibitemShut {NoStop}%
\bibitem [{\citenamefont {Aghanim}\ \emph {et~al.}(2018)\citenamefont {Aghanim}
  \emph {et~al.}}]{Aghanim:2018eyx}%
  \BibitemOpen
  \bibfield  {author} {\bibinfo {author} {\bibfnamefont {N.}~\bibnamefont
  {Aghanim}} \emph {et~al.} (\bibinfo {collaboration} {Planck}),\ }\href@noop
  {} {\  (\bibinfo {year} {2018})},\ \Eprint {http://arxiv.org/abs/1807.06209}
  {arXiv:1807.06209 [astro-ph.CO]} \BibitemShut {NoStop}%
\bibitem [{\citenamefont {Castorina}\ \emph {et~al.}(2019)\citenamefont
  {Castorina} \emph {et~al.}}]{Castorina:2019wmr}%
  \BibitemOpen
  \bibfield  {author} {\bibinfo {author} {\bibfnamefont {E.}~\bibnamefont
  {Castorina}} \emph {et~al.},\ }\href {\doibase 10.1088/1475-7516/2019/09/010}
  {\bibfield  {journal} {\bibinfo  {journal} {JCAP}\ }\textbf {\bibinfo
  {volume} {09}},\ \bibinfo {pages} {010} (\bibinfo {year} {2019})},\ \Eprint
  {http://arxiv.org/abs/1904.08859} {arXiv:1904.08859 [astro-ph.CO]}
  \BibitemShut {NoStop}%
\bibitem [{\citenamefont {Kennicutt}(1998)}]{Kennicutt:1998zb}%
  \BibitemOpen
  \bibfield  {author} {\bibinfo {author} {\bibfnamefont {J.}~\bibnamefont
  {Kennicutt}, \bibfnamefont {Robert~C.}},\ }\href {\doibase
  10.1146/annurev.astro.36.1.189} {\bibfield  {journal} {\bibinfo  {journal}
  {Ann. Rev. Astron. Astrophys.}\ }\textbf {\bibinfo {volume} {36}},\ \bibinfo
  {pages} {189} (\bibinfo {year} {1998})},\ \Eprint
  {http://arxiv.org/abs/astro-ph/9807187} {arXiv:astro-ph/9807187} \BibitemShut
  {NoStop}%
\bibitem [{\citenamefont {Robertson}\ \emph {et~al.}(2010)\citenamefont
  {Robertson}, \citenamefont {Ellis}, \citenamefont {Dunlop}, \citenamefont
  {McLure},\ and\ \citenamefont {Stark}}]{Robertson:2010an}%
  \BibitemOpen
  \bibfield  {author} {\bibinfo {author} {\bibfnamefont {B.~E.}\ \bibnamefont
  {Robertson}}, \bibinfo {author} {\bibfnamefont {R.~S.}\ \bibnamefont
  {Ellis}}, \bibinfo {author} {\bibfnamefont {J.~S.}\ \bibnamefont {Dunlop}},
  \bibinfo {author} {\bibfnamefont {R.~J.}\ \bibnamefont {McLure}}, \ and\
  \bibinfo {author} {\bibfnamefont {D.~P.}\ \bibnamefont {Stark}},\ }\href
  {\doibase 10.1038/nature09527} {\bibfield  {journal} {\bibinfo  {journal}
  {Nature}\ }\textbf {\bibinfo {volume} {468}},\ \bibinfo {pages} {49}
  (\bibinfo {year} {2010})},\ \Eprint {http://arxiv.org/abs/1011.0727}
  {arXiv:1011.0727 [astro-ph.CO]} \BibitemShut {NoStop}%
\bibitem [{\citenamefont {{Liu}}\ \emph {et~al.}(2016)\citenamefont {{Liu}},
  \citenamefont {{Mutch}}, \citenamefont {{Angel}}, \citenamefont {{Duffy}},
  \citenamefont {{Geil}}, \citenamefont {{Poole}}, \citenamefont {{Mesinger}},\
  and\ \citenamefont {{Wyithe}}}]{2016MNRAS.462..235L}%
  \BibitemOpen
  \bibfield  {author} {\bibinfo {author} {\bibfnamefont {C.}~\bibnamefont
  {{Liu}}}, \bibinfo {author} {\bibfnamefont {S.~J.}\ \bibnamefont {{Mutch}}},
  \bibinfo {author} {\bibfnamefont {P.~W.}\ \bibnamefont {{Angel}}}, \bibinfo
  {author} {\bibfnamefont {A.~R.}\ \bibnamefont {{Duffy}}}, \bibinfo {author}
  {\bibfnamefont {P.~M.}\ \bibnamefont {{Geil}}}, \bibinfo {author}
  {\bibfnamefont {G.~B.}\ \bibnamefont {{Poole}}}, \bibinfo {author}
  {\bibfnamefont {A.}~\bibnamefont {{Mesinger}}}, \ and\ \bibinfo {author}
  {\bibfnamefont {J.~S.~B.}\ \bibnamefont {{Wyithe}}},\ }\href {\doibase
  10.1093/mnras/stw1015} {\bibfield  {journal} {\bibinfo  {journal} {MNRAS}\
  }\textbf {\bibinfo {volume} {462}},\ \bibinfo {pages} {235} (\bibinfo {year}
  {2016})}\BibitemShut {NoStop}%
\bibitem [{\citenamefont {Tacchella}\ \emph {et~al.}(2018)\citenamefont
  {Tacchella}, \citenamefont {Bose}, \citenamefont {Conroy}, \citenamefont
  {Eisenstein},\ and\ \citenamefont {Johnson}}]{Tacchella:2018qny}%
  \BibitemOpen
  \bibfield  {author} {\bibinfo {author} {\bibfnamefont {S.}~\bibnamefont
  {Tacchella}}, \bibinfo {author} {\bibfnamefont {S.}~\bibnamefont {Bose}},
  \bibinfo {author} {\bibfnamefont {C.}~\bibnamefont {Conroy}}, \bibinfo
  {author} {\bibfnamefont {D.~J.}\ \bibnamefont {Eisenstein}}, \ and\ \bibinfo
  {author} {\bibfnamefont {B.~D.}\ \bibnamefont {Johnson}},\ }\href {\doibase
  10.3847/1538-4357/aae8e0} {\bibfield  {journal} {\bibinfo  {journal}
  {Astrophys. J.}\ }\textbf {\bibinfo {volume} {868}},\ \bibinfo {pages} {92}
  (\bibinfo {year} {2018})},\ \Eprint {http://arxiv.org/abs/1806.03299}
  {arXiv:1806.03299 [astro-ph.GA]} \BibitemShut {NoStop}%
\bibitem [{\citenamefont {Yung}\ \emph {et~al.}(2018)\citenamefont {Yung},
  \citenamefont {Somerville}, \citenamefont {Finkelstein}, \citenamefont
  {Popping},\ and\ \citenamefont {Dav\'{e}}}]{Yung_2018}%
  \BibitemOpen
  \bibfield  {author} {\bibinfo {author} {\bibfnamefont {L.~Y.~A.}\
  \bibnamefont {Yung}}, \bibinfo {author} {\bibfnamefont {R.~S.}\ \bibnamefont
  {Somerville}}, \bibinfo {author} {\bibfnamefont {S.~L.}\ \bibnamefont
  {Finkelstein}}, \bibinfo {author} {\bibfnamefont {G.}~\bibnamefont
  {Popping}}, \ and\ \bibinfo {author} {\bibfnamefont {R.}~\bibnamefont
  {Dav\'{e}}},\ }\href {\doibase 10.1093/mnras/sty3241} {\bibfield  {journal}
  {\bibinfo  {journal} {Monthly Notices of the Royal Astronomical Society}\
  }\textbf {\bibinfo {volume} {483}},\ \bibinfo {pages} {2983–3006} (\bibinfo
  {year} {2018})}\BibitemShut {NoStop}%
\bibitem [{\citenamefont {Jenkins}\ \emph {et~al.}(2001)\citenamefont
  {Jenkins}, \citenamefont {Frenk}, \citenamefont {White}, \citenamefont
  {Colberg}, \citenamefont {Cole}, \citenamefont {Evrard}, \citenamefont
  {Couchman},\ and\ \citenamefont {Yoshida}}]{Jenkins:2000bv}%
  \BibitemOpen
  \bibfield  {author} {\bibinfo {author} {\bibfnamefont {A.}~\bibnamefont
  {Jenkins}}, \bibinfo {author} {\bibfnamefont {C.}~\bibnamefont {Frenk}},
  \bibinfo {author} {\bibfnamefont {S.~D.}\ \bibnamefont {White}}, \bibinfo
  {author} {\bibfnamefont {J.}~\bibnamefont {Colberg}}, \bibinfo {author}
  {\bibfnamefont {S.}~\bibnamefont {Cole}}, \bibinfo {author} {\bibfnamefont
  {A.~E.}\ \bibnamefont {Evrard}}, \bibinfo {author} {\bibfnamefont
  {H.}~\bibnamefont {Couchman}}, \ and\ \bibinfo {author} {\bibfnamefont
  {N.}~\bibnamefont {Yoshida}},\ }\href {\doibase
  10.1046/j.1365-8711.2001.04029.x} {\bibfield  {journal} {\bibinfo  {journal}
  {Mon. Not. Roy. Astron. Soc.}\ }\textbf {\bibinfo {volume} {321}},\ \bibinfo
  {pages} {372} (\bibinfo {year} {2001})},\ \Eprint
  {http://arxiv.org/abs/astro-ph/0005260} {arXiv:astro-ph/0005260} \BibitemShut
  {NoStop}%
\bibitem [{\citenamefont {Reed}\ \emph {et~al.}(2007)\citenamefont {Reed},
  \citenamefont {Bower}, \citenamefont {Frenk}, \citenamefont {Jenkins},\ and\
  \citenamefont {Theuns}}]{Reed:2006rw}%
  \BibitemOpen
  \bibfield  {author} {\bibinfo {author} {\bibfnamefont {D.}~\bibnamefont
  {Reed}}, \bibinfo {author} {\bibfnamefont {R.}~\bibnamefont {Bower}},
  \bibinfo {author} {\bibfnamefont {C.}~\bibnamefont {Frenk}}, \bibinfo
  {author} {\bibfnamefont {A.}~\bibnamefont {Jenkins}}, \ and\ \bibinfo
  {author} {\bibfnamefont {T.}~\bibnamefont {Theuns}},\ }\href {\doibase
  10.1111/j.1365-2966.2006.11204.x} {\bibfield  {journal} {\bibinfo  {journal}
  {Mon. Not. Roy. Astron. Soc.}\ }\textbf {\bibinfo {volume} {374}},\ \bibinfo
  {pages} {2} (\bibinfo {year} {2007})},\ \Eprint
  {http://arxiv.org/abs/astro-ph/0607150} {arXiv:astro-ph/0607150} \BibitemShut
  {NoStop}%
\bibitem [{\citenamefont {Sheth}\ and\ \citenamefont
  {Tormen}(2002)}]{Sheth:2001dp}%
  \BibitemOpen
  \bibfield  {author} {\bibinfo {author} {\bibfnamefont {R.~K.}\ \bibnamefont
  {Sheth}}\ and\ \bibinfo {author} {\bibfnamefont {G.}~\bibnamefont {Tormen}},\
  }\href {\doibase 10.1046/j.1365-8711.2002.04950.x} {\bibfield  {journal}
  {\bibinfo  {journal} {Mon. Not. Roy. Astron. Soc.}\ }\textbf {\bibinfo
  {volume} {329}},\ \bibinfo {pages} {61} (\bibinfo {year} {2002})},\ \Eprint
  {http://arxiv.org/abs/astro-ph/0105113} {arXiv:astro-ph/0105113} \BibitemShut
  {NoStop}%
\bibitem [{\citenamefont {Wechsler}\ and\ \citenamefont
  {Tinker}(2018)}]{Wechsler:2018pic}%
  \BibitemOpen
  \bibfield  {author} {\bibinfo {author} {\bibfnamefont {R.~H.}\ \bibnamefont
  {Wechsler}}\ and\ \bibinfo {author} {\bibfnamefont {J.~L.}\ \bibnamefont
  {Tinker}},\ }\href {\doibase 10.1146/annurev-astro-081817-051756} {\bibfield
  {journal} {\bibinfo  {journal} {Ann. Rev. Astron. Astrophys.}\ }\textbf
  {\bibinfo {volume} {56}},\ \bibinfo {pages} {435} (\bibinfo {year} {2018})},\
  \Eprint {http://arxiv.org/abs/1804.03097} {arXiv:1804.03097 [astro-ph.GA]}
  \BibitemShut {NoStop}%
\bibitem [{\citenamefont {Trenti}\ \emph {et~al.}(2010)\citenamefont {Trenti},
  \citenamefont {Stiavelli}, \citenamefont {Bouwens}, \citenamefont {Oesch},
  \citenamefont {Shull}, \citenamefont {Illingworth}, \citenamefont {Bradley},\
  and\ \citenamefont {Carollo}}]{Trenti:2010sz}%
  \BibitemOpen
  \bibfield  {author} {\bibinfo {author} {\bibfnamefont {M.}~\bibnamefont
  {Trenti}}, \bibinfo {author} {\bibfnamefont {M.}~\bibnamefont {Stiavelli}},
  \bibinfo {author} {\bibfnamefont {R.}~\bibnamefont {Bouwens}}, \bibinfo
  {author} {\bibfnamefont {P.}~\bibnamefont {Oesch}}, \bibinfo {author}
  {\bibfnamefont {J.}~\bibnamefont {Shull}}, \bibinfo {author} {\bibfnamefont
  {G.}~\bibnamefont {Illingworth}}, \bibinfo {author} {\bibfnamefont
  {L.}~\bibnamefont {Bradley}}, \ and\ \bibinfo {author} {\bibfnamefont
  {C.}~\bibnamefont {Carollo}},\ }\href {\doibase 10.1088/2041-8205/714/2/L202}
  {\bibfield  {journal} {\bibinfo  {journal} {Astrophys. J. Lett.}\ }\textbf
  {\bibinfo {volume} {714}},\ \bibinfo {pages} {L202} (\bibinfo {year}
  {2010})},\ \Eprint {http://arxiv.org/abs/1004.0384} {arXiv:1004.0384
  [astro-ph.CO]} \BibitemShut {NoStop}%
\bibitem [{\citenamefont {Sitwell}\ \emph {et~al.}(2014)\citenamefont
  {Sitwell}, \citenamefont {Mesinger}, \citenamefont {Ma},\ and\ \citenamefont
  {Sigurdson}}]{Sitwell:2013fpa}%
  \BibitemOpen
  \bibfield  {author} {\bibinfo {author} {\bibfnamefont {M.}~\bibnamefont
  {Sitwell}}, \bibinfo {author} {\bibfnamefont {A.}~\bibnamefont {Mesinger}},
  \bibinfo {author} {\bibfnamefont {Y.-Z.}\ \bibnamefont {Ma}}, \ and\ \bibinfo
  {author} {\bibfnamefont {K.}~\bibnamefont {Sigurdson}},\ }\href {\doibase
  10.1093/mnras/stt2392} {\bibfield  {journal} {\bibinfo  {journal} {Mon. Not.
  Roy. Astron. Soc.}\ }\textbf {\bibinfo {volume} {438}},\ \bibinfo {pages}
  {2664} (\bibinfo {year} {2014})},\ \Eprint {http://arxiv.org/abs/1310.0029}
  {arXiv:1310.0029 [astro-ph.CO]} \BibitemShut {NoStop}%
\bibitem [{\citenamefont {Mason}\ \emph {et~al.}(2015)\citenamefont {Mason},
  \citenamefont {Trenti},\ and\ \citenamefont {Treu}}]{Mason:2015cna}%
  \BibitemOpen
  \bibfield  {author} {\bibinfo {author} {\bibfnamefont {C.}~\bibnamefont
  {Mason}}, \bibinfo {author} {\bibfnamefont {M.}~\bibnamefont {Trenti}}, \
  and\ \bibinfo {author} {\bibfnamefont {T.}~\bibnamefont {Treu}},\ }\href
  {\doibase 10.1088/0004-637X/813/1/21} {\bibfield  {journal} {\bibinfo
  {journal} {Astrophys. J.}\ }\textbf {\bibinfo {volume} {813}},\ \bibinfo
  {pages} {21} (\bibinfo {year} {2015})},\ \Eprint
  {http://arxiv.org/abs/1508.01204} {arXiv:1508.01204 [astro-ph.GA]}
  \BibitemShut {NoStop}%
\bibitem [{\citenamefont {Sun}\ and\ \citenamefont
  {Furlanetto}(2016)}]{Sun_2016}%
  \BibitemOpen
  \bibfield  {author} {\bibinfo {author} {\bibfnamefont {G.}~\bibnamefont
  {Sun}}\ and\ \bibinfo {author} {\bibfnamefont {S.~R.}\ \bibnamefont
  {Furlanetto}},\ }\href {\doibase 10.1093/mnras/stw980} {\bibfield  {journal}
  {\bibinfo  {journal} {Monthly Notices of the Royal Astronomical Society}\
  }\textbf {\bibinfo {volume} {460}},\ \bibinfo {pages} {417–433} (\bibinfo
  {year} {2016})}\BibitemShut {NoStop}%
\bibitem [{\citenamefont {Meurer}\ \emph {et~al.}(1999)\citenamefont {Meurer},
  \citenamefont {Heckman},\ and\ \citenamefont {Calzetti}}]{Meurer:1999jj}%
  \BibitemOpen
  \bibfield  {author} {\bibinfo {author} {\bibfnamefont {G.~R.}\ \bibnamefont
  {Meurer}}, \bibinfo {author} {\bibfnamefont {T.~M.}\ \bibnamefont {Heckman}},
  \ and\ \bibinfo {author} {\bibfnamefont {D.}~\bibnamefont {Calzetti}},\
  }\href {\doibase 10.1086/307523} {\bibfield  {journal} {\bibinfo  {journal}
  {Astrophys. J.}\ }\textbf {\bibinfo {volume} {521}},\ \bibinfo {pages} {64}
  (\bibinfo {year} {1999})},\ \Eprint {http://arxiv.org/abs/astro-ph/9903054}
  {arXiv:astro-ph/9903054} \BibitemShut {NoStop}%
\bibitem [{\citenamefont {Smit}\ \emph {et~al.}(2012)\citenamefont {Smit} \emph
  {et~al.}}]{Smit:2012nf}%
  \BibitemOpen
  \bibfield  {author} {\bibinfo {author} {\bibfnamefont {R.}~\bibnamefont
  {Smit}} \emph {et~al.},\ }\href {\doibase 10.1088/0004-637X/756/1/14}
  {\bibfield  {journal} {\bibinfo  {journal} {Astrophys. J.}\ }\textbf
  {\bibinfo {volume} {756}},\ \bibinfo {pages} {14} (\bibinfo {year} {2012})},\
  \Eprint {http://arxiv.org/abs/1204.3626} {arXiv:1204.3626 [astro-ph.CO]}
  \BibitemShut {NoStop}%
\bibitem [{\citenamefont {Bouwens}\ \emph {et~al.}(2014)\citenamefont {Bouwens}
  \emph {et~al.}}]{Bouwens:2013hxa}%
  \BibitemOpen
  \bibfield  {author} {\bibinfo {author} {\bibfnamefont {R.}~\bibnamefont
  {Bouwens}} \emph {et~al.},\ }\href {\doibase 10.1088/0004-637X/793/2/115}
  {\bibfield  {journal} {\bibinfo  {journal} {Astrophys. J.}\ }\textbf
  {\bibinfo {volume} {793}},\ \bibinfo {pages} {115} (\bibinfo {year}
  {2014})},\ \Eprint {http://arxiv.org/abs/1306.2950} {arXiv:1306.2950
  [astro-ph.CO]} \BibitemShut {NoStop}%
\bibitem [{\citenamefont {Trenti}\ \emph {et~al.}(2015)\citenamefont {Trenti},
  \citenamefont {Perna},\ and\ \citenamefont {Jimenez}}]{Trenti:2014hka}%
  \BibitemOpen
  \bibfield  {author} {\bibinfo {author} {\bibfnamefont {M.}~\bibnamefont
  {Trenti}}, \bibinfo {author} {\bibfnamefont {R.}~\bibnamefont {Perna}}, \
  and\ \bibinfo {author} {\bibfnamefont {R.}~\bibnamefont {Jimenez}},\ }\href
  {\doibase 10.1088/0004-637X/802/2/103} {\bibfield  {journal} {\bibinfo
  {journal} {Astrophys. J.}\ }\textbf {\bibinfo {volume} {802}},\ \bibinfo
  {pages} {103} (\bibinfo {year} {2015})},\ \Eprint
  {http://arxiv.org/abs/1406.1503} {arXiv:1406.1503 [astro-ph.GA]} \BibitemShut
  {NoStop}%
\bibitem [{\citenamefont {Bouwens}\ \emph {et~al.}(2012)\citenamefont {Bouwens}
  \emph {et~al.}}]{Bouwens:2011yy}%
  \BibitemOpen
  \bibfield  {author} {\bibinfo {author} {\bibfnamefont {R.}~\bibnamefont
  {Bouwens}} \emph {et~al.},\ }\href {\doibase 10.1088/0004-637X/754/2/83}
  {\bibfield  {journal} {\bibinfo  {journal} {Astrophys. J.}\ }\textbf
  {\bibinfo {volume} {754}},\ \bibinfo {pages} {83} (\bibinfo {year} {2012})},\
  \Eprint {http://arxiv.org/abs/1109.0994} {arXiv:1109.0994 [astro-ph.CO]}
  \BibitemShut {NoStop}%
\bibitem [{\citenamefont {Maizy}\ \emph {et~al.}(2010)\citenamefont {Maizy},
  \citenamefont {Richard}, \citenamefont {De~Leo}, \citenamefont {Pello},\ and\
  \citenamefont {Kneib}}]{Maizy:2009df}%
  \BibitemOpen
  \bibfield  {author} {\bibinfo {author} {\bibfnamefont {A.}~\bibnamefont
  {Maizy}}, \bibinfo {author} {\bibfnamefont {J.}~\bibnamefont {Richard}},
  \bibinfo {author} {\bibfnamefont {M.~A.}\ \bibnamefont {De~Leo}}, \bibinfo
  {author} {\bibfnamefont {R.}~\bibnamefont {Pello}}, \ and\ \bibinfo {author}
  {\bibfnamefont {J.~P.}\ \bibnamefont {Kneib}},\ }\href {\doibase
  10.1051/0004-6361/200911829} {\bibfield  {journal} {\bibinfo  {journal}
  {Astron. Astrophys.}\ }\textbf {\bibinfo {volume} {509}},\ \bibinfo {pages}
  {A105} (\bibinfo {year} {2010})},\ \Eprint {http://arxiv.org/abs/0910.4910}
  {arXiv:0910.4910 [astro-ph.CO]} \BibitemShut {NoStop}%
\bibitem [{\citenamefont {Williams}\ \emph {et~al.}(2018)\citenamefont
  {Williams} \emph {et~al.}}]{Williams_2018}%
  \BibitemOpen
  \bibfield  {author} {\bibinfo {author} {\bibfnamefont {C.~C.}\ \bibnamefont
  {Williams}} \emph {et~al.},\ }\href {\doibase 10.3847/1538-4365/aabcbb}
  {\bibfield  {journal} {\bibinfo  {journal} {The Astrophysical Journal
  Supplement Series}\ }\textbf {\bibinfo {volume} {236}},\ \bibinfo {pages}
  {33} (\bibinfo {year} {2018})}\BibitemShut {NoStop}%
\bibitem [{\citenamefont {Matsuoka}\ \emph {et~al.}(2018)\citenamefont
  {Matsuoka} \emph {et~al.}}]{Matsuoka:2017frx}%
  \BibitemOpen
  \bibfield  {author} {\bibinfo {author} {\bibfnamefont {Y.}~\bibnamefont
  {Matsuoka}} \emph {et~al.},\ }\href {\doibase 10.1093/pasj/psx046} {\bibfield
   {journal} {\bibinfo  {journal} {Publ. Astron. Soc. Jap.}\ }\textbf {\bibinfo
  {volume} {70}},\ \bibinfo {pages} {S35} (\bibinfo {year} {2018})},\ \Eprint
  {http://arxiv.org/abs/1704.05854} {arXiv:1704.05854 [astro-ph.GA]}
  \BibitemShut {NoStop}%
\bibitem [{\citenamefont {Ono}\ \emph {et~al.}(2018)\citenamefont {Ono} \emph
  {et~al.}}]{Ono:2017wjz}%
  \BibitemOpen
  \bibfield  {author} {\bibinfo {author} {\bibfnamefont {Y.}~\bibnamefont
  {Ono}} \emph {et~al.},\ }\href {\doibase 10.1093/pasj/psx103} {\bibfield
  {journal} {\bibinfo  {journal} {Publ. Astron. Soc. Jap.}\ }\textbf {\bibinfo
  {volume} {70}},\ \bibinfo {pages} {S10} (\bibinfo {year} {2018})},\ \Eprint
  {http://arxiv.org/abs/1704.06004} {arXiv:1704.06004 [astro-ph.GA]}
  \BibitemShut {NoStop}%
\bibitem [{\citenamefont {Young}\ and\ \citenamefont
  {Byrnes}(2013)}]{Young:2013oia}%
  \BibitemOpen
  \bibfield  {author} {\bibinfo {author} {\bibfnamefont {S.}~\bibnamefont
  {Young}}\ and\ \bibinfo {author} {\bibfnamefont {C.~T.}\ \bibnamefont
  {Byrnes}},\ }\href {\doibase 10.1088/1475-7516/2013/08/052} {\bibfield
  {journal} {\bibinfo  {journal} {JCAP}\ }\textbf {\bibinfo {volume} {08}},\
  \bibinfo {pages} {052} (\bibinfo {year} {2013})},\ \Eprint
  {http://arxiv.org/abs/1307.4995} {arXiv:1307.4995 [astro-ph.CO]} \BibitemShut
  {NoStop}%
\bibitem [{\citenamefont {Atal}\ and\ \citenamefont
  {Germani}(2019)}]{Atal:2018neu}%
  \BibitemOpen
  \bibfield  {author} {\bibinfo {author} {\bibfnamefont {V.}~\bibnamefont
  {Atal}}\ and\ \bibinfo {author} {\bibfnamefont {C.}~\bibnamefont {Germani}},\
  }\href {\doibase 10.1016/j.dark.2019.100275} {\bibfield  {journal} {\bibinfo
  {journal} {Phys. Dark Univ.}\ }\textbf {\bibinfo {volume} {24}},\ \bibinfo
  {pages} {100275} (\bibinfo {year} {2019})},\ \Eprint
  {http://arxiv.org/abs/1811.07857} {arXiv:1811.07857 [astro-ph.CO]}
  \BibitemShut {NoStop}%
\bibitem [{\citenamefont {Atal}\ \emph {et~al.}(2020)\citenamefont {Atal},
  \citenamefont {Cid}, \citenamefont {Escriv\`{a}},\ and\ \citenamefont
  {Garriga}}]{Atal:2019erb}%
  \BibitemOpen
  \bibfield  {author} {\bibinfo {author} {\bibfnamefont {V.}~\bibnamefont
  {Atal}}, \bibinfo {author} {\bibfnamefont {J.}~\bibnamefont {Cid}}, \bibinfo
  {author} {\bibfnamefont {A.}~\bibnamefont {Escriv\`{a}}}, \ and\ \bibinfo
  {author} {\bibfnamefont {J.}~\bibnamefont {Garriga}},\ }\href {\doibase
  10.1088/1475-7516/2020/05/022} {\bibfield  {journal} {\bibinfo  {journal}
  {JCAP}\ }\textbf {\bibinfo {volume} {05}},\ \bibinfo {pages} {022} (\bibinfo
  {year} {2020})},\ \Eprint {http://arxiv.org/abs/1908.11357} {arXiv:1908.11357
  [astro-ph.CO]} \BibitemShut {NoStop}%
\bibitem [{\citenamefont {De~Luca}\ \emph {et~al.}(2019)\citenamefont
  {De~Luca}, \citenamefont {Franciolini}, \citenamefont {Kehagias},
  \citenamefont {Peloso}, \citenamefont {Riotto},\ and\ \citenamefont
  {Unal}}]{DeLuca:2019qsy}%
  \BibitemOpen
  \bibfield  {author} {\bibinfo {author} {\bibfnamefont {V.}~\bibnamefont
  {De~Luca}}, \bibinfo {author} {\bibfnamefont {G.}~\bibnamefont
  {Franciolini}}, \bibinfo {author} {\bibfnamefont {A.}~\bibnamefont
  {Kehagias}}, \bibinfo {author} {\bibfnamefont {M.}~\bibnamefont {Peloso}},
  \bibinfo {author} {\bibfnamefont {A.}~\bibnamefont {Riotto}}, \ and\ \bibinfo
  {author} {\bibfnamefont {C.}~\bibnamefont {Unal}},\ }\href {\doibase
  10.1088/1475-7516/2019/07/048} {\bibfield  {journal} {\bibinfo  {journal}
  {JCAP}\ }\textbf {\bibinfo {volume} {07}},\ \bibinfo {pages} {048} (\bibinfo
  {year} {2019})},\ \Eprint {http://arxiv.org/abs/1904.00970} {arXiv:1904.00970
  [astro-ph.CO]} \BibitemShut {NoStop}%
\bibitem [{\citenamefont {Blas}\ \emph {et~al.}(2011)\citenamefont {Blas},
  \citenamefont {Lesgourgues},\ and\ \citenamefont {Tram}}]{Blas:2011rf}%
  \BibitemOpen
  \bibfield  {author} {\bibinfo {author} {\bibfnamefont {D.}~\bibnamefont
  {Blas}}, \bibinfo {author} {\bibfnamefont {J.}~\bibnamefont {Lesgourgues}}, \
  and\ \bibinfo {author} {\bibfnamefont {T.}~\bibnamefont {Tram}},\ }\href
  {\doibase 10.1088/1475-7516/2011/07/034} {\bibfield  {journal} {\bibinfo
  {journal} {JCAP}\ }\textbf {\bibinfo {volume} {07}},\ \bibinfo {pages} {034}
  (\bibinfo {year} {2011})},\ \Eprint {http://arxiv.org/abs/1104.2933}
  {arXiv:1104.2933 [astro-ph.CO]} \BibitemShut {NoStop}%
\bibitem [{\citenamefont {Juszkiewicz}\ \emph {et~al.}(1995)\citenamefont
  {Juszkiewicz}, \citenamefont {Weinberg}, \citenamefont {Amsterdamski},
  \citenamefont {Chodorowski},\ and\ \citenamefont
  {Bouchet}}]{Juszkiewicz:1993hm}%
  \BibitemOpen
  \bibfield  {author} {\bibinfo {author} {\bibfnamefont {R.}~\bibnamefont
  {Juszkiewicz}}, \bibinfo {author} {\bibfnamefont {D.~H.}\ \bibnamefont
  {Weinberg}}, \bibinfo {author} {\bibfnamefont {P.}~\bibnamefont
  {Amsterdamski}}, \bibinfo {author} {\bibfnamefont {M.}~\bibnamefont
  {Chodorowski}}, \ and\ \bibinfo {author} {\bibfnamefont {F.}~\bibnamefont
  {Bouchet}},\ }\href@noop {} {\bibfield  {journal} {\bibinfo  {journal}
  {Astrophys. J.}\ }\textbf {\bibinfo {volume} {442}},\ \bibinfo {pages} {39}
  (\bibinfo {year} {1995})},\ \Eprint {http://arxiv.org/abs/astro-ph/9308012}
  {arXiv:astro-ph/9308012} \BibitemShut {NoStop}%
\bibitem [{\citenamefont {Bernardeau}\ and\ \citenamefont
  {Kofman}(1995)}]{Bernardeau:1994aq}%
  \BibitemOpen
  \bibfield  {author} {\bibinfo {author} {\bibfnamefont {F.}~\bibnamefont
  {Bernardeau}}\ and\ \bibinfo {author} {\bibfnamefont {L.}~\bibnamefont
  {Kofman}},\ }\href {\doibase 10.1086/175542} {\bibfield  {journal} {\bibinfo
  {journal} {Astrophys. J.}\ }\textbf {\bibinfo {volume} {443}},\ \bibinfo
  {pages} {479} (\bibinfo {year} {1995})},\ \Eprint
  {http://arxiv.org/abs/astro-ph/9403028} {arXiv:astro-ph/9403028} \BibitemShut
  {NoStop}%
\bibitem [{\citenamefont {Behroozi}\ \emph {et~al.}(2020)\citenamefont
  {Behroozi} \emph {et~al.}}]{Behroozi:2020jhj}%
  \BibitemOpen
  \bibfield  {author} {\bibinfo {author} {\bibfnamefont {P.}~\bibnamefont
  {Behroozi}} \emph {et~al.},\ }\href@noop {} {\  (\bibinfo {year} {2020})},\
  \Eprint {http://arxiv.org/abs/2007.04988} {arXiv:2007.04988 [astro-ph.GA]}
  \BibitemShut {NoStop}%
\bibitem [{\citenamefont {Beckwith}\ \emph {et~al.}(2006)\citenamefont
  {Beckwith} \emph {et~al.}}]{Beckwith:2006qi}%
  \BibitemOpen
  \bibfield  {author} {\bibinfo {author} {\bibfnamefont {S.~V.}\ \bibnamefont
  {Beckwith}} \emph {et~al.},\ }\href {\doibase 10.1086/507302} {\bibfield
  {journal} {\bibinfo  {journal} {Astron. J.}\ }\textbf {\bibinfo {volume}
  {132}},\ \bibinfo {pages} {1729} (\bibinfo {year} {2006})},\ \Eprint
  {http://arxiv.org/abs/astro-ph/0607632} {arXiv:astro-ph/0607632} \BibitemShut
  {NoStop}%
\bibitem [{\citenamefont {Gardner}\ \emph {et~al.}(2006)\citenamefont {Gardner}
  \emph {et~al.}}]{Gardner:2006ky}%
  \BibitemOpen
  \bibfield  {author} {\bibinfo {author} {\bibfnamefont {J.~P.}\ \bibnamefont
  {Gardner}} \emph {et~al.},\ }\href {\doibase 10.1007/s11214-006-8315-7}
  {\bibfield  {journal} {\bibinfo  {journal} {Space Sci. Rev.}\ }\textbf
  {\bibinfo {volume} {123}},\ \bibinfo {pages} {485} (\bibinfo {year}
  {2006})},\ \Eprint {http://arxiv.org/abs/astro-ph/0606175}
  {arXiv:astro-ph/0606175} \BibitemShut {NoStop}%
\bibitem [{\citenamefont {Tacchella}\ \emph {et~al.}(2020)\citenamefont
  {Tacchella}, \citenamefont {Forbes},\ and\ \citenamefont
  {Caplar}}]{Tacchella_2020}%
  \BibitemOpen
  \bibfield  {author} {\bibinfo {author} {\bibfnamefont {S.}~\bibnamefont
  {Tacchella}}, \bibinfo {author} {\bibfnamefont {J.~C.}\ \bibnamefont
  {Forbes}}, \ and\ \bibinfo {author} {\bibfnamefont {N.}~\bibnamefont
  {Caplar}},\ }\href {\doibase 10.1093/mnras/staa1838} {\bibfield  {journal}
  {\bibinfo  {journal} {Monthly Notices of the Royal Astronomical Society}\
  }\textbf {\bibinfo {volume} {497}},\ \bibinfo {pages} {698–725} (\bibinfo
  {year} {2020})}\BibitemShut {NoStop}%
\bibitem [{\citenamefont {Spergel}\ \emph {et~al.}(2015)\citenamefont {Spergel}
  \emph {et~al.}}]{Spergel:2015sza}%
  \BibitemOpen
  \bibfield  {author} {\bibinfo {author} {\bibfnamefont {D.}~\bibnamefont
  {Spergel}} \emph {et~al.},\ }\href@noop {} {\  (\bibinfo {year} {2015})},\
  \Eprint {http://arxiv.org/abs/1503.03757} {arXiv:1503.03757 [astro-ph.IM]}
  \BibitemShut {NoStop}%
\bibitem [{\citenamefont {Trenti}\ and\ \citenamefont
  {Stiavelli}(2008)}]{Trenti:2007dh}%
  \BibitemOpen
  \bibfield  {author} {\bibinfo {author} {\bibfnamefont {M.}~\bibnamefont
  {Trenti}}\ and\ \bibinfo {author} {\bibfnamefont {M.}~\bibnamefont
  {Stiavelli}},\ }\href {\doibase 10.1086/528674} {\bibfield  {journal}
  {\bibinfo  {journal} {Astrophys. J.}\ }\textbf {\bibinfo {volume} {676}},\
  \bibinfo {pages} {767} (\bibinfo {year} {2008})},\ \Eprint
  {http://arxiv.org/abs/0712.0398} {arXiv:0712.0398 [astro-ph]} \BibitemShut
  {NoStop}%
\bibitem [{\citenamefont {Foreman-Mackey}\ \emph {et~al.}(2013)\citenamefont
  {Foreman-Mackey}, \citenamefont {Hogg}, \citenamefont {Lang},\ and\
  \citenamefont {Goodman}}]{ForemanMackey:2012ig}%
  \BibitemOpen
  \bibfield  {author} {\bibinfo {author} {\bibfnamefont {D.}~\bibnamefont
  {Foreman-Mackey}}, \bibinfo {author} {\bibfnamefont {D.~W.}\ \bibnamefont
  {Hogg}}, \bibinfo {author} {\bibfnamefont {D.}~\bibnamefont {Lang}}, \ and\
  \bibinfo {author} {\bibfnamefont {J.}~\bibnamefont {Goodman}},\ }\href
  {\doibase 10.1086/670067} {\bibfield  {journal} {\bibinfo  {journal} {Publ.
  Astron. Soc. Pac.}\ }\textbf {\bibinfo {volume} {125}},\ \bibinfo {pages}
  {306} (\bibinfo {year} {2013})},\ \Eprint {http://arxiv.org/abs/1202.3665}
  {arXiv:1202.3665 [astro-ph.IM]} \BibitemShut {NoStop}%
\bibitem [{\citenamefont {Foreman-Mackey}(2016)}]{corner}%
  \BibitemOpen
  \bibfield  {author} {\bibinfo {author} {\bibfnamefont {D.}~\bibnamefont
  {Foreman-Mackey}},\ }\href {\doibase 10.21105/joss.00024} {\bibfield
  {journal} {\bibinfo  {journal} {The Journal of Open Source Software}\
  }\textbf {\bibinfo {volume} {1}},\ \bibinfo {pages} {24} (\bibinfo {year}
  {2016})}\BibitemShut {NoStop}%
\bibitem [{\citenamefont {Vogelsberger}\ \emph {et~al.}(2020)\citenamefont
  {Vogelsberger} \emph {et~al.}}]{Vogelsberger_dust2020}%
  \BibitemOpen
  \bibfield  {author} {\bibinfo {author} {\bibfnamefont {M.}~\bibnamefont
  {Vogelsberger}} \emph {et~al.},\ }\href {\doibase 10.1093/mnras/staa137}
  {\bibfield  {journal} {\bibinfo  {journal} {Monthly Notices of the Royal
  Astronomical Society}\ }\textbf {\bibinfo {volume} {492}},\ \bibinfo {pages}
  {5167–5201} (\bibinfo {year} {2020})},\ \Eprint
  {http://arxiv.org/abs/1904.07238} {arXiv:1904.07238 [astro-ph.GA]}
  \BibitemShut {NoStop}%
\bibitem [{\citenamefont {Siana}\ \emph {et~al.}(2009)\citenamefont {Siana}
  \emph {et~al.}}]{Siana:2009um}%
  \BibitemOpen
  \bibfield  {author} {\bibinfo {author} {\bibfnamefont {B.}~\bibnamefont
  {Siana}} \emph {et~al.},\ }\href {\doibase 10.1088/0004-637X/698/2/1273}
  {\bibfield  {journal} {\bibinfo  {journal} {Astrophys. J.}\ }\textbf
  {\bibinfo {volume} {698}},\ \bibinfo {pages} {1273} (\bibinfo {year}
  {2009})},\ \Eprint {http://arxiv.org/abs/0904.1742} {arXiv:0904.1742
  [astro-ph.CO]} \BibitemShut {NoStop}%
\bibitem [{\citenamefont {Casey}\ \emph {et~al.}(2014)\citenamefont {Casey}
  \emph {et~al.}}]{Casey:2014cqa}%
  \BibitemOpen
  \bibfield  {author} {\bibinfo {author} {\bibfnamefont {C.}~\bibnamefont
  {Casey}} \emph {et~al.},\ }\href {\doibase 10.1088/0004-637X/796/2/95}
  {\bibfield  {journal} {\bibinfo  {journal} {Astrophys. J.}\ }\textbf
  {\bibinfo {volume} {796}},\ \bibinfo {pages} {95} (\bibinfo {year} {2014})},\
  \Eprint {http://arxiv.org/abs/1410.0702} {arXiv:1410.0702 [astro-ph.GA]}
  \BibitemShut {NoStop}%
\bibitem [{\citenamefont {Castellano}\ \emph {et~al.}(2014)\citenamefont
  {Castellano} \emph {et~al.}}]{Castellano:2014lua}%
  \BibitemOpen
  \bibfield  {author} {\bibinfo {author} {\bibfnamefont {M.}~\bibnamefont
  {Castellano}} \emph {et~al.},\ }\href {\doibase 10.1051/0004-6361/201322704}
  {\bibfield  {journal} {\bibinfo  {journal} {Astron. Astrophys.}\ }\textbf
  {\bibinfo {volume} {566}},\ \bibinfo {pages} {A19} (\bibinfo {year}
  {2014})},\ \Eprint {http://arxiv.org/abs/1403.0743} {arXiv:1403.0743
  [astro-ph.GA]} \BibitemShut {NoStop}%
\bibitem [{\citenamefont {Reddy}\ \emph {et~al.}(2015)\citenamefont {Reddy}
  \emph {et~al.}}]{reddy2015mosdef}%
  \BibitemOpen
  \bibfield  {author} {\bibinfo {author} {\bibfnamefont {N.~A.}\ \bibnamefont
  {Reddy}} \emph {et~al.},\ }\href@noop {} {\  (\bibinfo {year} {2015})},\
  \Eprint {http://arxiv.org/abs/1504.02782} {arXiv:1504.02782 [astro-ph.GA]}
  \BibitemShut {NoStop}%
\bibitem [{\citenamefont {J.~Bouwens}\ \emph {et~al.}(2016)\citenamefont
  {J.~Bouwens} \emph {et~al.}}]{J_Bouwens_2016}%
  \BibitemOpen
  \bibfield  {author} {\bibinfo {author} {\bibfnamefont {R.}~\bibnamefont
  {J.~Bouwens}} \emph {et~al.},\ }\href {\doibase 10.3847/1538-4357/833/1/72}
  {\bibfield  {journal} {\bibinfo  {journal} {The Astrophysical Journal}\
  }\textbf {\bibinfo {volume} {833}},\ \bibinfo {pages} {72} (\bibinfo {year}
  {2016})},\ \Eprint {http://arxiv.org/abs/1606.05280} {arXiv:1606.05280
  [astro-ph.GA]} \BibitemShut {NoStop}%
\bibitem [{\citenamefont {Mu\~{n}oz}\ \emph {et~al.}(2020)\citenamefont
  {Mu\~{n}oz}, \citenamefont {Dvorkin},\ and\ \citenamefont
  {Cyr-Racine}}]{Munoz:2019hjh}%
  \BibitemOpen
  \bibfield  {author} {\bibinfo {author} {\bibfnamefont {J.~B.}\ \bibnamefont
  {Mu\~{n}oz}}, \bibinfo {author} {\bibfnamefont {C.}~\bibnamefont {Dvorkin}},
  \ and\ \bibinfo {author} {\bibfnamefont {F.-Y.}\ \bibnamefont {Cyr-Racine}},\
  }\href {\doibase 10.1103/PhysRevD.101.063526} {\bibfield  {journal} {\bibinfo
   {journal} {Phys. Rev. D}\ }\textbf {\bibinfo {volume} {101}},\ \bibinfo
  {pages} {063526} (\bibinfo {year} {2020})},\ \Eprint
  {http://arxiv.org/abs/1911.11144} {arXiv:1911.11144 [astro-ph.CO]}
  \BibitemShut {NoStop}%
\bibitem [{\citenamefont {Machacek}\ \emph {et~al.}(2001)\citenamefont
  {Machacek}, \citenamefont {Bryan},\ and\ \citenamefont
  {Abel}}]{Machacek:2000us}%
  \BibitemOpen
  \bibfield  {author} {\bibinfo {author} {\bibfnamefont {M.~E.}\ \bibnamefont
  {Machacek}}, \bibinfo {author} {\bibfnamefont {G.~L.}\ \bibnamefont {Bryan}},
  \ and\ \bibinfo {author} {\bibfnamefont {T.}~\bibnamefont {Abel}},\ }\href
  {\doibase 10.1086/319014} {\bibfield  {journal} {\bibinfo  {journal}
  {Astrophys. J.}\ }\textbf {\bibinfo {volume} {548}},\ \bibinfo {pages} {509}
  (\bibinfo {year} {2001})},\ \Eprint {http://arxiv.org/abs/astro-ph/0007198}
  {arXiv:astro-ph/0007198} \BibitemShut {NoStop}%
\bibitem [{\citenamefont {Oh}\ and\ \citenamefont {Haiman}(2002)}]{Oh:2001ex}%
  \BibitemOpen
  \bibfield  {author} {\bibinfo {author} {\bibfnamefont {S.}~\bibnamefont
  {Oh}}\ and\ \bibinfo {author} {\bibfnamefont {Z.}~\bibnamefont {Haiman}},\
  }\href {\doibase 10.1086/339393} {\bibfield  {journal} {\bibinfo  {journal}
  {Astrophys. J.}\ }\textbf {\bibinfo {volume} {569}},\ \bibinfo {pages} {558}
  (\bibinfo {year} {2002})},\ \Eprint {http://arxiv.org/abs/astro-ph/0108071}
  {arXiv:astro-ph/0108071} \BibitemShut {NoStop}%
\bibitem [{\citenamefont {Yoshida}\ \emph {et~al.}(2003)\citenamefont
  {Yoshida}, \citenamefont {Abel}, \citenamefont {Hernquist},\ and\
  \citenamefont {Sugiyama}}]{Yoshida:2003rw}%
  \BibitemOpen
  \bibfield  {author} {\bibinfo {author} {\bibfnamefont {N.}~\bibnamefont
  {Yoshida}}, \bibinfo {author} {\bibfnamefont {T.}~\bibnamefont {Abel}},
  \bibinfo {author} {\bibfnamefont {L.}~\bibnamefont {Hernquist}}, \ and\
  \bibinfo {author} {\bibfnamefont {N.}~\bibnamefont {Sugiyama}},\ }\href
  {\doibase 10.1086/375810} {\bibfield  {journal} {\bibinfo  {journal}
  {Astrophys. J.}\ }\textbf {\bibinfo {volume} {592}},\ \bibinfo {pages} {645}
  (\bibinfo {year} {2003})},\ \Eprint {http://arxiv.org/abs/astro-ph/0301645}
  {arXiv:astro-ph/0301645} \BibitemShut {NoStop}%
\bibitem [{\citenamefont {Fialkov}\ \emph {et~al.}(2013)\citenamefont
  {Fialkov}, \citenamefont {Barkana}, \citenamefont {Visbal}, \citenamefont
  {Tseliakhovich},\ and\ \citenamefont {Hirata}}]{Fialkov:2012su}%
  \BibitemOpen
  \bibfield  {author} {\bibinfo {author} {\bibfnamefont {A.}~\bibnamefont
  {Fialkov}}, \bibinfo {author} {\bibfnamefont {R.}~\bibnamefont {Barkana}},
  \bibinfo {author} {\bibfnamefont {E.}~\bibnamefont {Visbal}}, \bibinfo
  {author} {\bibfnamefont {D.}~\bibnamefont {Tseliakhovich}}, \ and\ \bibinfo
  {author} {\bibfnamefont {C.~M.}\ \bibnamefont {Hirata}},\ }\href {\doibase
  10.1093/mnras/stt650} {\bibfield  {journal} {\bibinfo  {journal} {Mon. Not.
  Roy. Astron. Soc.}\ }\textbf {\bibinfo {volume} {432}},\ \bibinfo {pages}
  {2909} (\bibinfo {year} {2013})},\ \Eprint {http://arxiv.org/abs/1212.0513}
  {arXiv:1212.0513 [astro-ph.CO]} \BibitemShut {NoStop}%
\bibitem [{\citenamefont {{Wouthuysen}}(1952)}]{Wout}%
  \BibitemOpen
  \bibfield  {author} {\bibinfo {author} {\bibfnamefont {S.~A.}\ \bibnamefont
  {{Wouthuysen}}},\ }\href {\doibase 10.1086/106661} {\bibfield  {journal}
  {\bibinfo  {journal} {Astronomical Journal}\ }\textbf {\bibinfo {volume}
  {57}},\ \bibinfo {pages} {31} (\bibinfo {year} {1952})}\BibitemShut {NoStop}%
\bibitem [{\citenamefont {{Field}}(1959)}]{Field}%
  \BibitemOpen
  \bibfield  {author} {\bibinfo {author} {\bibfnamefont {G.~B.}\ \bibnamefont
  {{Field}}},\ }\href {\doibase 10.1086/146653} {\bibfield  {journal} {\bibinfo
   {journal} {Astrophys. J.}\ }\textbf {\bibinfo {volume} {129}},\ \bibinfo
  {pages} {536} (\bibinfo {year} {1959})}\BibitemShut {NoStop}%
\bibitem [{\citenamefont {Hirata}(2006)}]{Hirata:2005mz}%
  \BibitemOpen
  \bibfield  {author} {\bibinfo {author} {\bibfnamefont {C.~M.}\ \bibnamefont
  {Hirata}},\ }\href {\doibase 10.1111/j.1365-2966.2005.09949.x} {\bibfield
  {journal} {\bibinfo  {journal} {Mon. Not. Roy. Astron. Soc.}\ }\textbf
  {\bibinfo {volume} {367}},\ \bibinfo {pages} {259} (\bibinfo {year}
  {2006})},\ \Eprint {http://arxiv.org/abs/astro-ph/0507102}
  {arXiv:astro-ph/0507102} \BibitemShut {NoStop}%
\bibitem [{\citenamefont {Pritchard}\ and\ \citenamefont
  {Loeb}(2012)}]{Pritchard:2011xb}%
  \BibitemOpen
  \bibfield  {author} {\bibinfo {author} {\bibfnamefont {J.~R.}\ \bibnamefont
  {Pritchard}}\ and\ \bibinfo {author} {\bibfnamefont {A.}~\bibnamefont
  {Loeb}},\ }\href {\doibase 10.1088/0034-4885/75/8/086901} {\bibfield
  {journal} {\bibinfo  {journal} {Rept. Prog. Phys.}\ }\textbf {\bibinfo
  {volume} {75}},\ \bibinfo {pages} {086901} (\bibinfo {year} {2012})},\
  \Eprint {http://arxiv.org/abs/1109.6012} {arXiv:1109.6012 [astro-ph.CO]}
  \BibitemShut {NoStop}%
\bibitem [{\citenamefont {Loeb}\ and\ \citenamefont
  {Furlanetto}(2013)}]{loeb2013first}%
  \BibitemOpen
  \bibfield  {author} {\bibinfo {author} {\bibfnamefont {A.}~\bibnamefont
  {Loeb}}\ and\ \bibinfo {author} {\bibfnamefont {S.~R.}\ \bibnamefont
  {Furlanetto}},\ }\href@noop {} {\emph {\bibinfo {title} {The first galaxies
  in the universe}}}\ (\bibinfo  {publisher} {Princeton University Press},\
  \bibinfo {year} {2013})\BibitemShut {NoStop}%
\bibitem [{\citenamefont {Mu\~{n}oz}(2019{\natexlab{a}})}]{Munoz:2019rhi}%
  \BibitemOpen
  \bibfield  {author} {\bibinfo {author} {\bibfnamefont {J.~B.}\ \bibnamefont
  {Mu\~{n}oz}},\ }\href {\doibase 10.1103/PhysRevD.100.063538} {\bibfield
  {journal} {\bibinfo  {journal} {Phys. Rev. D}\ }\textbf {\bibinfo {volume}
  {100}},\ \bibinfo {pages} {063538} (\bibinfo {year} {2019}{\natexlab{a}})},\
  \Eprint {http://arxiv.org/abs/1904.07881} {arXiv:1904.07881 [astro-ph.CO]}
  \BibitemShut {NoStop}%
\bibitem [{\citenamefont {Mu\~{n}oz}(2019{\natexlab{b}})}]{Munoz:2019fkt}%
  \BibitemOpen
  \bibfield  {author} {\bibinfo {author} {\bibfnamefont {J.~B.}\ \bibnamefont
  {Mu\~{n}oz}},\ }\href {\doibase 10.1103/PhysRevLett.123.131301} {\bibfield
  {journal} {\bibinfo  {journal} {Phys. Rev. Lett.}\ }\textbf {\bibinfo
  {volume} {123}},\ \bibinfo {pages} {131301} (\bibinfo {year}
  {2019}{\natexlab{b}})},\ \Eprint {http://arxiv.org/abs/1904.07868}
  {arXiv:1904.07868 [astro-ph.CO]} \BibitemShut {NoStop}%
\bibitem [{\citenamefont {Mesinger}\ and\ \citenamefont
  {Furlanetto}(2007)}]{Mesinger:2007pd}%
  \BibitemOpen
  \bibfield  {author} {\bibinfo {author} {\bibfnamefont {A.}~\bibnamefont
  {Mesinger}}\ and\ \bibinfo {author} {\bibfnamefont {S.}~\bibnamefont
  {Furlanetto}},\ }\href {\doibase 10.1086/521806} {\bibfield  {journal}
  {\bibinfo  {journal} {Astrophys. J.}\ }\textbf {\bibinfo {volume} {669}},\
  \bibinfo {pages} {663} (\bibinfo {year} {2007})},\ \Eprint
  {http://arxiv.org/abs/0704.0946} {arXiv:0704.0946 [astro-ph]} \BibitemShut
  {NoStop}%
\bibitem [{\citenamefont {Mesinger}\ \emph {et~al.}(2011)\citenamefont
  {Mesinger}, \citenamefont {Furlanetto},\ and\ \citenamefont
  {Cen}}]{Mesinger:2010ne}%
  \BibitemOpen
  \bibfield  {author} {\bibinfo {author} {\bibfnamefont {A.}~\bibnamefont
  {Mesinger}}, \bibinfo {author} {\bibfnamefont {S.}~\bibnamefont
  {Furlanetto}}, \ and\ \bibinfo {author} {\bibfnamefont {R.}~\bibnamefont
  {Cen}},\ }\href {\doibase 10.1111/j.1365-2966.2010.17731.x} {\bibfield
  {journal} {\bibinfo  {journal} {Mon. Not. Roy. Astron. Soc.}\ }\textbf
  {\bibinfo {volume} {411}},\ \bibinfo {pages} {955} (\bibinfo {year}
  {2011})},\ \Eprint {http://arxiv.org/abs/1003.3878} {arXiv:1003.3878
  [astro-ph.CO]} \BibitemShut {NoStop}%
\bibitem [{\citenamefont {Greig}\ and\ \citenamefont
  {Mesinger}(2015)}]{Greig:2015qca}%
  \BibitemOpen
  \bibfield  {author} {\bibinfo {author} {\bibfnamefont {B.}~\bibnamefont
  {Greig}}\ and\ \bibinfo {author} {\bibfnamefont {A.}~\bibnamefont
  {Mesinger}},\ }\href {\doibase 10.1093/mnras/stv571} {\bibfield  {journal}
  {\bibinfo  {journal} {Mon. Not. Roy. Astron. Soc.}\ }\textbf {\bibinfo
  {volume} {449}},\ \bibinfo {pages} {4246} (\bibinfo {year} {2015})},\ \Eprint
  {http://arxiv.org/abs/1501.06576} {arXiv:1501.06576} \BibitemShut {NoStop}%
\bibitem [{\citenamefont {Qin}\ \emph {et~al.}(2020)\citenamefont {Qin},
  \citenamefont {Mesinger}, \citenamefont {Park}, \citenamefont {Greig},\ and\
  \citenamefont {Mu\~{n}oz}}]{Qin:2020xyh}%
  \BibitemOpen
  \bibfield  {author} {\bibinfo {author} {\bibfnamefont {Y.}~\bibnamefont
  {Qin}}, \bibinfo {author} {\bibfnamefont {A.}~\bibnamefont {Mesinger}},
  \bibinfo {author} {\bibfnamefont {J.}~\bibnamefont {Park}}, \bibinfo {author}
  {\bibfnamefont {B.}~\bibnamefont {Greig}}, \ and\ \bibinfo {author}
  {\bibfnamefont {J.~B.}\ \bibnamefont {Mu\~{n}oz}},\ }\href {\doibase
  10.1093/mnras/staa1131} {\bibfield  {journal} {\bibinfo  {journal} {Mon. Not.
  Roy. Astron. Soc.}\ }\textbf {\bibinfo {volume} {495}},\ \bibinfo {pages}
  {123} (\bibinfo {year} {2020})},\ \Eprint {http://arxiv.org/abs/2003.04442}
  {arXiv:2003.04442 [astro-ph.CO]} \BibitemShut {NoStop}%
\bibitem [{\citenamefont {Bowman}\ \emph {et~al.}(2018)\citenamefont {Bowman},
  \citenamefont {Rogers}, \citenamefont {Monsalve}, \citenamefont {Mozdzen},\
  and\ \citenamefont {Mahesh}}]{Bowman:2018yin}%
  \BibitemOpen
  \bibfield  {author} {\bibinfo {author} {\bibfnamefont {J.~D.}\ \bibnamefont
  {Bowman}}, \bibinfo {author} {\bibfnamefont {A.~E.~E.}\ \bibnamefont
  {Rogers}}, \bibinfo {author} {\bibfnamefont {R.~A.}\ \bibnamefont
  {Monsalve}}, \bibinfo {author} {\bibfnamefont {T.~J.}\ \bibnamefont
  {Mozdzen}}, \ and\ \bibinfo {author} {\bibfnamefont {N.}~\bibnamefont
  {Mahesh}},\ }\href {\doibase 10.1038/nature25792} {\bibfield  {journal}
  {\bibinfo  {journal} {Nature}\ }\textbf {\bibinfo {volume} {555}},\ \bibinfo
  {pages} {67} (\bibinfo {year} {2018})}\BibitemShut {NoStop}%
\bibitem [{\citenamefont {Singh}\ \emph {et~al.}(2017)\citenamefont {Singh}
  \emph {et~al.}}]{Singh:2017syr}%
  \BibitemOpen
  \bibfield  {author} {\bibinfo {author} {\bibfnamefont {S.}~\bibnamefont
  {Singh}} \emph {et~al.},\ }\href@noop {} {\  (\bibinfo {year} {2017})},\
  \Eprint {http://arxiv.org/abs/1710.01101} {arXiv:1710.01101} \BibitemShut
  {NoStop}%
\bibitem [{\citenamefont {{Price}}\ \emph {et~al.}(2017)\citenamefont {{Price}}
  \emph {et~al.}}]{LEDA}%
  \BibitemOpen
  \bibfield  {author} {\bibinfo {author} {\bibfnamefont {D.~C.}\ \bibnamefont
  {{Price}}} \emph {et~al.},\ }\href@noop {} {\  (\bibinfo {year} {2017})},\
  \Eprint {http://arxiv.org/abs/1709.09313} {arXiv:1709.09313} \BibitemShut
  {NoStop}%
\bibitem [{\citenamefont {Voytek}\ \emph {et~al.}(2014)\citenamefont {Voytek},
  \citenamefont {Natarajan}, \citenamefont {J\'{a}uregui~Garc\'{i}a},
  \citenamefont {Peterson},\ and\ \citenamefont
  {López-Cruz}}]{Voytek:2013nua}%
  \BibitemOpen
  \bibfield  {author} {\bibinfo {author} {\bibfnamefont {T.~C.}\ \bibnamefont
  {Voytek}}, \bibinfo {author} {\bibfnamefont {A.}~\bibnamefont {Natarajan}},
  \bibinfo {author} {\bibfnamefont {J.~M.}\ \bibnamefont
  {J\'{a}uregui~Garc\'{i}a}}, \bibinfo {author} {\bibfnamefont {J.~B.}\
  \bibnamefont {Peterson}}, \ and\ \bibinfo {author} {\bibfnamefont
  {O.}~\bibnamefont {López-Cruz}},\ }\href {\doibase
  10.1088/2041-8205/782/1/L9} {\bibfield  {journal} {\bibinfo  {journal}
  {Astrophys. J.}\ }\textbf {\bibinfo {volume} {782}},\ \bibinfo {pages} {L9}
  (\bibinfo {year} {2014})},\ \Eprint {http://arxiv.org/abs/1311.0014}
  {arXiv:1311.0014 [astro-ph.CO]} \BibitemShut {NoStop}%
\bibitem [{\citenamefont {{Philip}}\ \emph {et~al.}(2019)\citenamefont
  {{Philip}} \emph {et~al.}}]{PRIZM}%
  \BibitemOpen
  \bibfield  {author} {\bibinfo {author} {\bibfnamefont {L.}~\bibnamefont
  {{Philip}}} \emph {et~al.},\ }\href {\doibase 10.1142/S2251171719500041}
  {\bibfield  {journal} {\bibinfo  {journal} {Journal of Astronomical
  Instrumentation}\ }\textbf {\bibinfo {volume} {8}},\ \bibinfo {eid} {1950004}
  (\bibinfo {year} {2019})},\ \Eprint {http://arxiv.org/abs/1806.09531}
  {arXiv:1806.09531 [astro-ph.IM]} \BibitemShut {NoStop}%
\bibitem [{\citenamefont {Pober}\ \emph {et~al.}(2013)\citenamefont {Pober}
  \emph {et~al.}}]{Pober:2013ig}%
  \BibitemOpen
  \bibfield  {author} {\bibinfo {author} {\bibfnamefont {J.~C.}\ \bibnamefont
  {Pober}} \emph {et~al.},\ }\href {\doibase 10.1088/2041-8205/768/2/L36}
  {\bibfield  {journal} {\bibinfo  {journal} {Astrophys. J.}\ }\textbf
  {\bibinfo {volume} {768}},\ \bibinfo {pages} {L36} (\bibinfo {year}
  {2013})},\ \Eprint {http://arxiv.org/abs/1301.7099} {arXiv:1301.7099
  [astro-ph.CO]} \BibitemShut {NoStop}%
\bibitem [{\citenamefont {DeBoer}\ \emph {et~al.}(2017)\citenamefont {DeBoer}
  \emph {et~al.}}]{DeBoer:2016tnn}%
  \BibitemOpen
  \bibfield  {author} {\bibinfo {author} {\bibfnamefont {D.~R.}\ \bibnamefont
  {DeBoer}} \emph {et~al.},\ }\href {\doibase 10.1088/1538-3873/129/974/045001}
  {\bibfield  {journal} {\bibinfo  {journal} {Publ. Astron. Soc. Pac.}\
  }\textbf {\bibinfo {volume} {129}},\ \bibinfo {pages} {045001} (\bibinfo
  {year} {2017})},\ \Eprint {http://arxiv.org/abs/1606.07473}
  {arXiv:1606.07473} \BibitemShut {NoStop}%
\bibitem [{\citenamefont {van Haarlem}\ \emph {et~al.}(2013)\citenamefont {van
  Haarlem} \emph {et~al.}}]{vanHaarlem:2013dsa}%
  \BibitemOpen
  \bibfield  {author} {\bibinfo {author} {\bibfnamefont {M.~P.}\ \bibnamefont
  {van Haarlem}} \emph {et~al.},\ }\href {\doibase 10.1051/0004-6361/201220873}
  {\bibfield  {journal} {\bibinfo  {journal} {Astron. Astrophys.}\ }\textbf
  {\bibinfo {volume} {556}},\ \bibinfo {pages} {A2} (\bibinfo {year} {2013})},\
  \Eprint {http://arxiv.org/abs/1305.3550} {arXiv:1305.3550} \BibitemShut
  {NoStop}%
\bibitem [{\citenamefont {Koopmans}\ \emph {et~al.}(2015)\citenamefont
  {Koopmans} \emph {et~al.}}]{Koopmans:2015sua}%
  \BibitemOpen
  \bibfield  {author} {\bibinfo {author} {\bibfnamefont {L.~V.~E.}\
  \bibnamefont {Koopmans}} \emph {et~al.},\ }\href {\doibase
  10.22323/1.215.0001} {\bibfield  {journal} {\bibinfo  {journal} {PoS}\
  }\textbf {\bibinfo {volume} {AASKA14}},\ \bibinfo {pages} {001} (\bibinfo
  {year} {2015})},\ \Eprint {http://arxiv.org/abs/1505.07568} {arXiv:1505.07568
  [astro-ph.CO]} \BibitemShut {NoStop}%
\end{thebibliography}%

\newpage

\phantomsection
\section{Appendices}
\label{app:App}
\setcounter{secnumdepth}{2}
\renewcommand{\thesubsection}{\Alph{subsection}}
\setcounter{subsection}{0}

\subsection{Comparison with Literature}
\label{app:comparison_previous_literature}

\begin{figure}[h!]
    \centering
    \includegraphics[width=\linewidth]{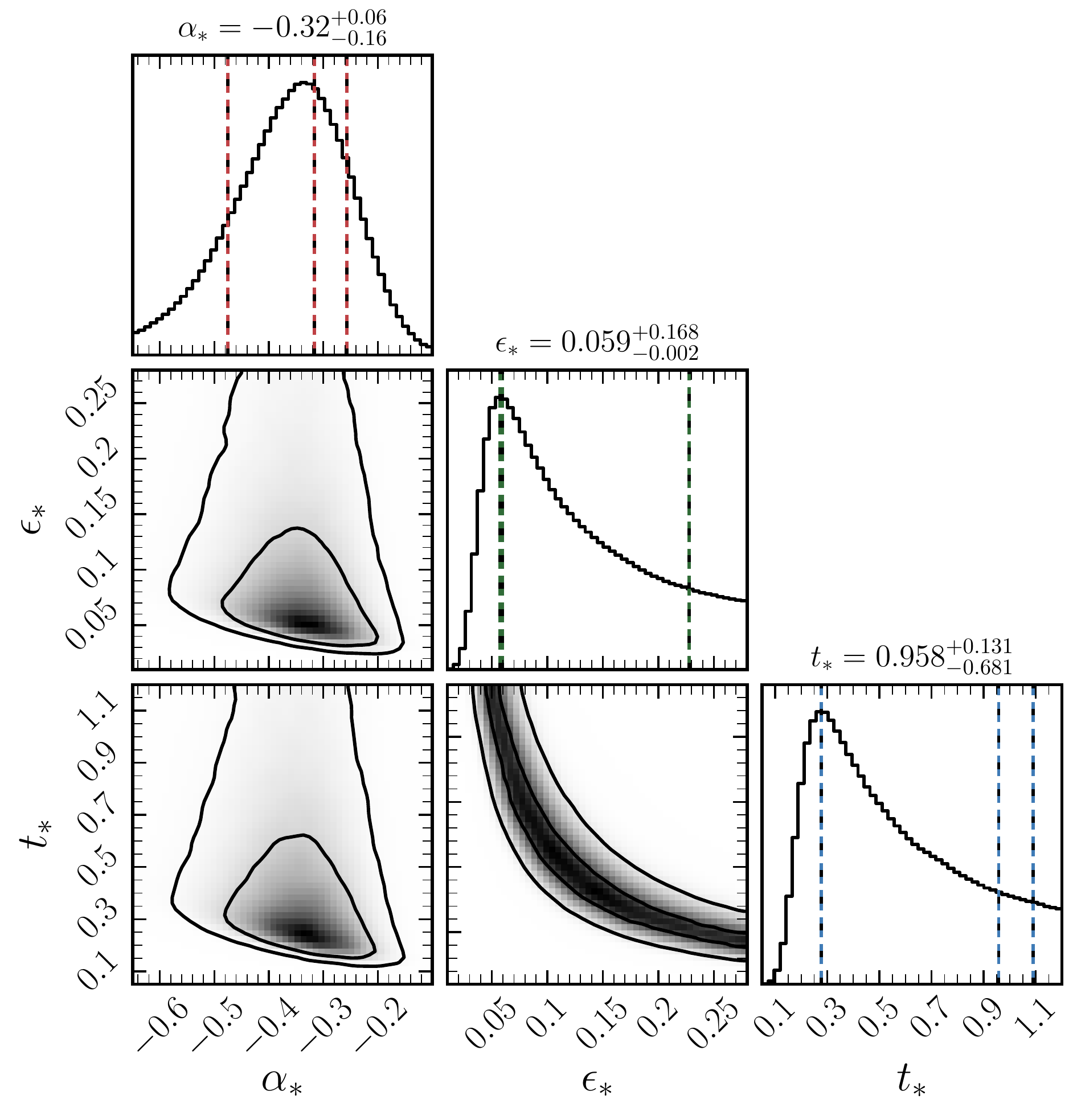}
    \caption{Posteriors on  $\alpha_*,\ \epsilon_*\ \mathrm{and}\ t_*$ using the `B+' benchmark data set specified in~\cite{Gillet:2019fjd}, which consists of data at $z = 6$ in the magnitude range $-20\leq M_\mathrm{UV}\leq -15$. The titles and vertical lines in the 1D posteriors depict the maximum-likelihood best fit (middle line) and the $\pm1\sigma$ quantiles (outer lines).
    }
    \label{fig:MesingerComparison}
\end{figure}

Here we perform a comparison of our results with~\cite{Gillet:2019fjd}.
We use the `B+' data set from~\cite{Gillet:2019fjd}, and compare against their result in Figure E2, where the UV LF data used is from~\cite{Bouwens_2017asdasd}, restricted to the magnitude range $-20\leq M_\mathrm{UV}\leq-15$. In this reference a single power-law is employed to describe the halo mass$-$stellar mass relation, instead of the double power-law in our approach (Eq.~\eqref{eq:Mh_Mstar_doublepower_approx}). In addition, while in Eq.~\eqref{eq:Mstardot_Mstar_relation} we fixed $t_* = 1$, we allow it to vary here. It means that for the purpose of comparison there are three free parameters: $\alpha_*$, $\epsilon_*$ (corresponding to $f_*$ in~\cite{Gillet:2019fjd}) and $t_*$. Note that we do not include the parameter $M_\mathrm{t}$, as it is only relevant for the smallest halos. Additionally, there is a minus sign difference between our definition of $\alpha_*$ and that of~\cite{Gillet:2019fjd}, and we translate their $f_*$ into $\epsilon_*$ in our framework using the following relation:
\begin{align}
\epsilon_* &= f_* \times \dfrac{\Omega_\mathrm{b}}{\Omega_\mathrm{m}} \left(\dfrac{M_\mathrm{c}}{10^{10}M_\odot}\right)^{-\alpha_*}\ ,
\end{align}
where $M_\mathrm{c} = 1.6\times 10^{11}\, M_\odot$. While this reference does not explicitly mention the best-fit values, a rough estimate can be obtained at the point where their 1D posterior is maximal. This gives $\alpha_*\approx 0.4$, $f_*\approx0.1$ (which translates into $\epsilon_*= 0.1 \times 0.157 \times 16^{0.4} \approx 0.048$) and $t_*\approx 1$ (since this parameter is highly degenerate with $f_*$). We performed an MCMC simulation and show the results in Figure~\ref{fig:MesingerComparison}. We find overall good agreement in both the best-fits and degeneracies between the parameters.
In particular, our contours for the power-law index $\alpha_*$, which is the least degenerate parameter, agree well with~\cite{Gillet:2019fjd}.
Note that here we considered a broader prior on $t_*=[0,2.5]$ than~\cite{Gillet:2019fjd} and did not assume a log-flat prior on $\epsilon_*$ (to keep consistency with the analysis in the main text).

\subsection{Fit for the Skewness}
\label{app:kappa3_fit}
The calculation of the skewness in Eq.~\eqref{eq:kappathree} is computationally expensive. Hence, we provide a fitting function that describes $\kappa_3$ for a wide range of halo masses, $M = 10^{5} - 10^{15} M_\odot$, and which we used in our main analysis with $k_\mathrm{cut} = 0.1\,\mathrm{Mpc}^{-1}$. It is given by:
\begin{align}
    \label{eq:kappa3_fit}
    \frac{\kappa_3(M)}{f_\mathrm{NL}} =&\ \left(A - B[\ln(M)]^C\right)\exp{\left(-\frac{D\, M}{\rho_\mathrm{M}\left(\frac{2\pi}{k_\mathrm{cut}}\right)^3}\right)}\times\nonumber\\
    &\times\frac{1}{E+F\tanh{\left(\frac{H\, M}{\rho_\mathrm{m}\left(\frac{2\pi}{k_\mathrm{cut}}\right)^3}\right)}}\ ,
\end{align}
where $\{A,B,C,D,E,F,H\} = \{-1.86\times10^{-9},\,-1.66\times10^{-13},\,2.59,\,-6.22,\,-3.95\times10^{-6},\,-132.59,\,1.56\times10^{-6}\}$. We emphasize that this fitting function is obtained using $k_\mathrm{cut} = 0.1\,\mathrm{Mpc}^{-1}$ and that some of these fitting parameters change with $k_\mathrm{cut}$. The fits for a number of different cut-off scales are shown in Figure~\ref{fig:kappa3_fit}. We also compared with the fitting function for $\kappa_3$ in~\cite{LoVerde:2011iz} (in the case of $k_\mathrm{cut} = 0$) and found good agreement.

\subsection{Higher-order Non-Gaussianity}
\label{app:higher_order_terms}
In this appendix we explore the impact of higher-order non-Gaussian terms on our bounds. The formalism employed in the main analysis expands both the gravitational potential (Eq.~\eqref{eq:phi_nonGaussian}) and the 1-point PDF (Eq.~\eqref{eq:PDF}) to first order in $f_\mathrm{NL}$. In principle, one should include higher-order terms proportional to $f_\mathrm{NL}^2,\, f_\mathrm{NL}^3,\, g_\mathrm{NL}\ \mathrm{and}\ \tau_\mathrm{NL}$ in both expansions. For the sake of simplicity, however, we only consider terms proportional to $f_\mathrm{NL}^2$ and $f_\mathrm{NL}^3$ in the PDF. Keeping in mind that $\kappa_3$ is proportional to $f_\mathrm{NL}$, the PDF is then given by:

\begin{align}
    \label{eq:pdf_higher_order}
    \rho(\nu, M) =& \frac{\exp(-\nu^2/2)}{(2\pi)^{1/2}}\left(1 + \frac{\kappa_3(M)H_3(\nu)}{6}\right.\nonumber\\
    +&\left.\frac{\kappa_3^2(M)H_6(\nu)}{72}+\frac{\kappa_3^3(M)H_9(\nu)}{1296}\right)\ ,
\end{align}

Just like before, this expression can be plugged in Eq.~\eqref{eq:collapse_fraction} to obtain the collapse fraction up to third order in $f_\mathrm{NL}$, which can be written as $F_\mathrm{NG}(M) = F_0(M) + F_1(M) + F_2(M) + F_3(M)$. The $F_i$'s read:
\begin{align}
    F_0 &= \frac{1}{2}\mathrm{erfc}\left(\frac{\nu_\mathrm{c}}{\sqrt{2}}\right)\\
    F_1 &= \frac{\exp(-\nu_\mathrm{c}^2/2)}{(2\pi)^{1/2}}\frac{\kappa_3H_2(\nu_\mathrm{c})}{6}\\
    F_2 &= \frac{\exp(-\nu_\mathrm{c}^2/2)}{(2\pi)^{1/2}}\frac{\kappa_3^2H_5(\nu_\mathrm{c})}{72}\\
    F_3 &= \frac{\exp(-\nu_\mathrm{c}^2/2)}{(2\pi)^{1/2}}\frac{\kappa_3^3H_8(\nu_\mathrm{c})}{1296}\ ,
\end{align}

and their derivatives are given by:
\begin{align}
    F_0' &= -\frac{\exp(-\nu_\mathrm{c}^2/2)}{(2\pi)^{1/2}}\nu_\mathrm{c}'\\
    \frac{F_1'}{F_0'} &= \frac{\kappa_3H_3(\nu_\mathrm{c})}{6} - \frac{H_2(\nu_\mathrm{c})}{6}\frac{\kappa_3'}{\nu_\mathrm{c}'}\\
    \frac{F_2'}{F_0'} &= \frac{\kappa_3^2H_6(\nu_\mathrm{c})}{72} - \frac{\kappa_3H_5(\nu_\mathrm{c})}{36}\frac{\kappa_3'}{\nu_\mathrm{c}'}\\
    \frac{F_3'}{F_0'} &= \frac{\kappa_3^3H_9(\nu_\mathrm{c})}{1296} - \frac{\kappa_3^2H_8(\nu_\mathrm{c})}{432}\frac{\kappa_3'}{\nu_\mathrm{c}'}\ .
\end{align}
Finally, the correction to the Gaussian mass function up to third order in $f_\mathrm{NL}$ is then given by:
\begin{align}
    \label{eq:higher_order_Edgeworth_correction}
    \frac{n_\mathrm{NG}'}{n_\mathrm{G}'}  \approx 1 + \frac{F_1' + F_2'+F_3'}{F_0'}\ .
\end{align}

\begin{figure}[h!]
    \centering
    \includegraphics[width=\linewidth]{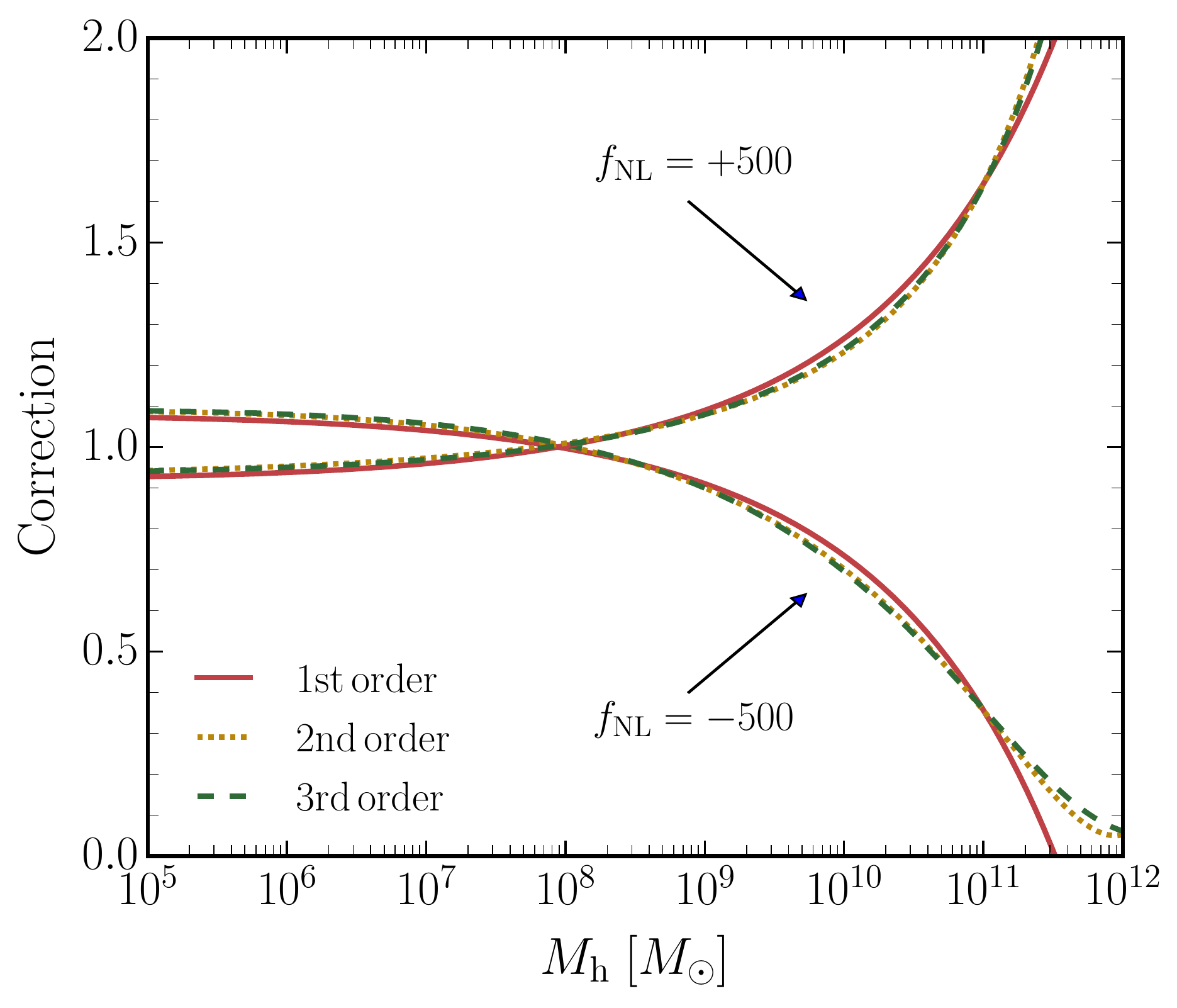}
    \caption{Corrections to the Gaussian halo mass function induced by small-scale primordial non-Gaussianity. A cut-off scale of $k_\mathrm{cut} = 0.1\,\Mpcinv$ is assumed in the bispectrum. The values $f_\mathrm{NL} = \pm500$ roughly encompass the $2\sigma$ bounds in the main analysis of this work (Section~\ref{sec:results}). The three different lines correspond to different order corrections in $f_\mathrm{NL}$ to the PDF (Eq.~\eqref{eq:pdf_higher_order}) and show that using an expansion to first order is already enough for obtaining bounds using Hubble galaxy luminosity function data.\vspace*{-0.4cm}}
    \label{fig:higher_order_terms}
\end{figure}

The individual contributions to this correction are shown in Figure~\ref{fig:higher_order_terms} for a benchmark case with $f_\mathrm{NL} = \pm 500$. These values encompass the $2\sigma$ bounds in our main analysis. It is clear that the higher-order terms only slightly alter the first-order correction to the halo mass function (and thus the UV luminosity function). Smaller values of $f_\mathrm{NL}$ lead to even smaller differences between the different curves. Moreover, we have performed a number of simulations using corrections with increasing order and found no shift in the bounds. This validates the expansion of both the bispectrum of the gravitational potential (Eq.~\eqref{eq:phi_nonGaussian_bispectrum}) and the PDF (Eq.~\eqref{eq:PDF}) only to first order in $f_\mathrm{NL}$ in our main analysis.

\subsection{Altered Dust Correction}
\label{app:dust}
The attenuation of the UV luminosity function due to dust extinction is mainly present at the bright end~\cite{Yung_2018, Vogelsberger_dust2020}. Therefore, it is conceivable that this effect and that of non-Gaussianity (lower right panel in Figure~\ref{fig:UVLF_params_dependence}) are degenerate to some extent. 
In this appendix we explore the impact of such possible degeneracy on our bounds. The main analysis in this work uses an empirical dust model that is calibrated using local galaxies and with an attenuation of the form~\cite{Smit:2012nf,Vogelsberger_dust2020}:

\begin{align}
    \label{eq:dust_general}
    \langle A_\mathrm{UV}\rangle = C_0 + 0.2\ln(10)\sigma_\beta^2C_1^2+C_1\langle\beta\rangle\ ,
\end{align}
where $C_0$ and $C_1$ are fitting parameters. We use the calibration from~\cite{Meurer:1999jj} (see Eq.~\eqref{eq:dust}), which is based on local star-forming galaxies and gives $C_0 = 4.43$ and $C_1 = 1.99$. 
Other references, e.g.~\cite{Siana:2009um,Casey:2014cqa, Castellano:2014lua, reddy2015mosdef, J_Bouwens_2016}, also use low-redshift probes and find slightly different fits. We have checked that our constraints do not change significantly when using different fits for $C_0$ and $C_1$. 
As such, instead of fixing the parameters $C_0$ and $C_1$ to different values, here we will perform a more general MCMC analysis and vary these along with $\alpha_*,\, \beta_*,\,  \epsilon_*$ and $f_\mathrm{NL}$. We allow negative values of $C_0$ and $C_1$, but avoid unrealistic physical situations by setting negative dust corrections equal to 0. 

\begin{figure}[h!]
    \centering
    \includegraphics[width=\linewidth]{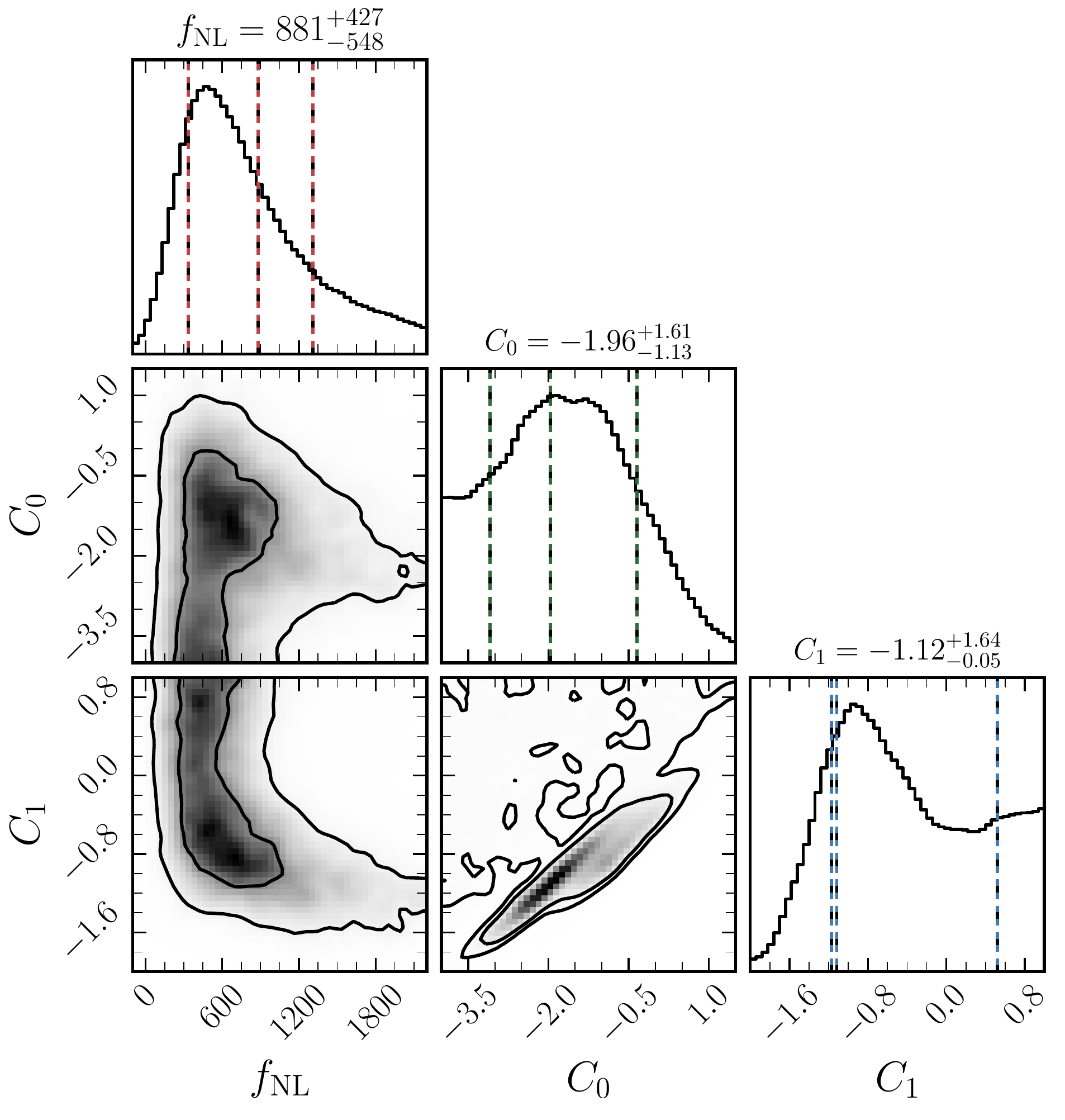}
    \caption{Posteriors on $f_\mathrm{NL}$ and the dust extinction parameters $C_0$ and $C_1$ (see Eq.~\eqref{eq:dust_general}). Data set 1 (Section~\ref{subsec:UVLF_data}) is used here. The titles and vertical lines in the 1D posteriors depict the maximum likelihood best-fit (middle line) and the $\pm1\sigma$ quantiles (outer lines).
    }
    \label{fig:dust}
\end{figure}

The results of our MCMC search are shown in Figure~\ref{fig:dust}, where only the posteriors for $f_\mathrm{NL},\, C_0$ and $C_1$ are displayed for clarity (although all parameters are varied). 
Altering the dust extinction in the analysis has a notable impact on our constraints and results in $f_\mathrm{NL} = 881^{+1114}_{-779}$ at $2\sigma$ (to be compared with $f_{\rm NL}=71^{+426}_{-237}$ and Figure~\ref{fig:MCMC_Bouwens2015zAll} where $C_0$ and $C_1$ are fixed).
However, the most striking feature of this analysis is that the preferred values for the extinction parameters are very different than those from local star-forming galaxies. 
The best-fit $C_0$ and $C_1$ are negative, which indicates that smaller galaxies suffer from more dust extinction.
This is in conflict with both local data-sets and physical intuition. We found that this is due to the fact that we are varying $f_\mathrm{NL}$ along with the astrophysical parameters. By performing an MCMC without including $f_\mathrm{NL}$, the best-fit values of $C_0$ and $C_1$ are more in line with those from low-redshift probes. If we do vary $f_\mathrm{NL}$ and restrict $C_0$ and $C_1$ to positive values, we obtain best-fits $C_0 = 2.9$ and $C_1 = 1.9$, in much better agreement with low-redshift observations. As a consequence, given that there already exist several observations of the extinction parameters, Figure~\ref{fig:dust} ought to be thought of as a worst-case scenario, in which no information on the dust extinction is known. As such, our results in the main text represent a more grounded picture.
Nevertheless, it is clear that an accurate observational determination of the attenuation is of importance when setting constraints on small-scale non-Gaussianity.

\vspace{-0.4cm}

\subsection{Results using HFF Data}
\label{app:alternative_dataset_results}

\begin{figure}[H]
    \centering
    \includegraphics[width=\linewidth]{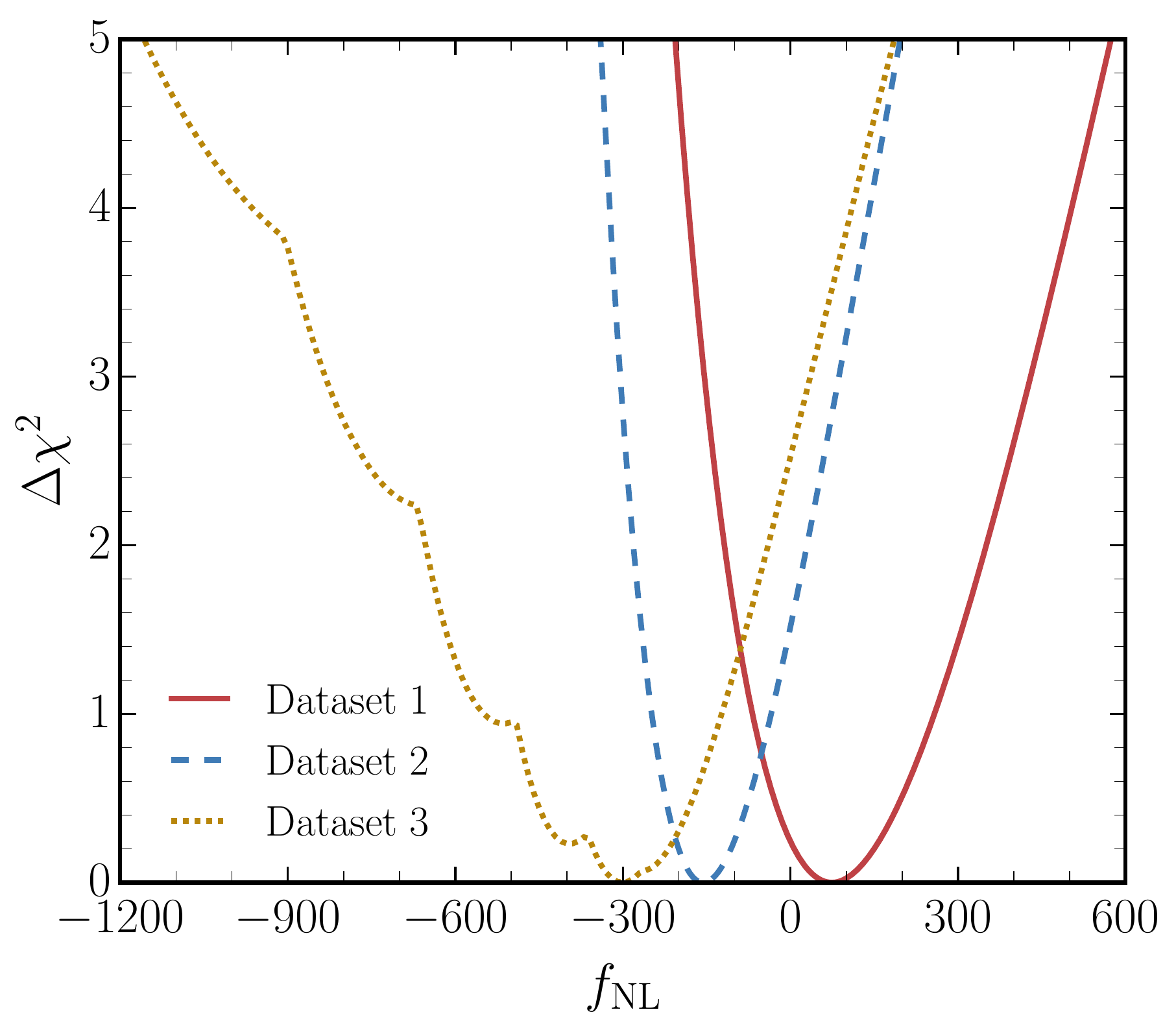}
    \caption{Marginalized $\Delta\chi^2$ as a function of $f_\mathrm{NL}$ for all three data sets described in Section~\ref{subsec:UVLF_data}.
    In all cases we have set $k_{\rm cut}=0.1\,\Mpcinv$.}
    \label{fig:chisq_constr_otherdatasets}
\end{figure}

Given that the main results of this work are based on data from the Hubble Legacy Fields catalog, we devote this appendix to constraints obtained with UV luminosity function observations from the Hubble Frontier Fields (data sets 2 and 3 described in Section~\ref{subsec:UVLF_data}). In Figure~\ref{fig:chisq_constr_otherdatasets} the accompanying marginalized $\chi^2$ to the constraints in Table~\ref{tab:HLF_bounds} are shown. In general, all three data sets give results that agree with each other to reasonable degree. The sudden jumps in the curve of data set 3 are a direct consequence of the mass cuts in the halo mass function correction (to prevent it from becoming negative). The reason why only this curve shows such behaviour is twofold: {\it i)} at higher redshift, the mass-cut scale is lower (i.e., the cut happens for smaller negative $f_\mathrm{NL}$) and {\it ii)} the lower-redshift data from~\cite{Livermore:2016mbs} has large errors, causing the $z=10$ data from~\cite{Oesch_2018} to have a more sizeable contribution to the overall $\chi^2$ (and thus making the jumps more prominent). The latter point can be proven by comparing $\chi^2_\text{best-fit} / g_\mathrm{dof}$ for data sets 2 and 3 (both share data from~\cite{Oesch_2018}), which gives a value of 1.3 and 0.4 respectively. The latter value is independent of the imposed 30\% minimum relative error on the data.\\

\subsection{Posteriors for JWST Forecast}
\label{app:JWST_posteriors}
\begin{figure*}[hbtp!]
    \centering
    \vspace{-0.2cm}
    \includegraphics[width=0.7\textwidth]{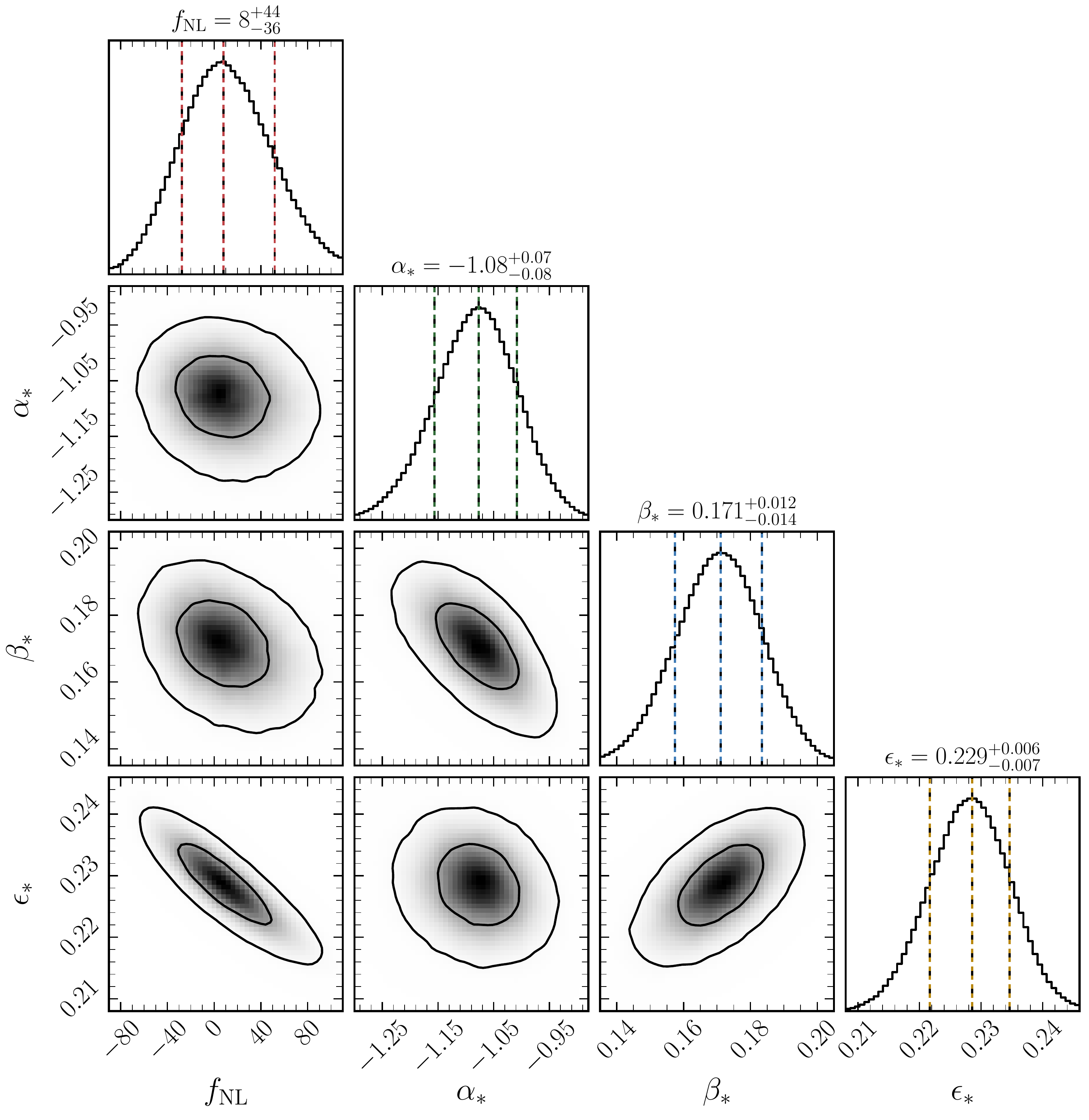}
    \vspace{-0.1cm}
    \caption{Posteriors for $\alpha_*,\, \beta_*,\,  \epsilon_*$ and $f_\mathrm{NL}$ using JWST mock data at $z = 4+5+6+7+8+9+10$. A minimal error of 10\% is assumed in the mock data points to account for cosmic variance. The cut-off scale in the bispectrum is set equal to $k_\mathrm{cut} = 0.1\,\mathrm{Mpc}^{-1}$. The 2D contours depict the $1\sigma$ and $2\sigma$ confidence levels. The titles and vertical lines in the 1D posteriors represent the maximum-likelihood best fit (middle line) and the $\pm1\sigma$ quantiles (outer lines).}
    \vspace{-0.18cm}
    \label{fig:MCMC_JWST}
\end{figure*}

In a completely analogous way as described in Section~\ref{sec:results}, we perform an MCMC simulation for the JWST forecast, using one realization of mock data according to the procedure detailed in Section~\ref{sec:forecasts}. The posteriors are shown in Figure~\ref{fig:MCMC_JWST}. We find that the bounds agree reasonably well with those in Table~\ref{tab:JWST} using the marginalized-$\chi^2$ method.

\subsection{Impact on Cosmic-dawn 21-cm}
\label{subsec:21cm_forecast}
We briefly study how the 21-cm signal during cosmic dawn would be altered in the presence of non-Gaussianities. 
This will provide a complementary approach to the UV luminosity functions, with different systematics and uncertainties.
We will not perform a full Fisher analysis of upcoming 21-cm data, but instead highlight the main effects of small-scale PNG on the 21-cm global signal and its fluctuations. 
This exploratory analysis, while less complete than our approach in Section~\ref{sec:forecasts}, will suffice to show an alternative way of probing small-scale PNG.\\
The same changes to the HMF that affect the UV LF at $z\sim4-10$ will leave a stronger imprint at the higher redshifts of cosmic dawn ($z\sim 10-30$). Moreover, as the star formation at such early times is dominated by lighter halos, we will have access to even smaller scales to probe PNG.
In that way our analysis here complements that of~\cite{Munoz:2019hjh}, as we study the impact of the small-scale bispectrum of fluctuations, as opposed to the power spectrum.\\
The physical picture is that the first stars formed in small halos and were able to cool gas through atomic- and molecular-line transitions~\cite{Machacek:2000us,Oh:2001ex,Yoshida:2003rw,Fialkov:2012su}.
These stars emitted UV photons and changed the state of the intergalactic medium through the Wouthuysen-Field effect~\cite{Wout,Field,Hirata:2005mz}, allowing neutral hydrogen to absorb 21-cm photons. Subsequently, the first galaxies heated up the hydrogen through X-rays, which lead to the emission of 21-cm radiation. Eventually, UV photons were able to reionize all of the hydrogen. For a detailed review of these processes see, e.g.,~\cite{Pritchard:2011xb,loeb2013first}.\\
Given the range of halo masses that are relevant for cosmic dawn, $M_\mathrm{h}\sim 10^5-10^8 \,\Msun$, and the accessibility of 21-cm cosmic-dawn observations to wavenumbers as large as $k\approx 50\,\Mpcinv$~\cite{Munoz:2019hjh}, we will impose a more restrictive cut on our bispectrum of $k_{\rm cut}=1\,\Mpcinv$ throughout this section.
Smaller values of $k_{\rm cut}$ will be at least similarly constrained. We choose $k_{\rm cut} = 1\,\Mpcinv$ also to showcase how the `anomaly' observed for this value in the HST UV LFs (see Section~\ref{subsec:other_kcuts}) would leave an observable signature on the 21-cm signal.\\
Altering the abundance of the first galaxies---by changing the HMF due to PNG---will shift the timing of all the events mentioned before and leave an imprint on the 21-cm signal.
We will use the HMF correction computed in Section~\ref{subsec:HMF_fNL_corrections} and apply it to {\tt 21cmvFAST} simulations~\cite{Munoz:2019rhi,Munoz:2019fkt}.
These are semi-numerical simulations based on the excursion-set approach of {\tt 21cmFAST}~\cite{Mesinger:2007pd,Mesinger:2010ne,Greig:2015qca,Qin:2020xyh}, modified however to include molecular cooling halos and the effect of the dark matter-baryon relative velocities (see~\cite{Munoz:2019rhi} for details on the implementation).
Here we just multiply the HMF in that code by Eq.~\eqref{eq:Edgeworth_correction} to include the effect of small-scale PNG and keep the same astrophysical parameters as in~\cite{Munoz:2019rhi}, with a simulation box of $L_{\rm box}=600$ Mpc and a cell size $R_{\rm cell}=3$ Mpc.
We emphasize that the large-scale initial conditions (both densities and relative velocities) of our simulations are kept Gaussian, as PNG only affects modes with $k>1\,\Mpcinv$.
Therefore, the only effect of small-scale PNGs will be to alter the abundance of the first galaxies.

\begin{figure}[t!]
    \centering
    \includegraphics[width=\linewidth]{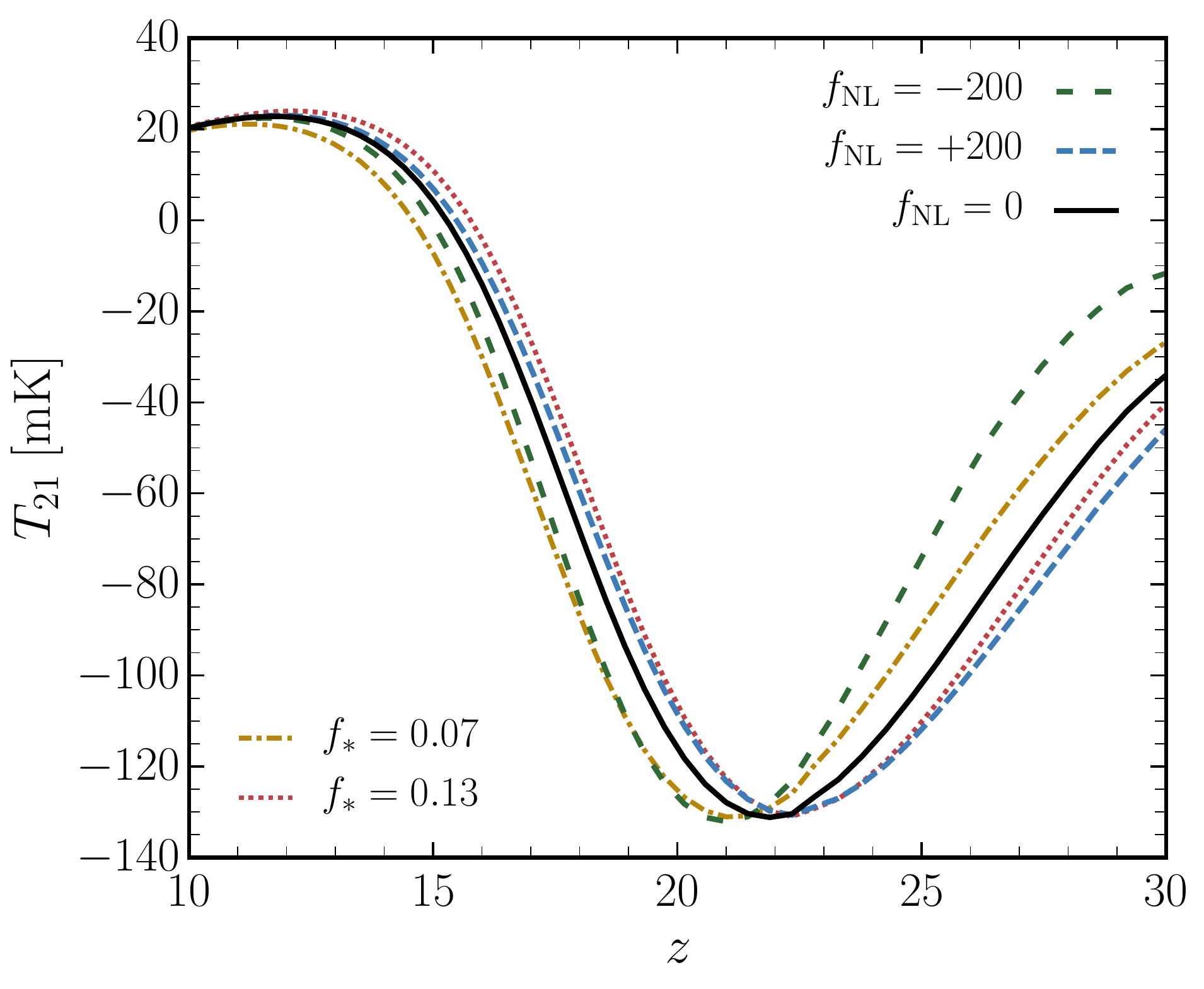}
    \vspace{-0.3cm}
    \caption{21-cm global signal as a function of redshift. The black line shows our fiducial case (with star fraction $f_*$ = 0.1), whereas the yellow dash-dotted and red dotted lines have 30\% lower or higher $f_*$.
    The green (blue) dashed line shows the global signal in the presence of negative (positive) small-scale PNG with $k_{\rm cut}=1\,\Mpcinv$, which delays (accelerates) the 21-cm landmarks.\vspace*{-0.3cm}
    }
    \label{fig:21cm_global}
\end{figure}

We show the 21-cm global signal in Figure~\ref{fig:21cm_global}, which corresponds to the total absorption or emission of 21-cm photons across the entire sky.
This signal shows the expected behavior: absorption ($T_{21}<0$) when the first stars form at $z\sim 25$, a trough at $z\sim20$ when the X-ray heating starts and emission ($T_{21}>0$) by $z\sim 15$ when the gas is fully heated.
As is clear from Figure~\ref{fig:21cm_global}, these landmarks are delayed for negative $\fNL$, as fewer halos form, and vice-versa for positive $\fNL$.
This delay of the 21-cm landmarks can, in principle, be mimicked to some degree by a change in the unknown astrophysical parameters of cosmic dawn.
As an example, we vary the fraction of gas that turns into stars $f_*$ (with a fiducial value of 0.1) and show the resulting 21-cm signal in Figure~\ref{fig:21cm_global}.
As expected, the effect of PNG is more marked at high $z$, showing that the effects of $f_\mathrm{NL}$ and $f_*$ can be disentangled to some extent.
This is promising for global-signal experiments like EDGES~\cite{Bowman:2018yin}, SARAS~\cite{Singh:2017syr},  LEDA~\cite{LEDA},
SCHI-HI~\cite{Voytek:2013nua} or PRIzM~\cite{PRIZM}.

\vspace{6pt}

\begin{figure}[b!]
    \vspace{-0.3cm}
    \centering
    \includegraphics[width=\linewidth]{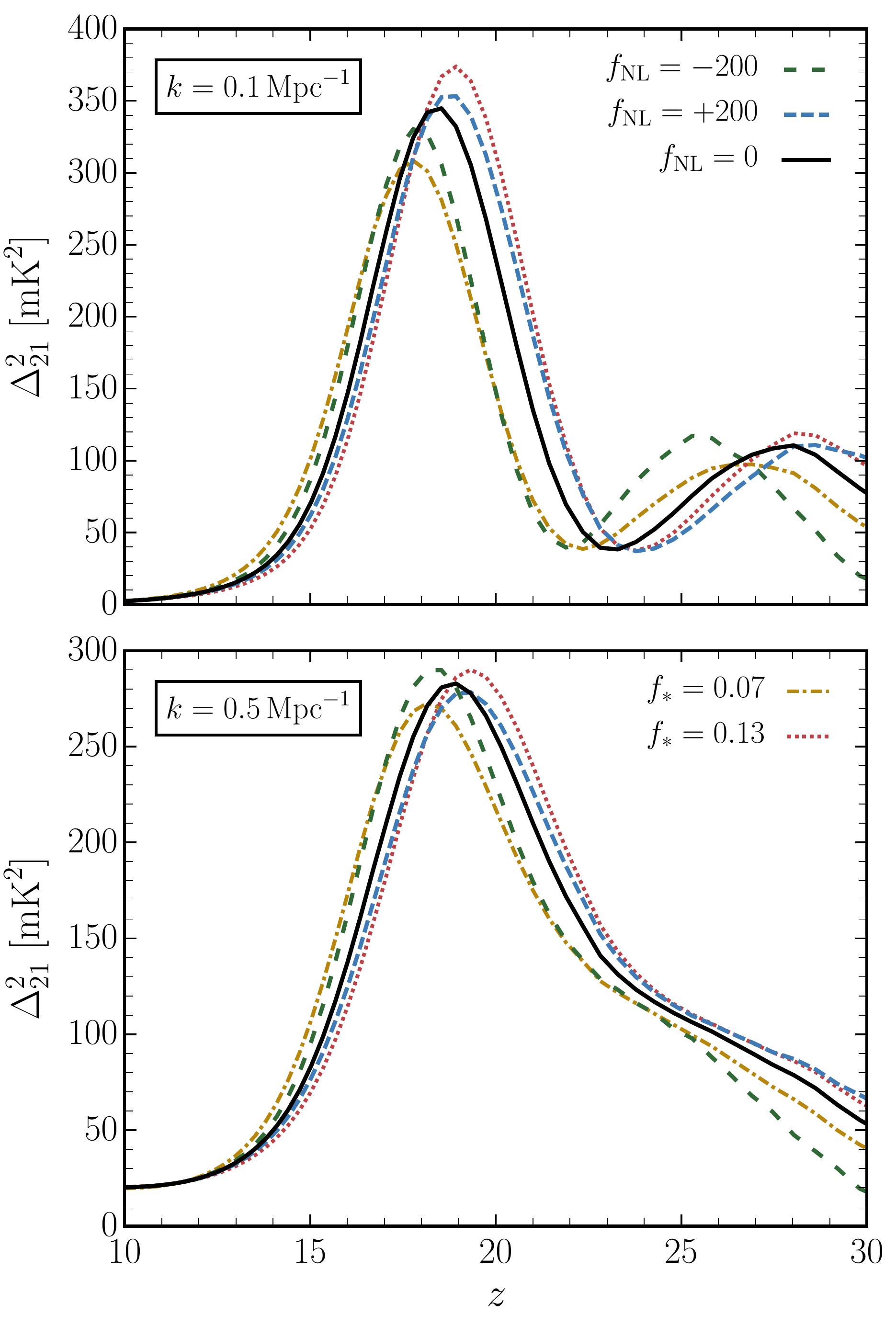}
    \caption{Evolution of the 21-cm power spectrum for wavenumbers $k = 0.1\,\mathrm{Mpc}^{-1}$ (\emph{top}) and $k = 0.5\,\mathrm{Mpc}^{-1}$ (\emph{bottom}). The different curves have the same meaning as in Figure~\ref{fig:21cm_global}. The PNG cut-off scale is $k_\mathrm{cut}=1\,\mathrm{Mpc}^{-1}$.
    }
    \label{fig:21cm_power}
\end{figure}

We now explore the effect of PNG on the 21-cm fluctuations.
We show in Figure~\ref{fig:21cm_power} the amplitude of 21-cm fluctuations $\Delta^2_{21}(k)$ at two wavenumbers, $k=0.1$ and 0.5 $\Mpcinv$.
These are chosen to be at the lowest edge of what can be observed due to foregrounds, and roughly in the middle where thermal noise dominates the HERA noise budget~\cite{Pober:2013ig}.
In both cases we see the same behaviour as in the global signal, with the 21-cm landmarks being either delayed or accelerated.
Here the difference between $\fNL$ and $f_*$ can be more readily seen, as even the heights of the curves differ.
This is potentially observable by upcoming interferometers, such as HERA~\cite{DeBoer:2016tnn} and LOFAR~\cite{vanHaarlem:2013dsa}, 
or in the future with SKA~\cite{Koopmans:2015sua}.

\vspace{6pt}

While detailed forecasts for each 21-cm experiment are beyond the scope of this work, our study shows that both the 21-cm global signal and its fluctuations 
appear to be sensitive to small-scale PNG at the level of $\fNL\sim 10^2$ (for $k_{\rm cut}=1\,\Mpcinv$).

\end{document}